\documentclass[]{aastex631}

\newcommand{\frb}[0]{FRB\,20180916B\,}
\def\gapp{\ifmmode\stackrel{>}{_{\sim}}\else$\stackrel{>}{_{\sim}}$\fi}

\shorttitle{Rate and morphology evolution of \frb}
\shortauthors{Sand et al.}
\graphicspath{{./}{figures/}}

\usepackage{soul}
\usepackage{upgreek}

\begin{document}

\title{A CHIME/FRB study of burst rate and morphological evolution of  the periodically repeating \frb}

\correspondingauthor{Ketan R. Sand}
\email{ketan.sand@mail.mcgill.ca}

\author[0000-0003-3154-3676]{Ketan R. Sand}
\affiliation{Department of Physics, McGill University, 3600 rue University, Montr\'eal, QC H3A 2T8, Canada}
\affiliation{Trottier Space Institute, McGill University, 3550 rue University, Montr\'eal, QC H3A 2A7, Canada}

\author[0000-0002-2349-3341]{Daniela Breitman}
\affiliation{Scuola Normale Superiore, Piazza dei Cavalieri 7, I-56126 Pisa, Italy}

\author[0000-0002-2551-7554]{Daniele Michilli}
\affiliation{MIT Kavli Institute for Astrophysics and Space Research, Massachusetts Institute of Technology, 77 Massachusetts Ave, Cambridge, MA 02139, USA}
\affiliation{Department of Physics, Massachusetts Institute of Technology, 77 Massachusetts Ave, Cambridge, MA 02139, USA}

\author[0000-0001-9345-0307]{Victoria M. Kaspi}
\affiliation{Department of Physics, McGill University, 3600 rue University, Montr\'eal, QC H3A 2T8, Canada}
\affiliation{Trottier Space Institute, McGill University, 3550 rue University, Montr\'eal, QC H3A 2A7, Canada}

\author[0000-0002-3426-7606]{Pragya Chawla}
\affiliation{Anton Pannekoek Institute for Astronomy, University of Amsterdam, Science Park 904, 1098 XH Amsterdam, The Netherlands}

\author[0000-0001-8384-5049]{Emmanuel Fonseca}
\affiliation{Center for Gravitational Waves and Cosmology, West Virginia University, Chestnut Ridge Research Building, Morgantown, WV 26505, USA}
\affiliation{Department of Physics and Astronomy, West Virginia University, P.O. Box 6315, Morgantown, WV 26506, USA}

\author[0000-0001-7348-6900]{Ryan Mckinven}
\affiliation{Department of Physics, McGill University, 3600 rue University, Montr\'eal, QC H3A 2T8, Canada}
\affiliation{Trottier Space Institute, McGill University, 3550 rue University, Montr\'eal, QC H3A 2A7, Canada}

\author[0000-0003-0510-0740]{Kenzie Nimmo}
\affiliation{MIT Kavli Institute for Astrophysics and Space Research, Massachusetts Institute of Technology, 77 Massachusetts Ave, Cambridge, MA 02139, USA}

\author[0000-0002-4795-697X]{Ziggy Pleunis}
\affiliation{Dunlap Institute for Astronomy \& Astrophysics, University of Toronto, 50 St.~George Street, Toronto, ON M5S 3H4, Canada}

\author[0000-0002-6823-2073]{Kaitlyn Shin}
\affiliation{MIT Kavli Institute for Astrophysics and Space Research, Massachusetts Institute of Technology, 77 Massachusetts Ave, Cambridge, MA 02139, USA}
\affiliation{Department of Physics, Massachusetts Institute of Technology, 77 Massachusetts Ave, Cambridge, MA 02139, USA}

\author[0000-0001-5908-3152]{Bridget C. Andersen}
\affiliation{Department of Physics, McGill University, 3600 rue University, Montr\'eal, QC H3A 2T8, Canada}
\affiliation{Trottier Space Institute, McGill University, 3550 rue University, Montr\'eal, QC H3A 2A7, Canada}

\author[0000-0002-3615-3514]{Mohit Bhardwaj}
\affiliation{Department of Physics, Carnegie Mellon University, 5000 Forbes Avenue, Pittsburgh, 15213, PA, USA}

\author[0000-0001-8537-9299]{P. J. Boyle}
\affiliation{Department of Physics, McGill University, 3600 rue University, Montr\'eal, QC H3A 2T8, Canada}
\affiliation{Trottier Space Institute, McGill University, 3550 rue University, Montr\'eal, QC H3A 2A7, Canada}

\author[0000-0002-1800-8233]{Charanjot Brar}
\affiliation{Department of Physics, McGill University, 3600 rue University, Montr\'eal, QC H3A 2T8, Canada}
\affiliation{Trottier Space Institute, McGill University, 3550 rue University, Montr\'eal, QC H3A 2A7, Canada}

\author[0000-0003-2047-5276]{Tomas Cassanelli}
\affiliation{Department of Electrical Engineering, Universidad de Chile, Av. Tupper 2007, Santiago 8370451, Chile}

\author[0000-0001-6422-8125]{Amanda M. Cook}
\affiliation{Dunlap Institute for Astronomy \& Astrophysics, University of Toronto, 50 St.~George Street, Toronto, ON M5S 3H4, Canada}
\affiliation{David A.~Dunlap Department of Astronomy \& Astrophysics, University of Toronto, 50 St.~George Street, Toronto, ON M5S 3H4, Canada}

\author[0000-0002-8376-1563]{Alice P. Curtin}
\affiliation{Department of Physics, McGill University, 3600 rue University, Montr\'eal, QC H3A 2T8, Canada}
\affiliation{Trottier Space Institute, McGill University, 3550 rue University, Montr\'eal, QC H3A 2A7, Canada}

\author[0000-0003-4098-5222]{Fengqiu Adam Dong}
\affiliation{Department of Physics and Astronomy, University of British Columbia, 6224 Agricultural Road, Vancouver, BC V6T 1Z1 Canada}

\author[0000-0003-3734-8177]{Gwendolyn M. Eadie}
\affiliation{David A.~Dunlap Department of Astronomy \& Astrophysics, University of Toronto, 50 St.~George Street, Toronto, ON M5S 3H4, Canada}
\affiliation{Department of Statistical Sciences, University of Toronto, 700 University Ave 9th Floor, Toronto, ON M5G 1X6, Canada}

\author[0000-0002-3382-9558]{B. M. Gaensler}
\affiliation{Dunlap Institute for Astronomy \& Astrophysics, University of Toronto, 50 St.~George Street, Toronto, ON M5S 3H4, Canada}
\affiliation{David A.~Dunlap Department of Astronomy \& Astrophysics, University of Toronto, 50 St.~George Street, Toronto, ON M5S 3H4, Canada}
\affiliation{Present address: Division of Physical and Biological Sciences, University of California Santa Cruz, Santa Cruz, CA 95064, USA}

\author[0000-0003-4810-7803]{Jane Kaczmarek}
\affiliation{CSIRO Space and Astronomy, Parkes Observatory, P.O. Box 276, Parkes NSW 2870, Australia}
\affiliation{Department of Computer Science, Math, Physics and Statistics, University of British Columbia, Kelowna, BC V1V 1V7, Canada}

\author[0000-0003-2116-3573]{Adam Lanman}
\affiliation{Department of Physics, McGill University, 3600 rue University, Montr\'eal, QC H3A 2T8, Canada}
\affiliation{Trottier Space Institute, McGill University, 3550 rue University, Montr\'eal, QC H3A 2A7, Canada}

\author[0000-0002-4209-7408]{Calvin Leung}
\affiliation{MIT Kavli Institute for Astrophysics and Space Research, Massachusetts Institute of Technology, 77 Massachusetts Ave, Cambridge, MA 02139, USA}
\affiliation{Department of Physics, Massachusetts Institute of Technology, 77 Massachusetts Ave, Cambridge, MA 02139, USA}
\affiliation{NHFP Einstein Fellow}

\author[0000-0002-4279-6946]{Kiyoshi W. Masui}
\affiliation{MIT Kavli Institute for Astrophysics and Space Research, Massachusetts Institute of Technology, 77 Massachusetts Ave, Cambridge, MA 02139, USA}
\affiliation{Department of Physics, Massachusetts Institute of Technology, 77 Massachusetts Ave, Cambridge, MA 02139, USA}

\author[0000-0003-1842-6096]{Mubdi Rahman}
\affiliation{Sidrat Research, 124 Merton Street, Suite 507, Toronto, ON M4S 2Z2, Canada}

\author[0000-0002-8897-1973]{Ayush Pandhi}
\affiliation{Dunlap Institute for Astronomy \& Astrophysics, University of Toronto, 50 St.~George Street, Toronto, ON M5S 3H4, Canada}
\affiliation{David A.~Dunlap Department of Astronomy \& Astrophysics, University of Toronto, 50 St.~George Street, Toronto, ON M5S 3H4, Canada}

\author[0000-0002-8912-0732]{Aaron B. Pearlman}
\affiliation{Department of Physics, McGill University, 3600 rue University, Montr\'eal, QC H3A 2T8, Canada}
\affiliation{Trottier Space Institute, McGill University, 3550 rue University, Montr\'eal, QC H3A 2A7, Canada}

\author[0000-0002-9822-8008]{Emily Petroff}
\affiliation{Perimeter Institute for Theoretical Physics, 31 Caroline Street N, Waterloo, ON N25 2YL, Canada}

\author[0000-0001-7694-6650]{Masoud Rafiei-Ravandi}
\affiliation{Department of Physics, McGill University, 3600 rue University, Montr\'eal, QC H3A 2T8, Canada}
\affiliation{Trottier Space Institute, McGill University, 3550 rue University, Montr\'eal, QC H3A 2A7, Canada}

\author[0000-0002-7374-7119]{Paul Scholz}
\affiliation{Dunlap Institute for Astronomy \& Astrophysics, University of Toronto, 50 St.~George Street, Toronto, ON M5S 3H4, Canada}
\affiliation{Department of Physics and Astronomy, York University, 4700 Keele Street, Toronto, Ontario, ON MJ3 1P3, Canada}

\author[0000-0002-4823-1946]{Vishwangi Shah}
\affiliation{Department of Physics, McGill University, 3600 rue University, Montr\'eal, QC H3A 2T8, Canada}
\affiliation{Trottier Space Institute, McGill University, 3550 rue University, Montr\'eal, QC H3A 2A7, Canada}

\author[0000-0002-2088-3125]{Kendrick Smith}
\affiliation{Perimeter Institute for Theoretical Physics, 31 Caroline Street N, Waterloo, ON N25 2YL, Canada}

\author[0000-0001-9784-8670]{Ingrid Stairs}
\affiliation{Department of Physics and Astronomy, University of British Columbia, 6224 Agricultural Road, Vancouver, BC V6T 1Z1 Canada}

\author[0000-0002-9761-4353]{David C. Stenning}
\affiliation{Department of Statistics and Actuarial Science, Simon Fraser University, 8888 University Dr, Burnaby, BC V5A 1S6, Canada}





\begin{abstract}

\frb is a repeating Fast Radio Burst (FRB) with a 16.3-day periodicity in its activity. In this study, we present morphological properties of 60 \frb bursts detected by CHIME/FRB between 2018 August and 2021 December. We recorded raw voltage data for 45 of these bursts, enabling microseconds time resolution in some cases. We studied variation of spectro-temporal properties with time and activity phase. We find that the variation in Dispersion Measure (DM) is $\lesssim$1 pc cm$^{-3}$ and that there is burst-to-burst variation in scattering time estimates ranging from $\sim$0.16 to over 2 ms, with no discernible trend with activity phase for either property. Furthermore, we find no DM and scattering variability corresponding to the recent change in rotation measure from the source, which has implications for the immediate environment of the source. We find that \frb has thus far shown no epochs of heightened activity as have been seen in other active repeaters by CHIME/FRB, with its burst count consistent with originating from a Poissonian process. We also observe no change in the value of the activity period over the duration of our observations and set a 1$\sigma$ upper limit of $1.5\times10^{-4}$ day day$^{-1}$ on the absolute period derivative. Finally, we discuss constraints on progenitor models yielded by our results, noting that our upper limits on changes in scattering and dispersion measure as a function of phase do not support models invoking a massive binary companion star as the origin of the 16.3-day periodicity.
 
\end{abstract}



\section{Introduction} \label{sec:intro}
Fast Radio Bursts (FRBs) are  millisecond-duration radio pulses of extragalactic origins (see \citealp{petroff_dawn_2022} for a recent review on FRBs). Their properties have inspired numerous theories to explain their origin (see \citealp{platts2019living} for review\footnote{\url{https://frbtheorycat.org/index.php/Main_Page}}). A luminous radio burst detected from SGR J1935+2154 \citep{andersen2020bright,bochenek2020fast} in 2020 April suggests that magnetars are likely progenitors for some of these sources. 
Some FRBs are seen to repeat i.e, multiple bursts from the same source. After the publication of the Canadian Hydrogen Intensity Mapping Experiment/Fast Radio Bursts (CHIME/FRB) first catalogue \citep{chimefrb_first_cat_2021}, the number of published FRBs has surpassed over 600. Out of which 51 show repeat bursts \citep{chimefrb_RN1_2019,Fonseca2020_RN2,RN3} in subsequent observations.


\frb is one of the first repeating sources discovered by the CHIME/FRB collaboration \citep{chimefrb_RN1_2019}, who later showed it exhibits a 16.33-day periodicity in its activity \citep{periodic_chime2020}. It has been localized to a nearby spiral galaxy approximately 150 Mpc away by \cite{mar2020}. The source is active for approximately five days in its 16.33-day cycle \citep{periodic_chime2020}. This activity is frequency-dependent, with higher frequency detections occurring at earlier phases \citep{pastormarazuela2020chromatic_180916,Pleunis_2021_r3}. The source has been detected over a wide frequency range, from  110 MHz by the Low-Frequency Array (LOFAR) \citep{pastormarazuela2020chromatic_180916,Pleunis_2021_r3,Gopinath_R3_2023} to 6 GHz by the Effelsberg telescope \citep{surya2022} at different activity phases. This chromatic periodicity has led to multiple progenitor scenarios involving precessing \cite[e.g.,][]{Li_emission_chromatic}, rotating \cite[e.g.,][]{Paz_period_magnetar} or binary systems \cite[e.g.,][]{Wada_bin_comb}. The first repeating source, FRB\,20121102A, has also shown a tentative activity period of around 160 days \citep{Rajwade_2020_121102_period,cruces_R1_period}.

Understanding the evolution of source properties with time can help further narrow down the models capable of explaining the activity period and emission origins. The distinctive ability of CHIME/FRB to monitor the entire Northern sky at a daily cadence makes it an ideal instrument for such a study. Rotation measure (RM) variation has been observed for FRB\,20121102A over 2.5 years, decreasing by almost 34\% its initial value \citep{mic2018,hilmarsson2021rotation}. This variation was accompanied by a four pc\,cm$^{-3}$ increase in the dispersion measure (DM). The RM of \frb is four orders of magnitude less than that of FRB\,20121102A and was fairly stable until 2021 April. \cite{mckinven_2022_r3} reported a linear increase in the RM of \frb in their recent CHIME/FRB polarimetric data analysis. No significant DM changes have been observed for the source so far. \cite{Gopinath_R3_2023} also observed these changes at LOFAR frequencies (150 MHz). This suggests a change in the magnetic field environment in the source proximity that is independent of the mechanism causing the periodic activity. 

Scattering is another crucial property useful in understanding the source environment.
Most FRB scattering times are longer than what is expected from their Galactic contribution \citep{Cordes_2016,cordes2019fast}. \cite{chawla_scattering_dm} showed that scattering in FRBs is likely due to their more extreme circumburst environments compared to those of pulsars in the Galactic plane. Contributions from the circumgalactic medium (CGM) of intervening galaxies can also add to the total measured scattering \citep{vedantham_phinney_2019}. \frb has a low Galactic latitude and most of its scattering, as determined using scintillation bandwidth measurements, has been attributed to the Milky Way  \citep{mar2020,surya2022}. However, bursts with scattering tails considerably longer than the expected value from the Milky Way intestellar medium (ISM) have been observed in low frequency observations, suggesting possible contributions from the immediate source environment \citep{Pleunis_2021_r3,sand2022,Gopinath_R3_2023}. Thus understanding the long-term and phase-dependent variation in scattering behavior of the source can in principle help better characterize its immediate environment. 

Repeating FRBs tend to have complex burst morphologies \citep{chimefrb_RN1_2019,Fonseca2020_RN2,RN3}. In particular, the downward-drifting ``sad trombone'' effect is a characteristic feature of their dynamic spectra \citep{Hes2019,pleunismorph}. This may sometimes prevent the accurate estimation of both DM and scattering values. \frb exhibits microstructure at 1.7 GHz as narrow as 3 $\upmu$s \citep{nimmo2021_micro} and down to 800 MHz as narrow as 30 $\upmu$s \citep{sand2022}. Such structures might remain hidden at lower time resolution and can contaminate scattering estimates. Moreover, studying such structures can constrain the source's emission models. Microstructures as narrow as 60 ns have been seen for FRB 20200120E \citep{nimmo2022burst} at 1.4 GHz, suggesting a magnetically powered emission mechanism that can generate energies comparable to FRBs at $\sim\,\upmu s$ timescales.  Past studies in the CHIME band \citep{periodic_chime2020,Pleunis_2021_r3} of this source have had insufficient time resolution to detect microstructure.  \frb's less turbulent immediate environment compared to other active repeaters such as FRB\,20121102A \citep{Pleunis_2021_r3} makes it an ideal source for such a study.  Identifying microsecond variations in the burst envelope to accurately estimate the DM and scattering requires high time resolution, available only with raw voltage data that can be coherently dedispersed. By `raw voltage' we mean the complex voltage induced by the incoming electromagnetic radiation in the telescope feed. This allows us to extract phase information. We can then use this to completely remove the effects of dispersion by coherent dedispersion to out best estimate of DM. The CHIME/FRB baseband system \citep{chime18_overview,Baseband} provides such functionality, enabling the comprehensive study of these ultra-fast emissions on a population scale and for individual repeating sources down to 2.56 $\upmu$s. 


Repetition rates of repeaters are useful in understanding their activity which can give clues into possible progenitors. For example, a rate of more than 100 bursts per hour has been observed for FRB\,20121102A \citep{FAST_121102A} at 1.3 GHz. Such high rates put limits on the energy budget of the source. Sources like FRB 20201124A \citep{lanman_20201124A} and FRB 20220912A \citep{Mckinven_R117} have shown sudden activity with repetition rates of tens to hundreds of bursts per hour, respectively. Although transiting daily in the CHIME/FRBs field of view, no burst above our detection threshold had been detected from these sources despite a lot of exposure before their heightened activity periods. \frb, on the other hand, has been detected continuously at the expected period since its discovery by CHIME/FRB in 2018. However, a comprehensive study on the evolution of its long-timescale emission rate until now has only been done at LOFAR frequencies \citep{Gopinath_R3_2023}.  

Here we present a morphology analysis of 45 bursts from \frb detected by the CHIME/FRB baseband system from its first detection in 2018 September 16 to 2021 December 31. We use the same baseband burst sample published by \cite{mckinven_2022_r3}, where they report the polarization properties. In addition, we present burst properties for 15 bursts that have been detected only with intensity data since the last published results for the source \citep{Pleunis_2021_r3}. We also analyze the evolution of the burst rate of the source and the burst count distribution with respect to its activity cycles in the aforementioned date range. In Section \ref{sec:obs}, we provide a brief outline of CHIME/FRB and the CHIME/FRB baseband system used for our observations. In Section \ref{sec:results}, we present the analysis and the results, providing a detailed outline of our Baseband morphology pipeline. In Section \ref{sec:disc}, we discuss the implications of our results. Lastly, we provide concluding remarks in Section \ref{sec:conc}.


\section{Observations} \label{sec:obs}

CHIME is an array of four stationary cylinders, each $\mathrm{100\,m\times20\,m}$ in dimensions, located near Penticton, British Columbia. Along the axis of each cylinder are hung 256 equispaced feeds recording dual polarization in the frequency range 400-800 MHz for the overhead sky \citep{chime_cosmo_overview}. The system was designed to map the distribution of neutral hydrogen in the redshift range 0.8-2.5 \citep{CHIME_stacking}. However, its daily Northern sky coverage and high sensitivity make it an excellent FRB detection machine. 

The CHIME/FRB backend uses the CHIME telescope to search for FRBs in real time as the sky transits \citep{chime18_overview}. This process involves a comprehensive set of stages, from dedispersion to Radio Frequency Interference (RFI) cleaning \citep{masoudrfi}, beam grouping, known source sifting and finally, data callbacks for signals deemed astrophysical. The system stores intensity data for all FRB sources at a resolution of 0.983 ms and 16384 frequency channels (i.e., 24.4 kHz in frequency resolution). A detailed analysis of 536 of these sources was published by \cite{chimefrb_first_cat_2021}. Optimized for FRB searching, this system is restricted in time resolution and does not store polarization information. 

For this reason, CHIME/FRB has a separate triggered baseband system \citep{chime18_overview}. This system records buffered raw voltage data when an FRB is detected above a specified S/N threshold. The system has a data buffer of 20 seconds, permitting the recording of CHIME's full bandwidth for a DM of upto 1000 pc cm$^{-3}$. This allows us to coherently dedisperse the voltage data around the detected DM and study the bursts at a time resolution of 2.56 $\upmu$s. More details on the pipeline can be found in \cite{Baseband}. The baseband data also permit analysis of Stokes parameters and extraction of polarization information, the specifics of which have been presented by \cite{mckinven_pol_pipe}. 

In this study, all the 45 baseband bursts from \frb were beamformed at the published VLBI position of the source \citep{mar2020}. Additionally, there are 15 bursts with only total intensity data. These were mostly low S/N and hence did not trigger the baseband system. 





\section{Analysis and Results} \label{sec:results}

All the new bursts in our dataset were identified using the clustering algorithm described by \cite{RN3}. Among the baseband burst sample, we re-analyze the morphological properties of 21 previously published bursts \citep{periodic_chime2020,Pleunis_2021_r3}. The remaining 24 bursts in the sample are new detections. 

Figures \ref{fig:BB1} and \ref{fig:BB2} in the Appendix show our burst sample with baseband data, with Table \ref{Baseband_Tab} listing their measured properties. Figure \ref{fig:Int} show the new detections with only intensity data with measured morphological properties listed in Table \ref{Intensity_Tab}.

\subsection{Baseband Morphology Pipeline } \label{sec: Base_morph}
The baseband morphology pipeline is designed to fit a model to the data with morphological parameters such as burst width, scattering, bandwidth, arrival time, and DM from bursts detected by the baseband system. This pipeline interacts with Stokes intensity (I) data products in the final beamformed file, and a single tied-array beam is formed at the best known sky position of the given source \citep{Baseband,RN12_basebandloc}. 

\begin{figure*}[ht!]
\centering
\includegraphics[scale=0.2]{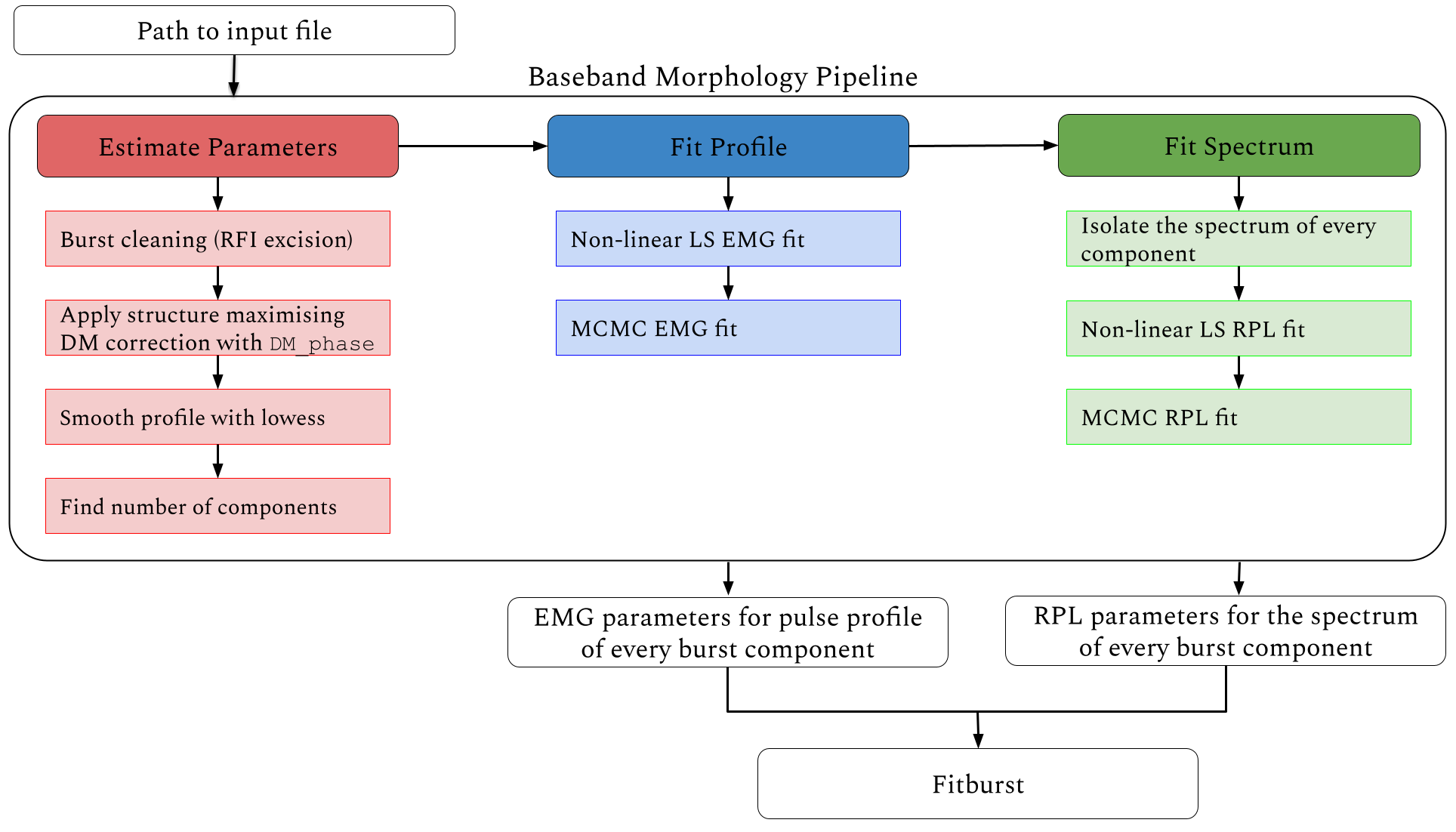}
\caption{Flowchart showing the Baseband Morphology Pipeline. Path on the left involves extracting the Stokes Intensity (I) data products from the beamformed file, dedispersing and cleaning the spectra, getting the burst profile and estimating the number of sub-bursts/peaks to fit for. The middle portion indicates the process of extracting initial guesses on the temporal profile of each sub-burst, first by a Least Square (LS) fit and then by MCMC sampling using an Exponential Modified Gaussian (EMG). As indicated in the right hand column we then estimate the spectral parameters using the same methods assuming a Running Power Law (RPL) model. The resulting datafile along with initial guesses are then provided to \texttt{fitburst}. }
\label{fig:pipeline}
\end{figure*}

The pipeline has three main parts. See Figure \ref{fig:pipeline} for a graphical representation of the pipeline workflow.

The first part starts with performing RFI excision using the functionalities described by \cite{Baseband}. The burst is then dedispersed at its structure-maximising DM, calculated using the \texttt{DM\_phase} package \citep{dm_phase}. The pulse profile is then smoothed by applying locally weighted scatterplot smoothing (\textit{LOWESS}) implemented by \texttt{statsmodels} \citep{seabold2010statsmodels}. \textit{LOWESS} is a non-parametric regression method that weights a data neighborhood with some kernel. Thus, the only parameter it requires is the neighborhood size for smoothing, also known as the smoothing fraction. A small neighborhood will smooth very little, while a large neighborhood will smooth a lot. The pipeline typically uses a neighborhood size of 1.5\% - 20\% of the profile estimated using an algorithm that considers burst properties such as S/N, time resolution, and pulse width. This smoothed profile is then used to estimate the number of burst components using \texttt{find\_peaks} implemented by \texttt{SciPy} \citep{scipy_Virtanen_2020}.

The second step focuses on getting initial guesses for the burst profile and its spectrum. We fit a sum of Exponentially Modified Gaussians (EMG) to the entire burst profile, with one term per sub-burst $i$, and  with the scattering timescale fixed for all sub-bursts. We define the EMG as follows:
\begin{equation}\label{emg_eq}
    \mathrm{EMG}(x;A_i;\upmu_{i};\sigma_{i};\tau) = \sum_{i=1}^{N}A_{i}\frac{1}{2\tau}\exp\left({\frac{1}{2\tau}\left(2\upmu_{i}+\frac{\sigma_{i}}{\tau^2}-2x\right)}\right)\cdot \mathrm{erfc} \left(\frac{\upmu_{i}^{2}+\frac{\sigma_{i}^2}{\tau}-x}{2\sqrt{\sigma_{i}}}\right),
\end{equation}
where $x$ is the input time series, $A$ is the amplitude, $\upmu$ is the Gaussian mean, $\sigma$ is the variance and $\tau$ is the scattering time scale for each of the $N$ components.

The 1D spectrum of each sub-burst is fit with a Running Power Law (RPL) defined as follows
\begin{equation}\label{rpl_eq}
    \mathrm{RPL}_{i}(f;A_{i};r_{i};\gamma_{i})=A_{i}(f/f_{o})^{-\gamma_{i}+r_{i}ln(f/f_{o})},
\end{equation}
where $f_{o}$ is an arbitrary reference frequency chosen to be the middle of the band at 600 MHz, $r$ is the spectral running, and $\gamma$ is the spectral index. The spectral model is identical to the one described by \cite{chimefrb_first_cat_2021}. 

Using these models, first, a non-linear least-squares algorithm \texttt{curve\_fit} from \texttt{SciPy} is used to get initial conditions for the steps (\textit{walkers}) of each of the Markov chains. A Markov Chain Monte Carlo (MCMC) sampling algorithm is then implemented using the \texttt{emcee} routine \citep{emcee} using independent wide uniform prior distributions for all parameters. We use a chi squared log-likelihood function. The MCMC process is quite fast (up to a few minutes, depending on the time resolution of the data), therefore it can be easily run multiple times until the posterior has converged. However, note that this depends on the quality of the data being fit. Very low S/N data may not converge quickly, in which case, we would increase the downsampling factor (i.e. reduce the time resolution) in order to increase the S/N. Once the downsampling has been increased enough, we can adjust the chain length and number of walkers to convergence.

In the third and final step, the EMG and RPL fit parameters obtained for the time series and the spectrum, respectively, are passed on to \texttt{fitburst}, that performs a 2-D fit on the dataset, as described in the following section.

\subsubsection{Fitburst} \label{sec: fitburst}
\texttt{Fitburst} is a least-squares optimization routine that models the burst morphology and dynamic spectrum. The model includes all the fundamental burst parameters:  DM, the time of arrival ($t_\mathrm{arr}$) in the given data file, the signal amplitude (\textit{A}), temporal width (\textit{$\sigma$}), power-law spectral index ($\gamma$), `running' of the spectral index (\textit{r}), and scattering timescale of the signal ($\tau$). For a signal with $N$ components, the DM and scattering timescale are set to be same for each of the sub-bursts, with $2 + 5N$ parameters being fitted through $\chi^2$ optimization. The resulting mathematical formulation can be found in \cite{chimefrb_first_cat_2021} (see Eqns. 1 and 2 of Section 3.3). 

The routine computes a noise weighted fit residual using following equation:
\begin{equation}\label{eqn_noise}
    h_{t,f}(\lambda) = \frac{d_{t,f}}{\sigma_f} - S_{t,f}(\lambda).
\end{equation}
Here $d_{t,f}$ is the data as a function of discrete time and frequency channel, $\lambda$ is the set of fit parameters described above, $\sigma_f$ is the standard deviation in the noise calculated for each frequency channel, and $S_{t,f}$ is the model. \texttt{Fitburst} then employs the \texttt{optimize.least\_squares} functionality of \texttt{scipy} to minimize $\chi^2(\lambda)=\Sigma_{t,f}[h_{t,f}(\lambda)]^2$ with respect to model parameters. The uncertainties on the resulting fit parameters are computed using the covariance in the resulting $\chi^2$ fit provided by the solver. This involves a Jacobian matrix comprised of partial derivatives of the fit equation at the minimized fit parameter values. More information about \texttt{fitburst} can be found in \cite{chimefrb_first_cat_2021} and \cite{pleunismorph}. A more detailed description will be forthcoming (Fonseca et al. \textit{in prep}), along with a public release of the codebase.

We report morphological parameters for the intensity and baseband bursts using \texttt{fitburst}, the only difference betwee the two being that the initial guesses for DM, burst width etc., for the intensity burst fits are computed using the methodology described by \cite{chimefrb_first_cat_2021}. For baseband bursts, we do not keep the DM fixed at its structure maximizing value during fitting by \texttt{fitburst} but rather permit an offset in DM, making it a free parameter. This was done to obtain an optimized model for the burst parameters. We report this optimized value in Table \ref{Baseband_Tab}. 


\subsection{Simulations} \label{sec: sim}



To obtain robust uncertainty estimates, we undertook simulations to better characterize the fitting results from \texttt{fitburst} for all the baseband bursts. These simulations consisted of synthetic bursts generated using the \texttt{simpulse} \footnote{\url{https://github.com/kmsmith137/simpulse}} routine \citep{marcus_injection}. The spectro-temporal properties of the simulated bursts mimic the measured parameters (see Table \ref{Baseband_Tab}). For each of the 45 bursts, we generated 50 synthetic bursts each integrated into an array of random noise. \texttt{fitburst} was then run on the entire sample, and the measured value for each successful run was saved. We calculated the standard deviations of the measured parameters and then calculated the relative uncertainty with respect to simulated values. Finally, we scaled the error measurements in scattering and width for every burst in our sample by the relative uncertainty computed from the simulations. We repeated the entire process three times to verify our results. This yielded robust uncertainties, resulting in better characterization of the temporal evolution of morphological properties as described below. 




\subsection{Period and Period Derivative Analysis} \label{sec:pdot}

\begin{figure*}[ht!]
\centering
\includegraphics[scale=0.6]{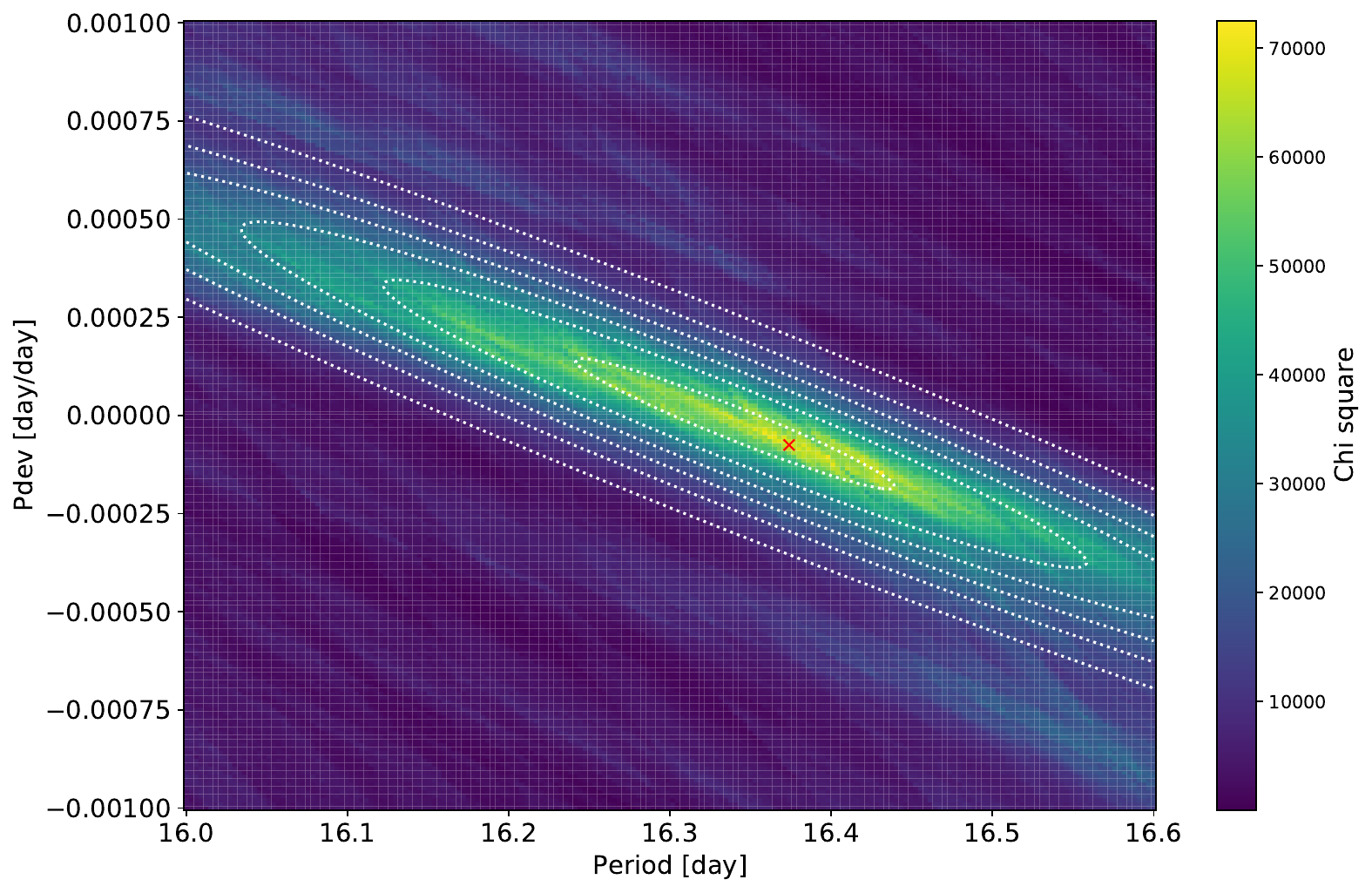}
\caption{$\chi^2$ distribution of different period (x-axis) and period derivative (y-axis) combinations. The red cross shows the maximum $\chi^2$ value. The white contours show the results of the 2D Gaussian fit. We thus find $\dot{P} = - 0.2 \pm 1.5 \times 10^{-4}$ day day$^{-1}$. We calculated the $\chi^2$ using the Pearson $\chi^2$ test described by \cite{periodic_chime2020}. See Section \ref{sec:pdot} for details.}\label{fig:pdot}
\end{figure*}

We recalculated the period of \frb using the arrival time of the new detections presented here and the updated exposure. The total on-sky exposure of the source from 2018 August 28 to 2021 December 31 was computed using the procedure described by \cite{chimefrb_first_cat_2021}. In summary, we record metrics that indicate the variation in up-time and sensitivity of the CHIME/FRB system for the given interval. These metrics are combined with the beam model \footnote{\url{https://chime-frb-open-data.github.io/beam-model/}} to generate exposure maps, which we query at the VLBI position of the source \citep{mar2020}. Note that we only calculate exposure if the sky location lies within the FWHM of the beam at 600 MHz. The total exposure on the source in the aforementioned time interval is 201 hours, in which we detected 94 bursts. 

We used the Pearson $\chi^2$ test described by \cite{periodic_chime2020} to estimate the best period. The updated period is $\mathrm{16.34}\pm\mathrm{0.07}$ days, assuming the period derivative to be zero. We have used this period in all of our phase calculations. The source period has thus not changed prior to 2021 December 31 and is similar to the one reported by \cite{periodic_chime2020} and \cite{Pleunis_2021_r3}. 
To constrain the period derivative, we modified our phase calculation term as follows

\begin{equation}
    \phi = \frac{t-t_o}{P} + \frac{1}{2}\left(\frac{-\dot{P}}{P^2}\right)(t-t_o)^2,
\end{equation}
where $\phi$ is the phase, $t_o$ is the reference Modified Julian Date (MJD) (in this case 58369.40), $P$ is the period and $\dot{P}$ is the period derivative. We then calculate $\chi^2$ values again for each period and period derivative combination. Figure \ref{fig:pdot} shows our $\chi^2$ distribution, and we perform a 2-D Gaussian fit on it using the \texttt{Astropy} package. In this way, we find $P=16.34\pm0.32$ days and $\dot{P} = -0.2 \pm 1.5 \times 10^{-4}$ day day$^{-1}$. This is consistent with the derivative being zero. The above analysis used 8 phase bins, but we find similar order of magnitude results for up to 200 bins. Longer-term monitoring may eventually reveal a non-zero $\dot{P}$.

\subsection{Burst Rate Analysis} \label{sec : rate}


\begin{figure*}[ht!]
\centering
\includegraphics[scale=0.7]{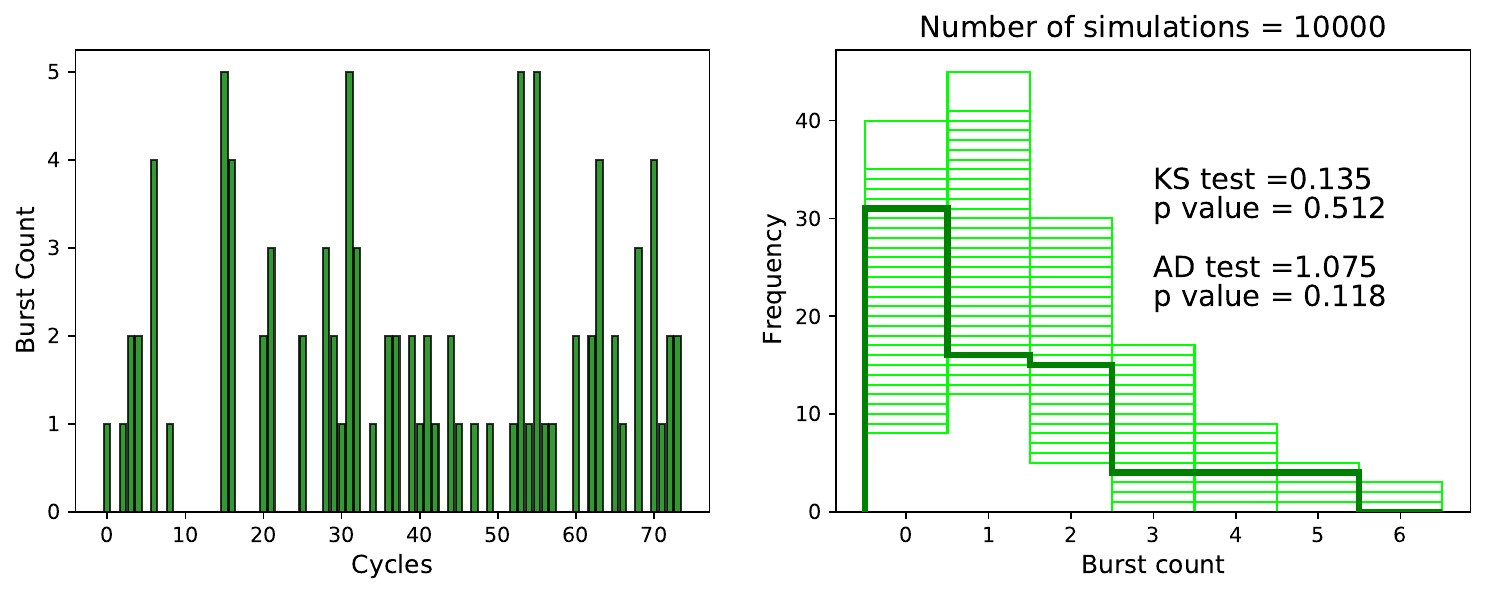}
\caption{Distribution of burst counts for \frb. The left panel shows the number of bursts detected during each activity cycle. One activity cycle is defined as a period of 16.34 days (starting August 2018). In total, we have 74 activity cycles observed by CHIME/FRB from 2018 August to 2021 December. The right panel shows the frequency distribution of the burst count in each activity cycle in dark green. We also show a distribution from 10000 simulations in light green sampled assuming a Poissionian process with mean rate equal to the value we have observed. We performed a KS and AD test between the observed and simulated distributions and report the median p-value, 0.512 and 0.12 respectively. See section \ref{sec : rate} for details.}
\label{fig:count}
\end{figure*}

Next we do an analysis of the rate of bursts from FRB 20180916B in the CHIME band. After excising the bursts detected during days of low sensitivity and those detected outside the FWHM at 600 MHz, we are left with 40 bursts in our rate analysis.

The exposure for each activity cycle was calculated by folding the daily exposure data using a period of 16.34 days. We have had 74 activity cycles since the source's discovery in 2018 September up to the end of 2021. Figure \ref{fig:count} (left) shows the burst count in each cycle. To test whether the counts are consistent with being drawn from a Poissonian distribution, we used a Kolomogrov-Smirnoff (KS) test \citep{massey1951kolmogorov} and an Anderson-Darling (AD) test \citep{scholz1987k}. Figure \ref{fig:count} (right) shows the frequency distribution of the counts in dark green. We simulated 10000 instances of a Poisson process, assuming a mean rate equal to the observed mean count of bursts, 1.27 events per activity period. We performed a KS and AD test for each distribution with our observed count distribution. Our median p-values, 0.512 (KS) and 0.12 (AD), shows that the observed counts distribution is consistent with being Poissonian at 95\% confidence.  

\begin{figure*}[ht!]
\centering
\includegraphics[scale=0.6]{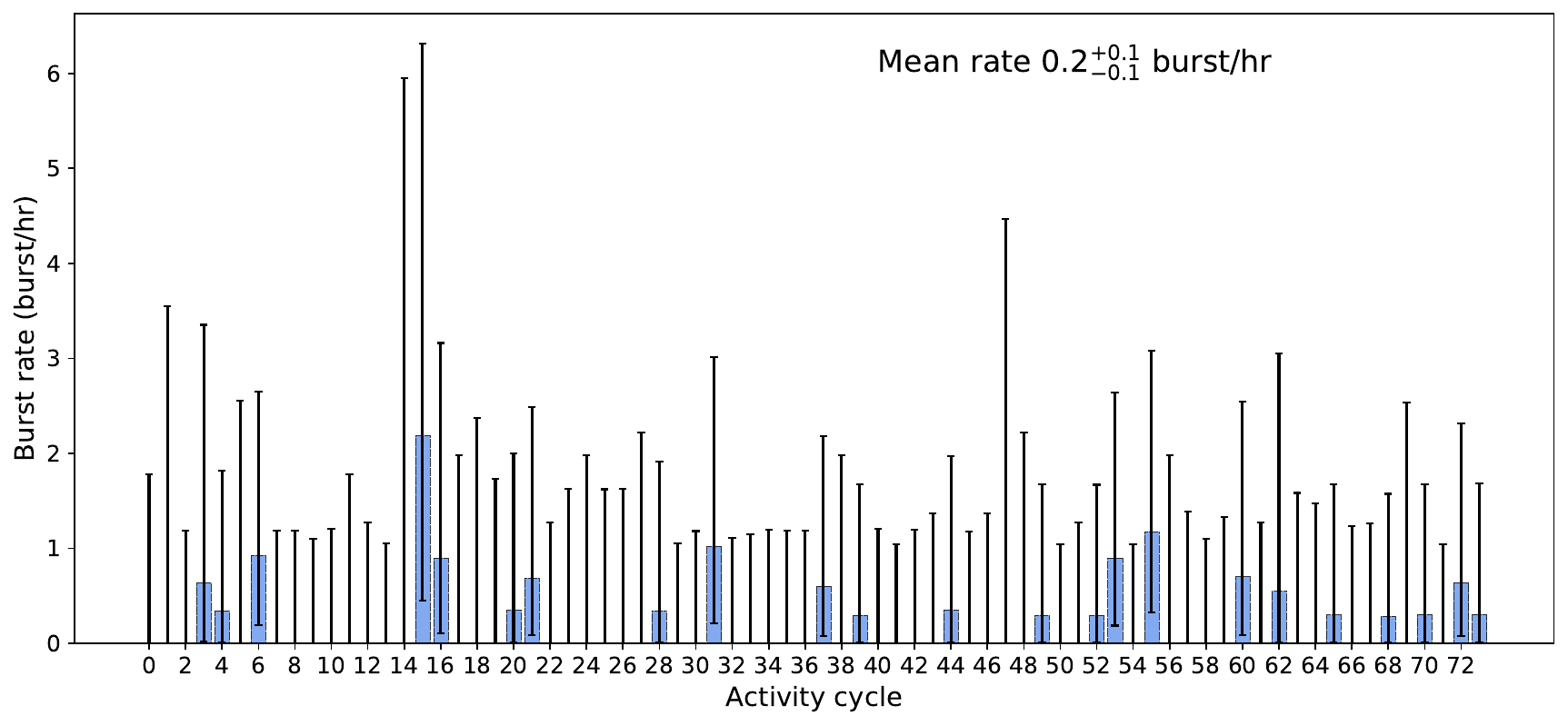}
\caption{Burst rate in different activity cycles. The burst rate of \frb (blue bars show detections) is steady within error bars, which denote 95\% confidence intervals assuming a Poissonian process. Many of these cycle have zero rates where we denote upper limits. This is because not all the bursts detected in our sample satisfy the exposure criteria (see Section \ref{sec : rate} for details).} 
\label{fig:rate vs cycle}
\end{figure*}

We then estimated the burst rate for each activity cycle, as shown in Figure \ref{fig:rate vs cycle}. There, error bars represent 95\% confidence uncertainties assuming a Poissonian process within each cycle. As the telescope sensitivity varies across cycles, the rate for each cycle is scaled to a fluence threshold of 5.2 Jy ms by using an index of $\gamma=-2.3$ \citep{periodic_chime2020}. This threshold was calculated using the methodology described by \cite{josephy2019chime}. We see that the source's activity is consistent with a single Poisson rate throughout the three years of this study. The mean rate throughout the observation period is $0.2^{+0.1}_{-0.1}$ bursts per hour. 

\begin{figure*}[ht!]
\centering
\includegraphics[scale=0.7]{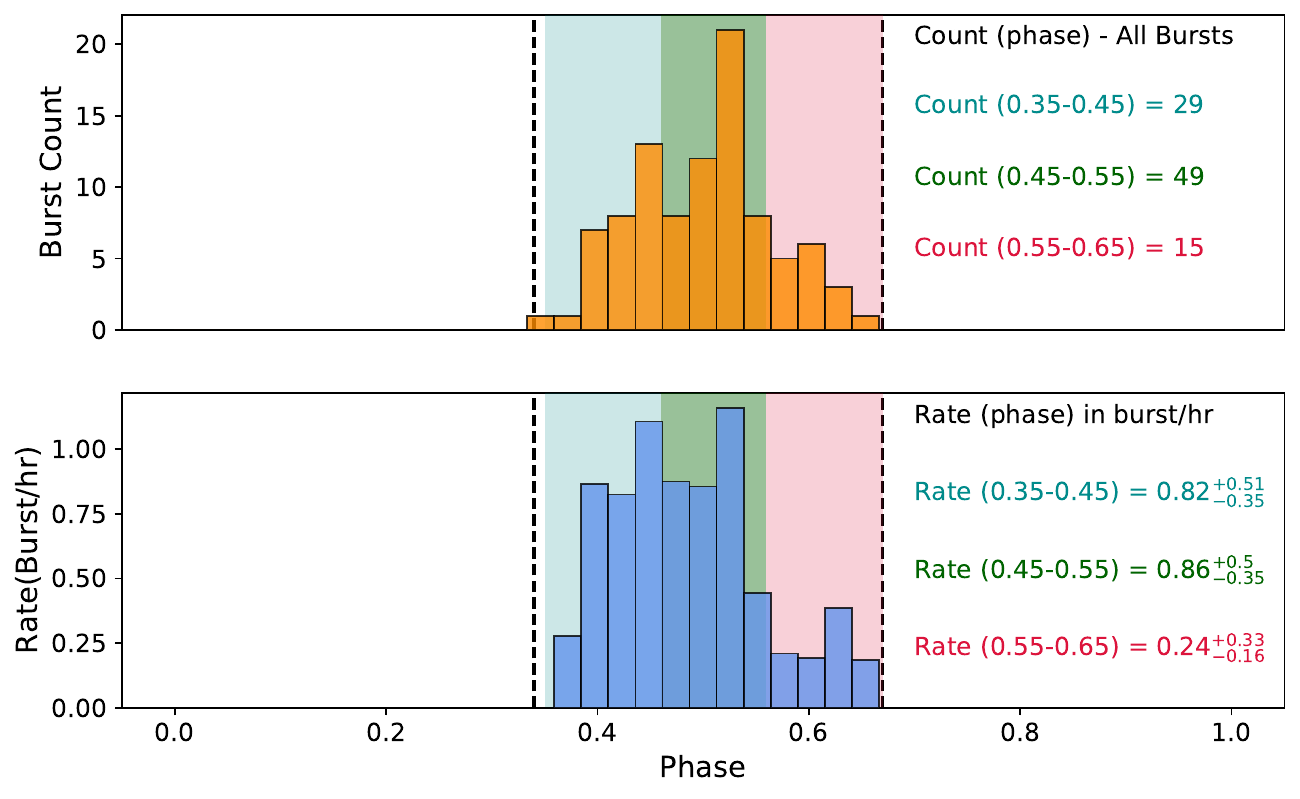}
\caption{Burst count and rate variation with phase. We folded our burst arrival times at a period of 16.34 days with reference MJD 58369.40. We did the same with our calculated exposure and estimated the burst rate of \frb in different phase bins. The top panel shows the burst count distribution of all the bursts detected by CHIME/FRB till 2021 December 31. We detect most of the bursts near the peak activity phase of CHIME/FRB i.e., 0.45-0.55. The bottom panel shows the rate. Here we have excised the bursts detected during days of low sensitivity and those detected outside the FWHM at 600 MHz to get true estimate on the rate. The rate is maximum in the phase range 0.45-0.55 (green). We also show the rates in phases earlier i.e 0.35-0.45 (blue) and later, i.e., 0.55-0.65 (red). The three rates are consistent within error bars (95\%) assuming a Poissonian process. See Section \ref{sec : rate} for details.}
\label{fig: rate vs phase}
\end{figure*}

The count and rate distribution with respect to phase is shown in Figure \ref{fig: rate vs phase}. We divide the active phase into three sub-intervals. The rate centered around the peak activity of the source (0.45--0.55) is $0.86^{+0.5}_{-0.35}$ burst per hour at a fluence threshold of 5.2 Jy ms. However the differences are not statiscally siginificant with rates during earlier phases (0.35--0.45) of $0.82^{+0.51}_{-0.35}$ or later phases (0.55--0.65) of $0.24^{+0.33}_{-0.16}$ burst per hour. We might observe significant differences in phase dependent rates with more detections in the future.

\subsection{Evolution of Properties} \label{sec:prop}

\begin{figure*}[ht!]
\centering
\includegraphics[scale=0.45]{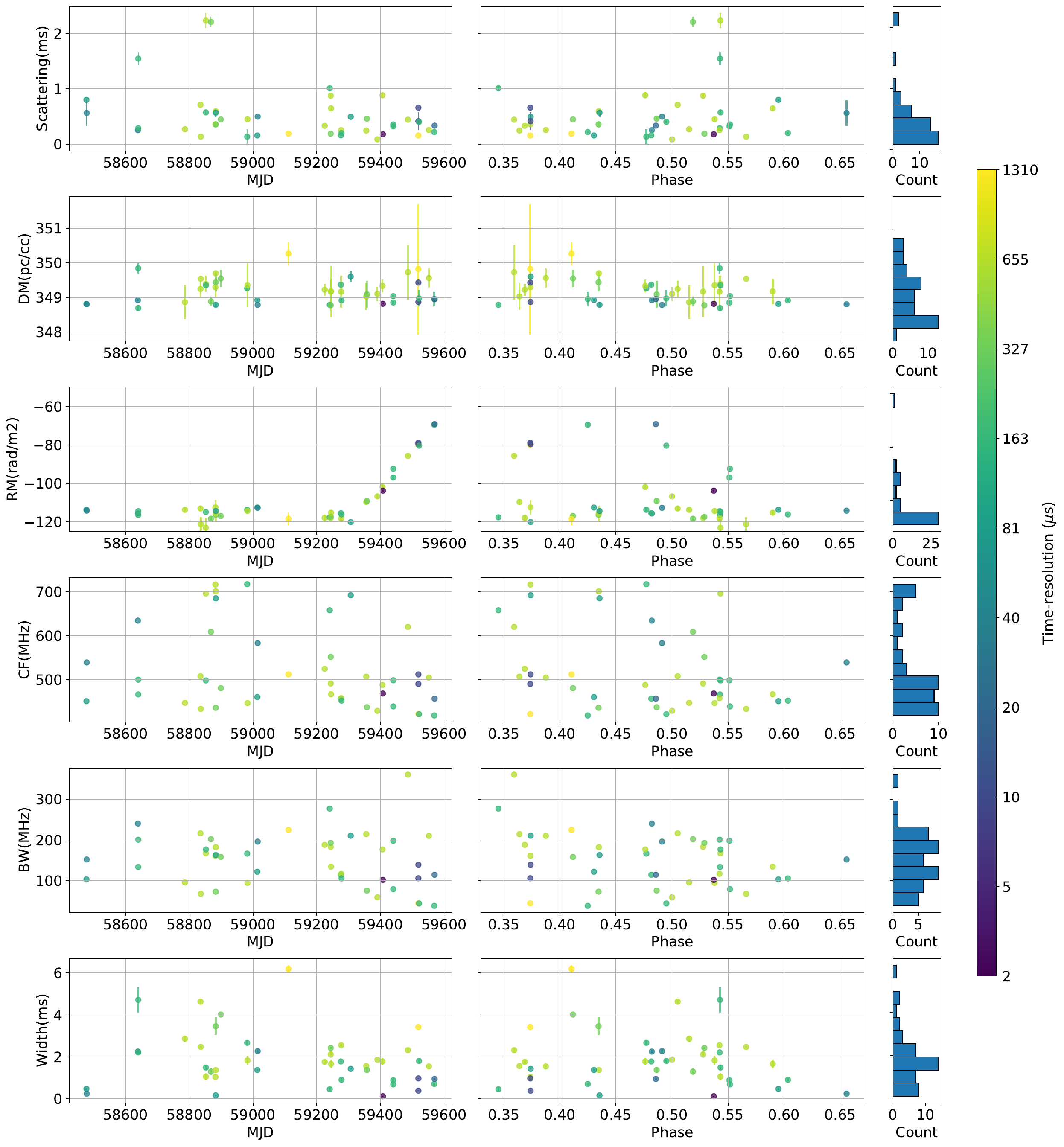}
\caption{The evolution of various spectro-temporal properties of bursts from \frb as a function of time (left) and phase of the 16.34 day cycle (reference MJD 58369.40). The histograms on the far right show overall distribution of the properties. We show here six prorperties - scattering, DM, RM, Central Frequency (CF), Bandwidth (BW) and combined width of all the sub-bursts (W). The color scale shows the time-resolution at which we measured the morphological parameters for each of the burst since not all bursts were bright enough to be studied at 2.56 $\upmu$s time resolution. We see the RM increase in the source around April 2021 as reported by \cite{mckinven_2022_r3}. We find no trend in any other properties in both time and phase. Only the bursts with baseband data have been shown here since we were able to better characterize their spectral and temporal properties. (See Sec.\ref{sec:prop} for details)}
\label{fig:money}
\end{figure*}

Figure \ref{fig:money} shows the evolution of various spectro-temporal properties of the source with respect to MJD and to the phase of the 16.34-day cycle for all the baseband bursts in our dataset. The color scale shows the resolution at which we measured the burst's morphological properties. The side-panel histogram presents the overall distribution. Only the bursts from the baseband sample were used in this analysis as their spectral and temporal properties are better constrained. 


\cite{mckinven_2022_r3} reported a systematic linear increase in RM as shown in third panel of Figure \ref{fig:money}. However, interestingly, we do not find any corresponding change in any other properties.

We performed an Augmented Dickey Fuller (ADF) test on DM values to test the time dependence. The ADF estimates a regression equation and checks if the regression coefficient applied to a lagged time series variable is significantly different from unity \citep{dickey1979distribution}. It tests the null hypothesis that a time series is non-stationary. With a p-value of $\sim 10^{-8}$, we find the DM is stationary and see no clear evolution with time, even during the corresponding RM evolution. The DM of the brightest burst in our dataset, B35, is 348.81 $\pm$ 0.01 pc\,cm$^{-3}$, which is 0.04 pc\,cm$^{-3}$ greater than what was reported by \cite{nimmo2021_micro} (348.772 $\pm$ 0.001 pc\,cm$^{-3}$) . The DM we measure for different bursts requiring lower time resolution can be affected by underlying unresolved structure. We constrain the DM variation to $\lesssim$ 1\,pc\,cm$^{-3}$. 


The scattering timescale shows significant variation from burst to burst. However, in our ADF test, we do not observe any trend in this variation (p-value $\sim 10^{-10}$). We also do not see any phase dependence. Similarly, the widths also show stochastic variation with no particular trend, with the broadest burst being 6 ms. However, there are caveats regarding our measurements:  weaker bursts are at lower time resolution. For bursts B4-B5 and B10-B11 (see Table \ref{Baseband_Tab}), we see significant changes in the scattering value within the same CHIME transit. However, the bursts are not bright enough to resolve structure narrower than $\sim$ 160 $\upmu s$, which could affect our scattering estimates. Such sudden changes have also been claimed for FRB 20190520B \citep{ocker_20190520B} and have been attributed to fluctuations in the turbulence in proximity to the circumburst environment. 

In terms of spectral properties, we find that most of the bursts are narrow-band, as is also seen for all other repeaters \citep{pleunismorph,kumar_narrow}. The average bandwidth in the CHIME band is $\sim$ 150 MHz, with the narrowest case being 40 MHz. Some bursts were detected at the top or bottom of the CHIME band so we cannot estimate their true bandwidth. However, for this source, we have not yet observed any multi-band detection even after $\sim$ 70 hours of observation \citep{sand2022,pastormarazuela2020chromatic_180916}. The central emission frequency of the bursts in our sample is distributed across the CHIME band and there is no obvious correlation between the phase and emitting frequency. With more bursts in the future, we may be able to detect chromaticity as a function of phase in the CHIME band. 

We also performed a structure-function analysis similar to the one presented by \cite{mckinven_2022_r3} for scattering, in order to search for preferred time scales for variation. We find no measurable trend at any timescale, suggesting that the variations are random and occur due to fluctuations in the propagation environment.

\subsection{Microstructure Analysis} \label{sec:micro}

\begin{figure}[ht!]
\centering
\includegraphics[scale=0.45]{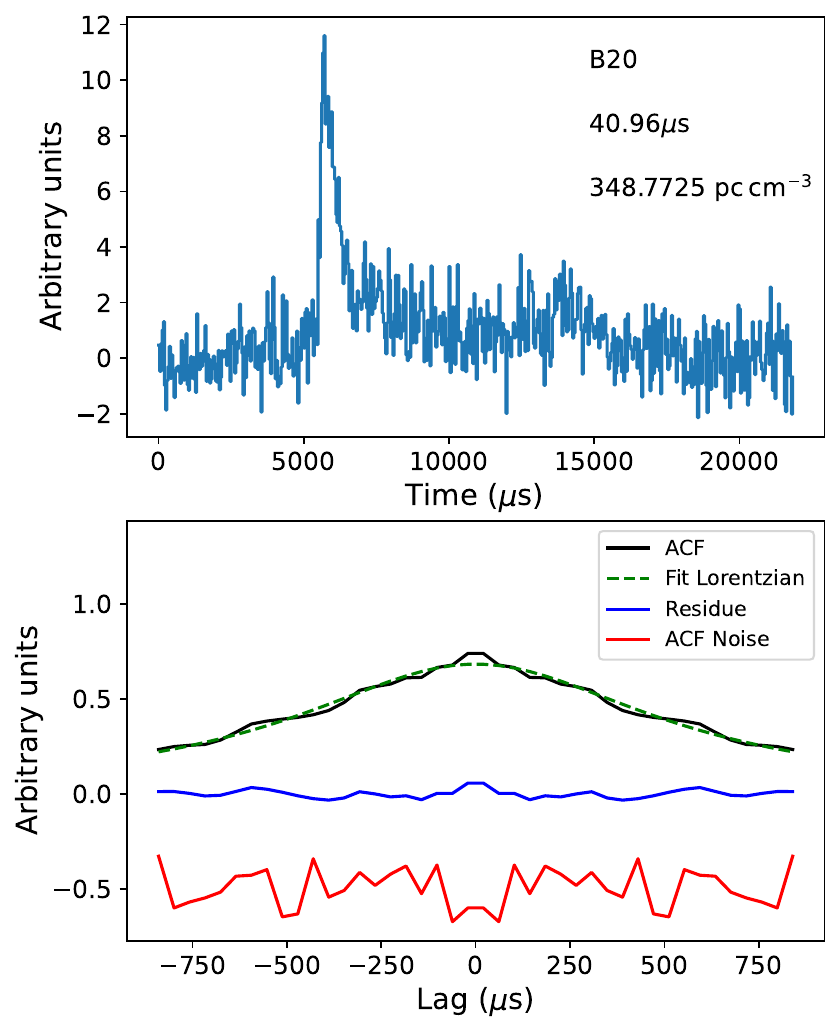}
\caption{Microstructure analysis of B20 from \frb. The top panel shows the burst profile, with the burst number, resolution at which it was extracted and the DM indicated in the top right. The bottom panel shows the temporal ACF (black), Lorentzian fit to the ACF (green dashed), the residual (blue) and ACF of the noise (red). We find a weak peak in the ACF with width of $\sim50 \upmu$s but cannot constrain the true nature of this structure with the present dataset. See Section \ref{sec:micro} for details. }
\label{fig:micro}
\end{figure}

Some FRBs have shown temporal microstructure. For \frb, we have seen structures as narrow as 3 $\upmu$s at 1.7 GHz \citep{nimmo2021_micro} and 30 $\upmu$s at 800 MHz \citep{sand2022}. The bursts need to be coherently dedispersed and bright enough at finer time resolution  for such narrow structure to be detectable.
We selected 11 bursts from our data that were visible at a resolution of 80 $\upmu$s or less (maximum downsample of 32, Table \ref{Baseband_Tab}).

We analysed the burst structure using a temporal auto-correlation function (ACF). ACF is a useful tool for studying these ultra-fast emission features. First, we summed over the sub-bands in which the burst was detected. We then performed the ACF in time of the pulse profile of these bursts. We flagged the zero-lag peak, which inhibits the detection of narrow features in the ACF. We performed a Lorentzian fit on the wider burst envelope, which in this case corresponds roughly to the width of the burst. 

Only one of the bright bursts showed interesting feature in our sample. Figure \ref{fig:micro} illustrates the analysis conducted for B20 within our sample. The error associated with the DM value for this burst was 0.009 pc\,cm$^{-3}$ (refer to Table \ref{Baseband_Tab}), corresponding to a smear time of approximately 80 $\upmu$s within CHIME band. To account for this effect, we measured the ACF of the time profile at 100 DM steps ranging from -0.009 to 0.009 pc\,cm$^{-3}$, centered around the optimal DM value (348.781 pc\,cm$^{-3}$). The figure displays the outcome at the DM value that maximizes our ACF, which is characterized by the greatest separation between the second Lorentzian peak and the fit to the broader envelope. It is to be noted that this DM value does not correspond to the true DM of the burst. We cannot reliable ascertain whether the peak is due to a microstructure or can be attributed to amplitude modulated noise which will create stronger power on the one bin level. We redid the analysis at higher time resolution and the burst was too weak to extract any reliable estimates. If real, detecting such features at 600 MHz opens the possibility of ultra-fast emission features in brighter detections from other sources in the CHIME band. The remaining bursts in our bright sample did not show any interesting features in their ACF.


\section{Discussion} \label{sec:disc}

\subsection{Burst rate evolution}

Active repeaters have shown variations in their burst rates over time. The rate can change from less than one burst per hour to more than 100 bursts per hour, as was observed for FRB\,20121102A by \cite{FAST_121102A} in their 60 hours of observations spanning 47 days at 1.2 GHz above a 7$\sigma$ threshold of 0.015 Jy ms. \cite{Gaj18} also reported clustered arrival times in their high frequency (4-8 GHz) study of the same source. \cite{nimmo_burst_storm} also observed this behavior from FRB\,20200120E, where they detected 53 bursts within 40 minutes, almost 80 bursts per hour. Before this epoch, the observed peak rate from the source was $0.4^{+2.0}_{-0.4}$ bursts per hour in their Effelsberg observations at 1.4 GHz above a 7$\sigma$ threshold of 0.05 Jy ms. These burst storms after a dormant period can be a characteristic feature of some repeating sources. Notably, these observations were at different frequencies and sensitivity thresholds, making it harder to compare the rates and energy distributions among active repeaters.  

CHIME/FRB, in its continuous monitoring, has been crucial in catching spontaneously active repeaters. FRB 20201124A is a prime example \citep{lanman_20201124A}. After its initial discovery in 2020 November, it was not detected above our sensitivity threshold \citep{chimefrb_first_cat_2021} and suddenly entered a high-activity period in 2021 April. Follow-up observations led to detections of thousands of bursts using multiple instruments, particularly by the Five hundred meter Aperture Spherical Telescope (FAST) \citep{FAST_R67}. Recently, FRB\,20220912A \citep{Mckinven_R117} has shown similar behavior, with no bursts detected prior to 2022 September followed by nine bursts in three CHIME/FRB transits, suggesting rates as high as 200-300 bursts per day. Higher frequency followups by GBT \citep{Feng_2023_R117} and FAST \citep{Zhang_2023_R117} confirmed this high burst rate for the source. Magnetars have also been known to exhibit burst storms at X-ray energies, with 100s of bursts per hour \cite[e.g.,][]{Younes_Sgr,2002gavriil}. SGR 1935+2154 is a prime example, but so far, we have only observed one radio burst from this or any magnetar with a luminosity comparable to that of FRBs \citep{andersen2020bright,bochenek2020fast}.

 
We do not see any episodes of heightened activity from \frb, even after $\sim$ 200 hours of observation by CHIME/FRB across the entire activity phase. Cycle-to-cycle variation has been observed within the same activity phase from the source at other frequencies \citep{surya2022,pastormarazuela2020chromatic_180916}. However, in our long-term monitoring of the source, we find such variations to be consistent with being Poissonian in their distribution (see Section \ref{sec : rate}). As shown in Figure \ref{fig:count}, the burst count can vary from zero to five bursts from cycle to cycle, but such behavior is expected from a Poisson process of a mean rate of around 1.3 bursts per activity cycle. Figure \ref{fig:rate vs cycle} shows that the burst rate of the source is steady within 95\% confidence uncertainty, assuming a Poissonian process in the observed 74 activity cycles. The source seems to have a steady rate, with no period of heightened activity, unlike what has been observed from other prolific repeaters. But the CHIME/FRB transit duration is $\sim$ 15 minutes; extended exposure with a sensitive telescope might provide more insights into whether \frb is prone to burst storms. In addition, the RM change observed in the source \citep{mckinven_2022_r3} does not coincide with any change in the activity of the source, suggesting that the secular rise observed likely has more to do with evolution in the propagation environment rather than variations in the intrinsic nature of the source. 

 \frb is known to be chromatic in its activity. Higher frequency bursts tend to be observed at earlier phases \citep{surya2022} and lower frequency bursts at later phases \citep{Pleunis_2021_r3}. This spectral dependence makes it difficult to estimate the full rate, across all frequencies, per cycle. In conjunction with the narrow band nature of the bursts, this can lead to differences in rates as a function of phase within a specific frequency range. In Figure \ref{fig: rate vs phase}, we see that more bursts are detected at phases 0.45--0.55, but the overall rate of the source is consistent among activity phase within the uncertainities. With more detections in the future, any difference in rate might become prominent in the CHIME/FRB band.
 
 Alongside rate, we find that the activity period of the source has not significantly varied with time since the source's discovery. Our estimate, based on data from 2018 August to 2021 December, the longest time span yet, is  16.34$\pm$0.07 days (see Section \ref{sec : rate}), consistent with the original measurement of 16.35$\pm$0.15 days \citep{periodic_chime2020}. This makes \frb unique compared to other prominent repeaters. A stable period with no episodes of heightened activity might suggest a distinct progenitor scenario to what has been proposed for other active sources. This will be discussed in detail in Section \ref{sec : progen}.

\subsection{Spectro-Temporal Properties} \label{sec:tempo}

FRBs can have complex structures in their temporal profiles, often well described as multiple sub-bursts. Repeaters generally show more such sub-bursts compared to apparent non-repeaters, as was first noticed in the first known repeater FRB\,20121102A \citep{Hes2019}. Bursts from repeating sources also tend to be wider and have narrower bandwidths than for apparent one-off FRBs \citep{pleunismorph}. Our burst sample from \frb exhibits the same behavior. Most of our widths lie in the range 1--3 ms (see Table \ref{Baseband_Tab}). Our bursts are wider than seen in high frequency detections at 6 GHz ($\sim$ 0.3 ms) \citep{surya2022} and narrower compared to LOFAR detections ($\sim$ 40 ms) \citep{Pleunis_2021_r3,pastormarazuela2020chromatic_180916}. Our average bandwidth of 150 MHz is narrower compared to bandwidth $>$ 500 MHz detections at high frequencies, a trend that has been seen for \frb \citep{surya2022} as well as FRB\,20121102A \citep{Gaj18,Hes2019}. We have observed sub-bursts as narrow as 78 $\upmu$s (B20), suggesting there might be more sub-ms emissions from repeaters even at low frequencies. Additionally, as shown in Figure \ref{fig:money}, we find no temporal or phase dependence in burst width over the three-year timescale. We also do not find any significant bimodality in the peak emitting frequency from the source, unlike what has been observed for FRB 20201124A \citep{lanman_20201124A}.

Scattering timescales trace inhomogeneities along the line of sight towards the source. \frb's low Galactic latitude (b $\sim$ 3.7$^{\circ}$) implies a major contribution from the Milky Way interstellar medium (ISM) to the scattering time. The expected scattering in the direction of \frb at 600 MHz according to the NE2001 model \citep{Cor02} for the Galactic electron distribution is $\sim$ 0.16 ms. Assuming that the frequency dependence of scattering time is a power law with index $-$4, we expect a scattering timescale of the order of $\sim$0.17 ms in the CHIME band, extrapolating from the 2.7-$\upmu$s scattering time reported by \cite{mar2020} at 1700 MHz using scintillation bandwidth measurements. \cite{chawla2020detection} report an upper limit of 1.7 ms at 350 MHz, corresponding to a scattering timescale $<$0.2 ms at 600 MHz. Our scattering values range from 0.08 ms to 2.2 ms, suggesting time-variability in the medium causing the scattering.  In our dataset, we observed bursts with scattering less than 0.16 ms (see Table \ref{Baseband_Tab}). However, our measured scattering might be contaminated by unresolved structure, since not all the bursts in our sample are bright enough to be studied down to 2.56 $\upmu$s (see Table \ref{Baseband_Tab}). Most of the scattering observed from \frb has been attributed to the Milky Way ISM \citep{sand2022,surya2022,mar2020} using scintillation bandwidth measurements. Hence, the observed scattering profile for the source, if real, may not always follow a power-law index of $-$4, as expected from a thin screen model. Such deviations in the index have also been observed for Galactic pulsars \citep{lewandowski2015,geyer_2017}. \cite{surya2022} also find tentative scattering variability from scintillation measurements; they attribute most of this to the refractive timescale of the Milky Way screen.  

Microstructures have been observed for many FRB sources, suggesting luminous ultra-fast emission \citep{nimmo2022burst}. For \frb, we have seen structures as narrow as 3 $\upmu s$ at 1700 MHz \citep{nimmo2021_micro} and 30 $\upmu s$ at 800 MHz \citep{sand2022} which is in agreement with the prediction of Milky Way scattering from NE2001. Although we did not find strong evidence for structure in this work, future studies should continue to search for short timescale structure from \frb and other FRBs, especially those at higher galactic latitudes where Galactic scattering should be even smaller. Detecting such structures will suggest an occasionally less turbulent environment that doesn't inhibit the propagation of $\upmu$s substructures by scattering, even at low frequencies. Microstructure also favors the emission origin to be magnetospheric instead of a shock wave interaction since the latter's larger physical scales do not naturally predict very short temporal behavior, whereas the former is inherently compact \citep{mms19,lu_emission}. 

\subsection{Local Environment} \label{sec : local}

Repeaters are a unique tool for understanding the immediate surroundings of FRBs, as they permit studies of short-term and long-term variability of DM, RM and scattering properties. The first repeating source FRB\,20121102A was proposed to be in a highly magnetized environment given its extremely high RM and association with a Persistent Radio Source (PRS) \citep{mic2018}. Since then, there has been a secular decrease in RM from the source with an increase in DM \citep{hilmarsson2021rotation}. This change can be explained by an FRB source within a magnetar wind nebula, a pulsar orbiting a supermassive black hole or an evolving Supernova Remnant(SNR) \citep{mar_metz_R1,zhang_R1}. RM variability has also been observed for FRB 20201124A \citep{FAST_R67,Hilmarsson_R67} on timescales of weeks. A more extreme case of this has been shown by FRB 20190520B, where a reversal in RM sign was observed \citep{dai_reversal_FRB20190520B,thomas_reversal_FRB20190520B}.  Such variability can be attributed to variation in the magnetic field configuration close to the source region. \cite{Wang_binary} suggest that these two sources might reside in a magnetar/Be binary star system, with interaction from the Be star decretion disc producing these RM variations. 

\frb has a nominal RM, which was reasonably stable with stochastic variations, until recently when \cite{mckinven_2022_r3} observed a secular rise in the value. We do not see any corresponding evolution in the DM of the source. \cite{mckinven_2022_r3} attribute this to changes in the $B_{\parallel}$ of the local environment and put a limit of $\leq 4$~pc on the size of the Faraday active medium. This is greater than the 1~pc upper limit on the compact radio counterpart of FRB\,20121102A \citep{plavin_r1}. \cite{Zhao_RM_reversal} interpret the RM variation of \frb as periastron passage of a freely precessing magnetar in an orbital period of 1600--16000 day,s with no DM contribution from stellar wind or disc. We do not see any changes in the scattering timescale that might hint at an interaction with a stellar wind during the RM variation epochs. However, the mass loss rate of the companion star may be too low to discern any significant changes.

We find no obvious trend in the long-term variations in scattering, although burst-to-burst measurements tend to vary. Interestingly for bursts B4-B5 and B10-B11, we observe large fluctuations in the measured scattering value within the same transit. \cite{ocker_20190520B} have also observed changes in scattering on short timescale ($\sim$ minutes) for FRB 20190520B. They attribute such behavior to a dynamic, inhomogeneous plasma in the circumsource environment and predicted similar variations in other FRB sources. Though 
our measured values might not be physical scattering due to the possibility of underlying structures, we can put definitive limits at the given time resolution for respective bursts. Overall this suggests that although \frb has a cleaner circumburst medium compared to other sources, we can still expect considerable variations due to discrete plasma patches in the circumstellar medium \citep{ocker_20190520B}. Similar to us \cite{Gopinath_R3_2023} also observed large scattering changes ($\sim$ 10 ms) in short timescale for \frb in their observation at 150 MHz using the LOFAR telescope. They also attributed these to variations in the turbulence in the local environment. They do not find any phase dependence in scattering. We also do not observe any periodic dependence in scattering variability (see Figure \ref{fig:money}). This seems at odds with strong binary wind interactions near the source if we assume that the 16-day periodicity is due to orbital motion. We also do not observe any correlation between changes in scattering with DM for \frb, unlike what has been seen for the Crab pulsar over its long-term monitoring \citep{Mckee_Crab}. These correlations have been attributed to seeing different sightlines through the nebula due to the pulsar’s proper motion, which traces the filamentary structure of the nebula. This suggests that \frb might not reside in a Crab-like remnant, or may be older, with the remnant so dissipated that it has lost its filamentary structure. 


\subsection{Constraints on Progenitors} \label{sec : progen}

The periodic nature of \frb has led to multiple progenitor theories. These can be broadly classified into two types. The first type involves FRB emission from a compact object, where the period is due to rotation, or to precession (either of an isolated object or of one in a binary system). The second broad class invokes a compact object in a binary orbit, where interactions with the companion wind or disk lead to FRB emission, and the orbital period explains the periodicity.

\subsubsection{Rotation or Precession as the Origin of the Periodicity}

The rotation of an ultra-long period magnetar has been suggested to explain the periodicity
\citep{Paz_period_magnetar}. In this case, an isolated magnetar is slowed down to a 16-day rotation period by episodic mass loss in particle winds, angular momentum kicks, or accretion due to a long-lasting fallback disk. The magnetar, in this case, would be of age comparable to that of typical Galactic magnetars ($\sim$1--10 kyr), which is consistent with the observed offset of the source from a star formation region in {\it Hubble} observations \citep{ten2021_60pc}. 
The stability of the period over the three years of our observations is consistent with this scenario; however, the constancy of the burst rate is somewhat at odds with the usually more episodic magnetar bursting behavior.
In this picture, we expect no phase-related DM variations and relatively constant polarization angles. However, we have not observed such long periods from any Galactic magnetars/neutron stars, though a possibility of a 6.7-hour period has been proposed for the unusal X-ray source 1E 161348$-$5055 in the SNR RCW 103 \citep{de2006long}.

Free precession in a flaring magnetar or luminous radio pulsar has also been suggested to explain the periodicity.
\citep{levin_precess,zanazzi_free_precess,katz21}. 
This model predicts an increase in the period as the neutron star spins down with time.  Our upper limit, $\dot{P} < 1.5 \times 10^{-4}$, implies a lower limit on the spin-down age
$\gapp$150~yr, or equivalently 
$B\gapp 5 \times 10^{14}$~G assuming standard magnetic braking
\citep{levin_precess,katz21}.
These constraints will improve with time, although as noted by \cite{katz21}, the lower limit on the (unknown) spin-down age is not constraining, since the latter is generally interpreted as an upper limit on the true age.
Additionally, in such models, a shorter timescale periodicity from the rotation of the compact object itself is expected to be present, possibly on the order of seconds as for Galactic magnetars \citep[e.g.][]{Olausen_and_Kaspi}.  This has not yet been seen, though detection could be hindered by timing noise or glitches in the case of a very young magnetar \citep[e.g.][]{klc00,Gavriil_2008}.

To explain the chromatic nature of the burst arrival times from \frb, \cite{Li_emission_chromatic}, suggest altitude-dependent emission from a magnetar, which can be precessing or have a long spin period. 
For a rotating magnetar, the model predicts a constant position angle at similar phases, but for a precessing system, the position angle can vary at fixed phase. 
In addition the spin down rate could help distinguish between the two possibilities, since it is expected to be much larger in the case of free precession \citep{Li_emission_chromatic}.

Recently \cite{Zhao_RM_reversal} proposed a precessing magnetar in orbit with a Be-star for \frb, to explain its latest RM evolution. In this case, the 16.3-day period is due to a precessing or rotating magnetar, but it is in a binary orbit of $\sim$1000--20,000 days. In the model, the magnetar is undergoing a periastron passage, resulting in a secular rise in RM due to interaction with the disk of the Be-star. Stochastic RM variations in this model are due to clumps in the stellar winds/disk interacting with the FRB. However,  DM and scattering variations correlated with RM are expected in this model, yet our observations show no evidence for this.  

Forced precession due to a fallback disk has also been proposed to explain the observed periodicity \citep{tong_forced_precess}. Here a shorter rotational periodicity from the compact object is also expected.  This model does not obviously account for chromatic periodicity \citep{Li_emission_chromatic}, and a range of polarization position angle values is expected, but has not been observed \citep{mckinven_2022_r3,nimmo2021_micro}.  
The forced precession fallback disk model is thus not favored by the data.
The model predicts that the observed period should increase as the disk dissipates, such that 
$\dot{P} = - P/t_{diss}$, where $t_{diss}$ is the unknown dissipation timescale \citep{katz21}.  Our constraint on $\dot{P}$ implies $t_{diss} \gapp 300$~yr.

Another model involves geodetic precession in a relativistic orbit. \cite{Yang_orbit_precess} propose a neutron star precessing in a tight orbit with a companion (orbital period $\sim$ 100--1000~s). This system would be short-lived, and a decrease in the period and a change in burst rate would be expected as the orbit decays. However, the timescale of this decay could range from hundreds to millions of years \citep{UCB_haaften, katz21}, depending strongly on the distance between objects and their masses. Our current period derivative limit of $1.5 \times 10^{-4}$ days/day suggests that we will start probing interesting phase space (e.g. mass ratios of unity and a neutron-star mass) in just a few years.

The last precessional model we consider here is a compact object in an ultra luminous X-ray binary (ULXB) undergoing super-Eddington mass transfer \citep{sridhar_xray_bin}. Here, the precession of the polar accretion funnel results in periodic FRB emission. In this case, a systematic variation in DM of $\sim$0.1--1~pc~cm$^{-3}$ and am RM of $\sim$0.5--1~rad~m$^{-2}$ with activity phase is expected, depending on the parameters for the quiescent jet, taking into consideration the optical depth needed for FRB to escape. As shown in Figure \ref{fig:money}, we constrain $\Delta$DM $\lessapprox$1 pc cm$^{-3}$ within the activity phase.  However, this model does not naturally explain the observed evolution in RM if the variation is due to changes in the immediate emission environment. Moreover, a `turn-off' in emission is predicted on a timescale of years, perhaps followed by the appearance of an optical/IR counterpart.

\subsubsection{Binary Motion as Origin of Periodicity}

The source of the radio bursts in \frb has been suggested to be in different types of binary orbits in order to explain the observed 16.34-day periodicity.  

\cite{lyutikov_ob_binary} suggested a pulsar in orbit with an O/B type star, and
\cite{Li_Be/X-ray_bin} suggested FRB emission from star quakes occurring due to stress on the neutron star surface by accreting material from a Be star companion. 
\cite{Ioka_comb} proposed a highly magnetised pulsar whose magnetic field is `combed' by a strong wind from a massive companion star, resulting in FRB emission. 
\cite{Wada_bin_comb} made additional alterations to the model to explain the chromatic periodicity, by proposing variations in emitting frequency with phase due to the influx of aurora particles at different phases of the orbit. 
\cite{deng_binary} consider an accreting system with a compact object generating FRBs by the synchrotron maser mechanism.
\cite{Wang_binary} explained the RM variations observed in FRB 20201124A and FRB 20190520B using an eccentric magnetar/Be star binary model. Here, a Be-star has a magnetized decretion disk, and when its binary companion magnetar approaches periastron, the radio waves interact with the disk, resulting in the observed RM variation.

Such models naturally predict DM and scattering time variations with phase,
which we do not see.
 We can look to Galactic neutron-star/OB binary systems for indications of the expected sizes of DM or scattering variations due to changes in location of the radio source within the companion wind.
\cite{jwn+01} studied the highly eccentric ($e=0.87$) 3.4-yr radio pulsar/Be star binary PSR B1259$-$63, which exhibited
DM variations of over $\sim$10~pc~cm$^{-3}$, and scattering times that ranged from 0.1 to nearly 10 ms at 1.5 GHz, both correlated with orbital phase.  Similarly, \cite{snowbird} studied the CHIME/FRB-discovered pulsar/OBe star binary PSR J2108+4516, which has an orbital period 269 days and eccentricity 0.09, showing that the pulsar exhibited DM variations of amplitude $\sim$4~pc~cm$^{-3}$, and scattering times ranging from well under 1 ms to over 20 ms in the CHIME band, both also correlated with orbital phase.  The latter system, with only modest eccentricity (0.09), has a projected semi-major axis of over 800 light seconds, implying only modest proximity of the two objects.

In \frb, by contrast, we do not observe any such correlated DM or scattering variations, in spite of strong modulation of the burst rate. On the whole, this argues against a binary orbit with a massive star as the origin of the periodicity in \frb's burst rate. 

We note that a massive companion with a weak wind might be invoked to explain the lack of scattering and DM variability. This has been observed for the pulsar binary PSR J0045$-$7319 in the Small Magellanic Cloud (SMC) \citep{kaspi_binary_wind}. Here the apparent low mass-loss rate of the companion, as implied by the absence of any DM variation in spite of a highly eccentric orbit and great proximity of the sources at periastron, is explained as a result of the low metallicity of the SMC, and the fact that OB stars have radiatively driven winds that make use of metals for the mass-loss mechanism.  However, the host galaxy of \frb has a metallicity similar to that of the Milky Way, which suggests that any binary companion to \frb would have a wind with Milky Way-type metal abundances \citep{mar2020}.

These models also predict a longer activity window at higher frequencies, which has not been observed and which also cannot easily explain the chromatic nature of the periodicity. 

Finally, a neutron star in orbit with a magnetized white dwarf has been argued to produce FRB emission by accretion \citep{Gu_NS_WD_binary,Chen_ULX_FRB}. This model requires a highly eccentric orbit to explain the duty cycle for \frb. 
\cite{Dai_asteroid} propose an old, slowly spinning pulsar captured by a star interacting with an extragalactic asteroid belt. Here interaction of asteroids with the magnetosphere of the neutron star leads to FRB emission. However, the recent dramatic long-term rise in RM is not naturally explained in this model.

Overall, the origin of the periodicity in \frb remains a mystery, with our observations providing key new constraints that further challenge existing models.

\section{Conclusion} \label{sec:conc}

In this study, we report a morphological analysis of 60 bursts observed from the periodic Fast Radio Burst source \frb, using data from CHIME/FRB from 2018 August to 2021 December. Out of these, 45 bursts were recorded with high-time resolution baseband data, which enabled us to study their properties in detail. We present a comprehensive outline of our pulse morphology pipeline, which has been integrated into the baseband processing system, and employ it to measure the DM and spectro-temporal properties of the bursts in our dataset.

Our analysis reveals that the variation in DM for \frb is constrained to be $\lessapprox$ 1\,pc\,cm$^{-3}$, in agreement with findings from other telescopes operating at different wavelengths. Furthermore, we measured scattering times, and found evidence for variation from burst to burst, ranging from $\sim$0.16 to over 2~ms, although in some cases unresolved microstructure may impact the measurement.  We find no evidence for a correlation of scattering time with the phase of the 16-day cycle. The lack of phase-dependent DM or scattering times seems at odds with models for the periodicity involving a binary orbit around a massive star. Additionally, we found no correlation between any morphological property and the reported secular RM increase of the source \citep{mckinven_2022_r3}. We also estimated the source activity period and provided constraints on the period derivative. Our analysis shows that the activity period of the source has remained unchanged, with a value of $16.34\pm0.07$ days, and the period derivative is constrained to be $-0.2 \pm 1.5 \times 10^{-4}$ day day$^{-1}$. The burst count of the source in each activity cycle was found to be consistent with a Poisson process, and the rate of the source showed no significant variation over our observation of 74 activity cycles. Unlike other prolific repeaters, \frb appears stable in its emission and has not exhibited any heightened activity above our detection threshold. We also discussed and provided constraints on various isolated and binary models proposed to explain the periodic nature of the source. \frb is an exciting puzzle, and further measurements of RM, DM, and scattering will hopefully provide insights into the evolution of the source environment and further constrain these models. Monitoring the source daily over an extended period will play a crucial role in this study, and CHIME/FRB is well-positioned to undertake this task.


\begin{acknowledgements}

CHIME is located on the traditional, ancestral, and unceded territory of the Syilx/Okanagan people.

We thank the Dominion Radio Astrophysical Observatory, operated by the National Research Council Canada, for gracious hospitality and expertise.

CHIME is funded by a grant from the Canada Foundation for Innovation (CFI) 2012 Leading Edge Fund (Project 31170) and by contributions from the provinces of British Columbia, Qu\'{e}bec and Ontario. The CHIME/FRB Project is funded by a grant from the CFI 2015 Innovation Fund (Project 33213) and by contributions from the provinces of British Columbia and Qu\'{e}bec, and by the Dunlap Institute for Astronomy and Astrophysics at the University of Toronto. Additional support was provided by the Canadian Institute for Advanced Research (CIFAR), McGill University and the Trottier Space Institute via the Trottier Family Foundation, and the University of British Columbia. The CHIME/FRB baseband system was funded in part by a CFI John R. Evans Leaders Fund grant to IHS. The Dunlap Institute is funded through an endowment established by the David Dunlap family and the University of Toronto. Research at Perimeter Institute is supported by the Government of Canada through Industry Canada and by the Province of Ontario through the Ministry of Research \& Innovation. The National Radio Astronomy Observatory is a facility of the National Science Foundation (NSF) operated under cooperative agreement by Associated Universities, Inc. FRB research at UBC is supported by an NSERC Discovery Grant and by the Canadian Institute for Advanced Research.  

K.R.S. acknowledges support from Murata Family Fellowship and Fonds de Recherche du Quebec -- Nature et Technologies~(FRQNT) Doctoral Research Award. V.M.K. holds the Lorne Trottier Chair in Astrophysics and Cosmology, a Distinguished James McGill Professorship, and receives support from an NSERC Discovery grant (RGPIN 228738-13), from an R. Howard Webster Foundation Fellowship from CIFAR, and from the FRQNT CRAQ. K.N. is a Kavli Fellow. Z.P. is a Dunlap Fellow. K.S. is supported by the NSF Graduate Research Fellowship Program. B.\,C.\,A. is supported by an FRQNT Doctoral Research Award. M.B. is a Mcwilliams fellow and an IAU Gruber fellow. A.M.C. is funded by an NSERC Doctoral Postgraduate Scholarship. A.P.C. is a Vanier Canada Graduate Scholar. F.A.D is supported by the UBC Four Year Fellowship. G.E. is supported by an NSERC Discovery Grant (RGPIN-2020-04554) and by a Canadian Statistical Sciences Institute (CANSSI) Collaborative Research Team Grant. B.M.G. is supported by an NSERC Discovery Grant (RGPIN-2022-03163), and by the Canada Research Chairs (CRC) program. K.\,W.\,M. holds the Adam J. Burgasser Chair in Astrophysics and is supported by NSF grants (2008031, 2018490). A.P. is funded by the NSERC Canada Graduate Scholarships -- Doctoral program. A.B.P. is a Banting Fellow, a McGill Space Institute~(MSI) Fellow, and a FRQNT postdoctoral fellow. D.C.S. is supported by an NSERC Discovery Grant (RGPIN-2021-03985) and by a Canadian Statistical Sciences Institute (CANSSI) Collaborative Research Team Grant.

\end{acknowledgements}

%

\vspace{5mm}
\facilities{CHIME}


\software{astropy \citep{astropy1,astropy2,astropy3}, bitshuffle \citep{mas17_bitshuffle}, DM Phase \citep{dm_phase}, emcee \citep{emcee}, HEALPix \citep{ghb+05_healpix}, healpy \citep{healpy}, hdf5 \citep{hdf5}, matplotlib \citep{matplotlib}, numpy \citep{numpy}, pandas \citep{pandas}, scipy \citep{scipy_Virtanen_2020}, statsmodel \citep{seabold2010statsmodels}.
          }



\appendix

\begin{figure}
\centering
\gridline{\fig{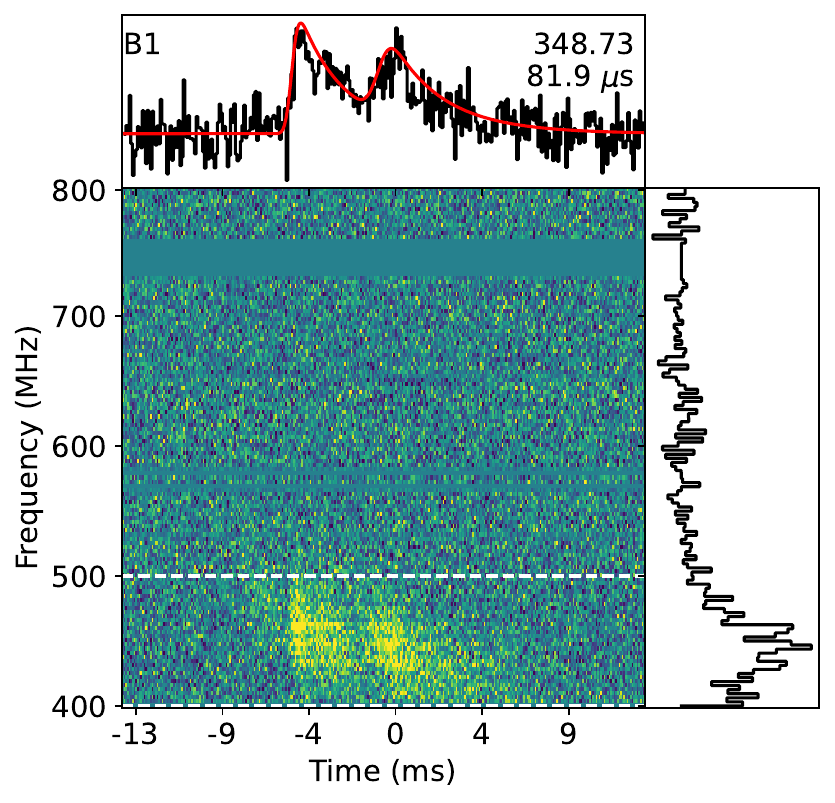}{0.2\textwidth}{}
          \fig{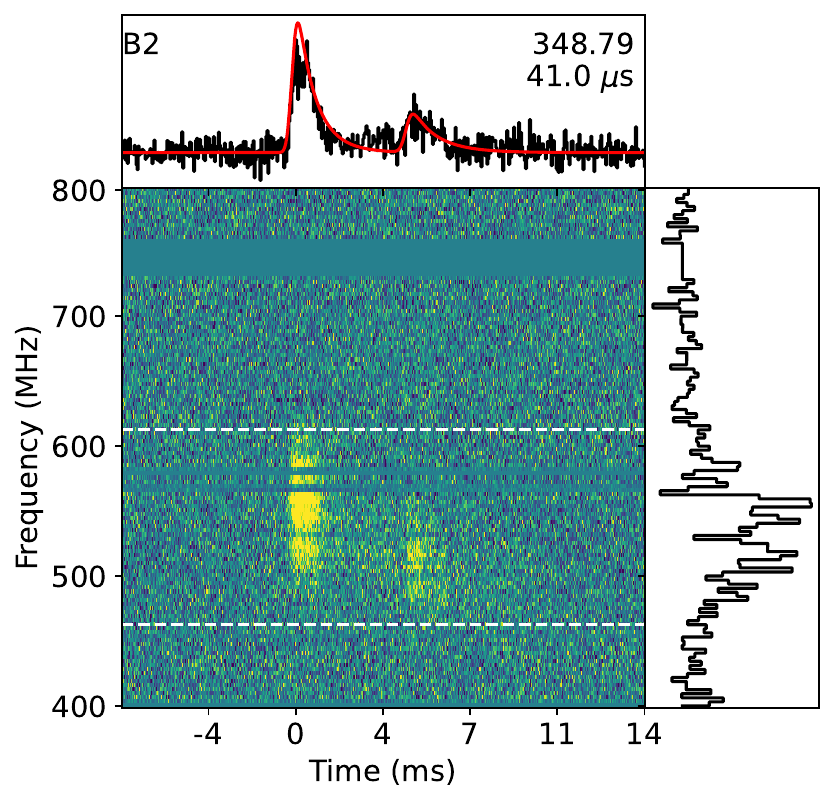}{0.2\textwidth}{}
          \fig{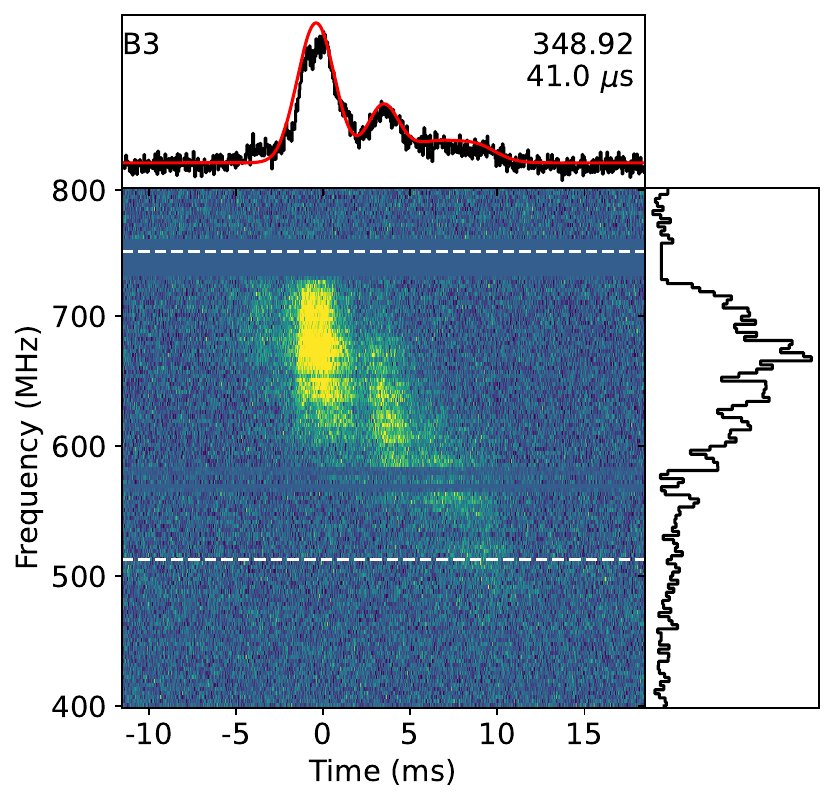}{0.2\textwidth}{}
          \fig{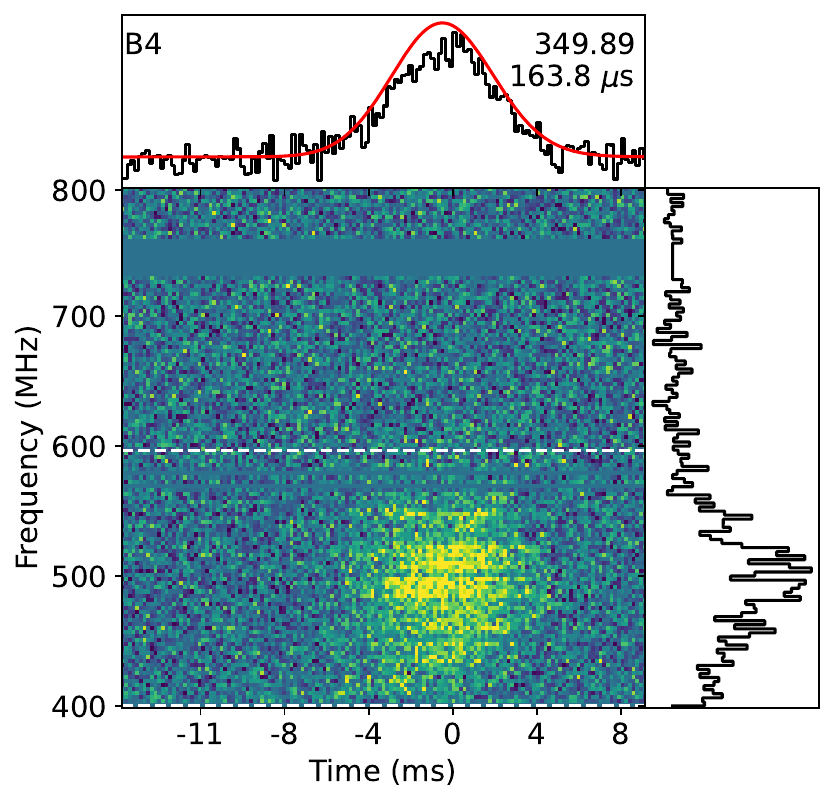}{0.2\textwidth}{}
          \fig{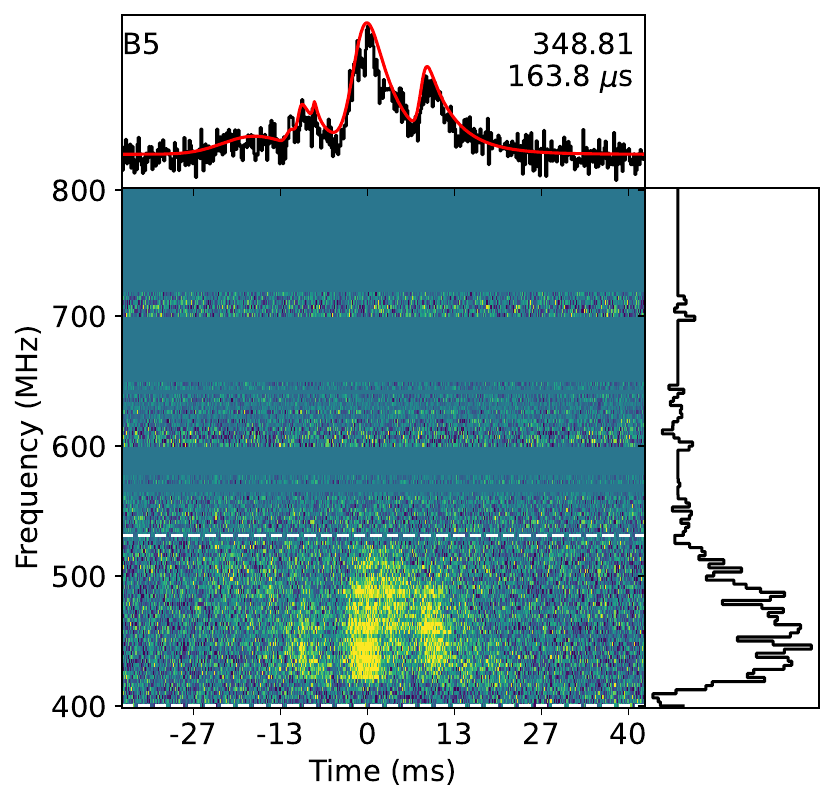}{0.2\textwidth}{}
          }
\gridline{\fig{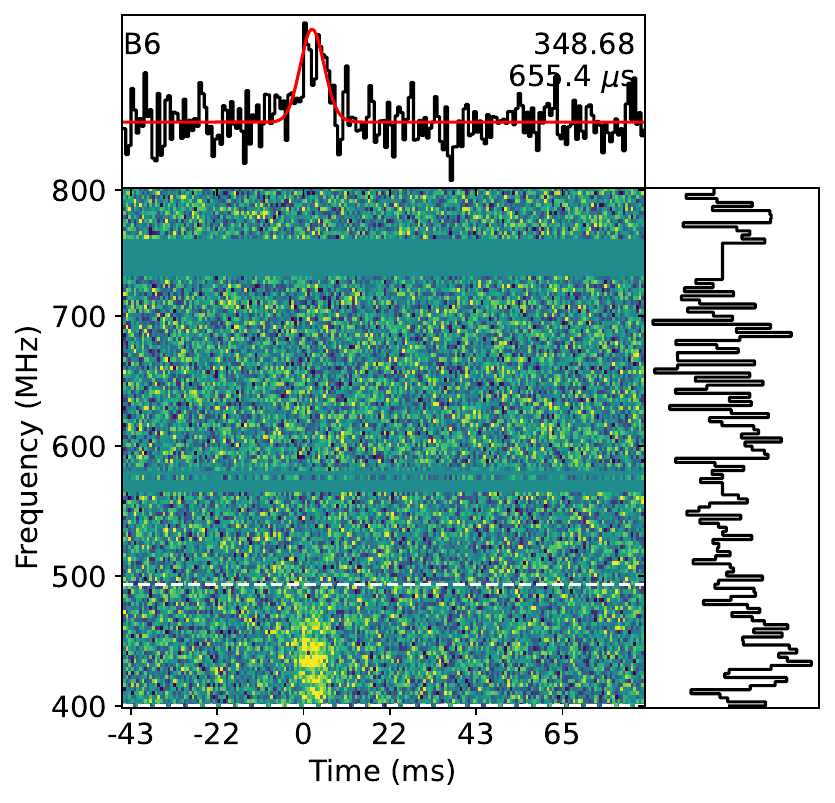}{0.2\textwidth}{}
          \fig{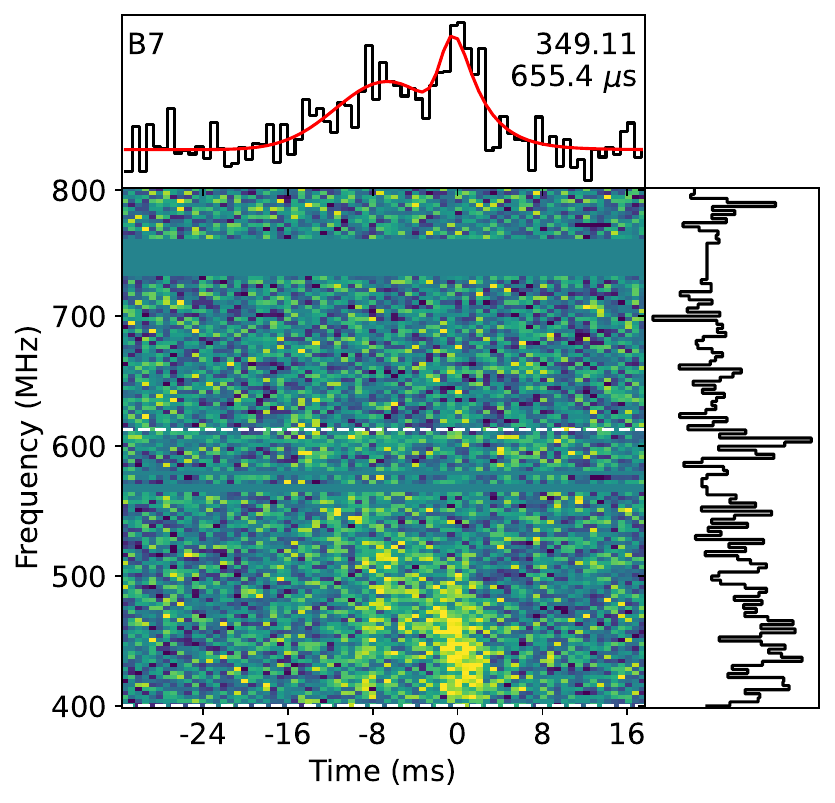}{0.2\textwidth}{}
          \fig{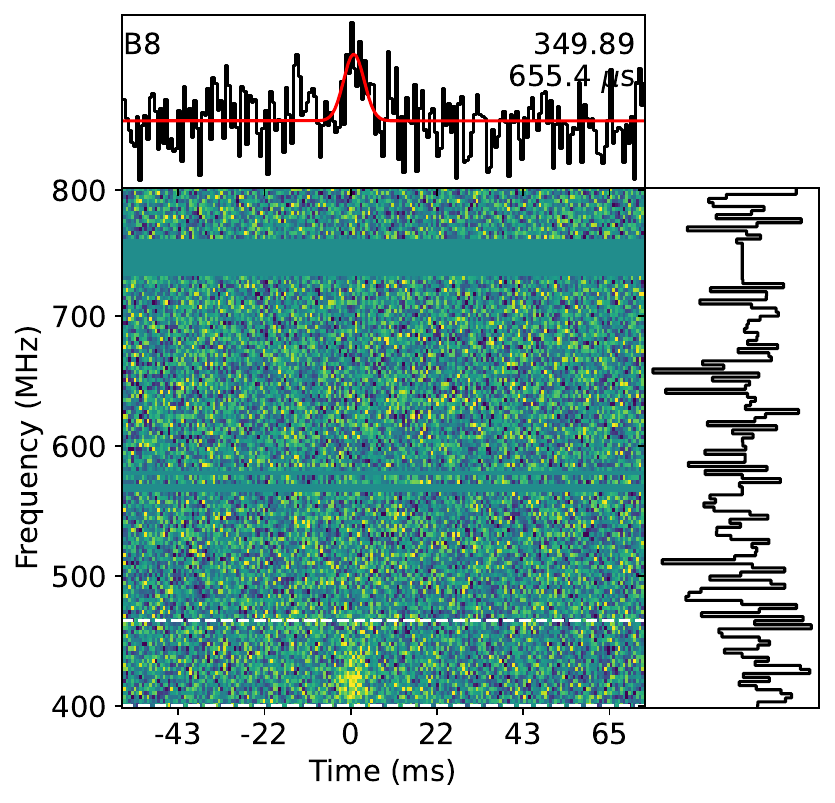}{0.2\textwidth}{}
          \fig{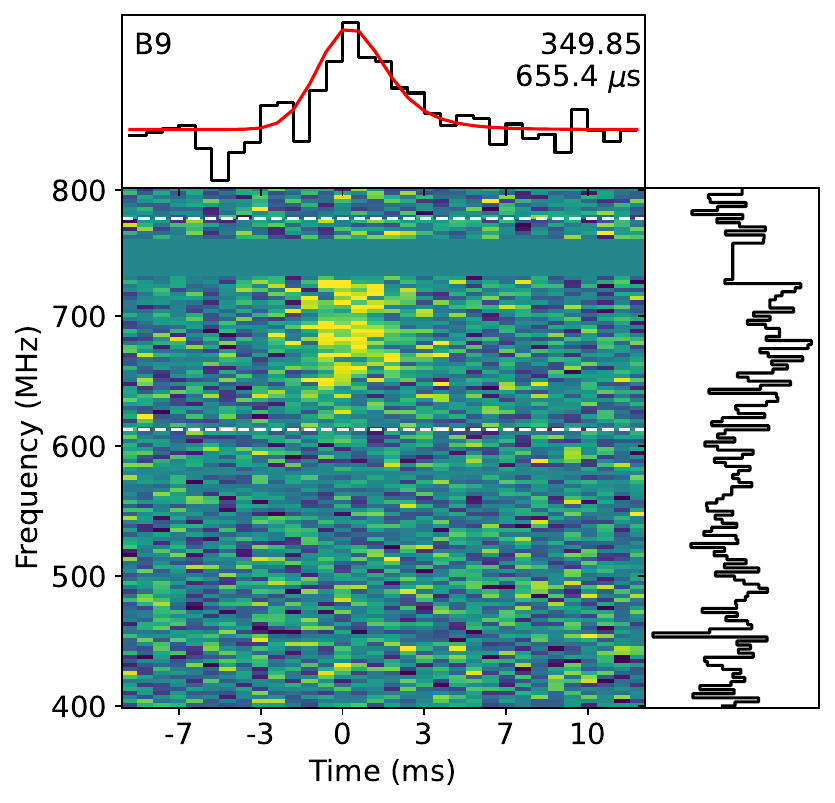}{0.2\textwidth}{}
          \fig{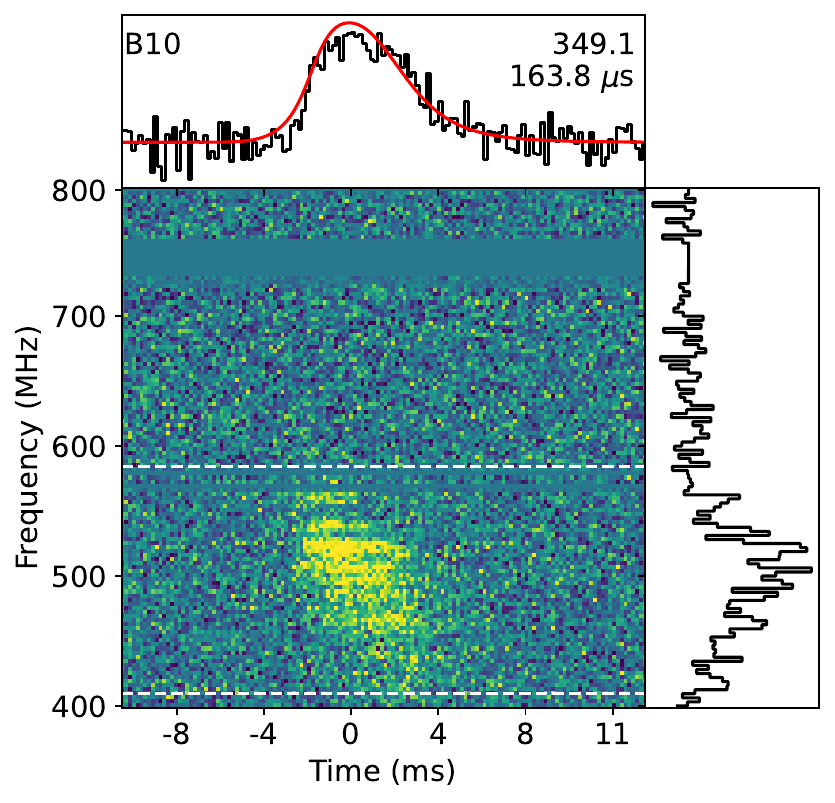}{0.2\textwidth}{}
          }
\gridline{\fig{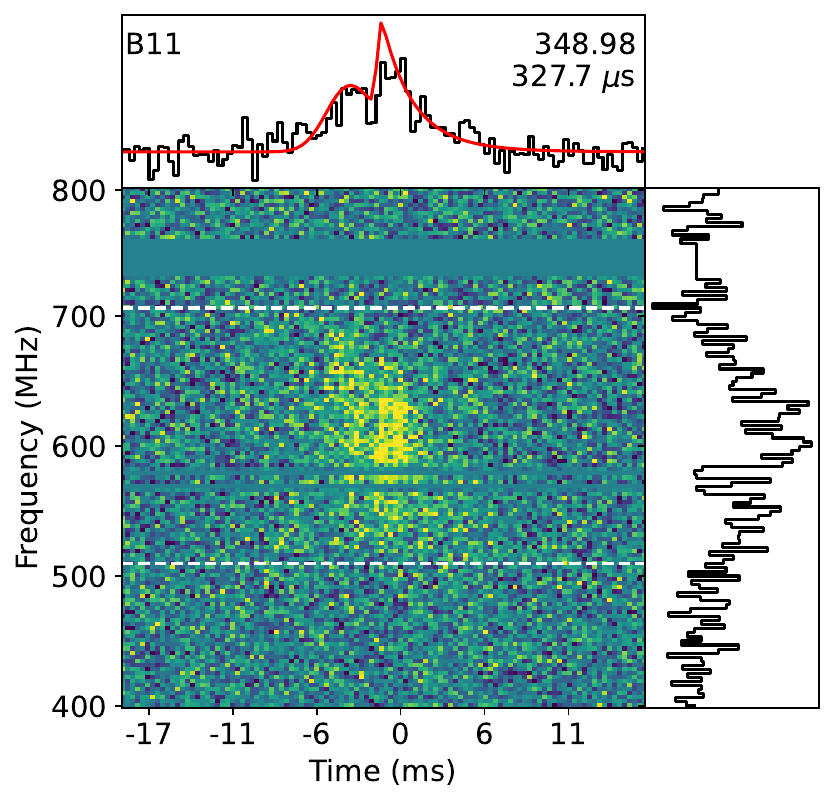}{0.2\textwidth}{}
          \fig{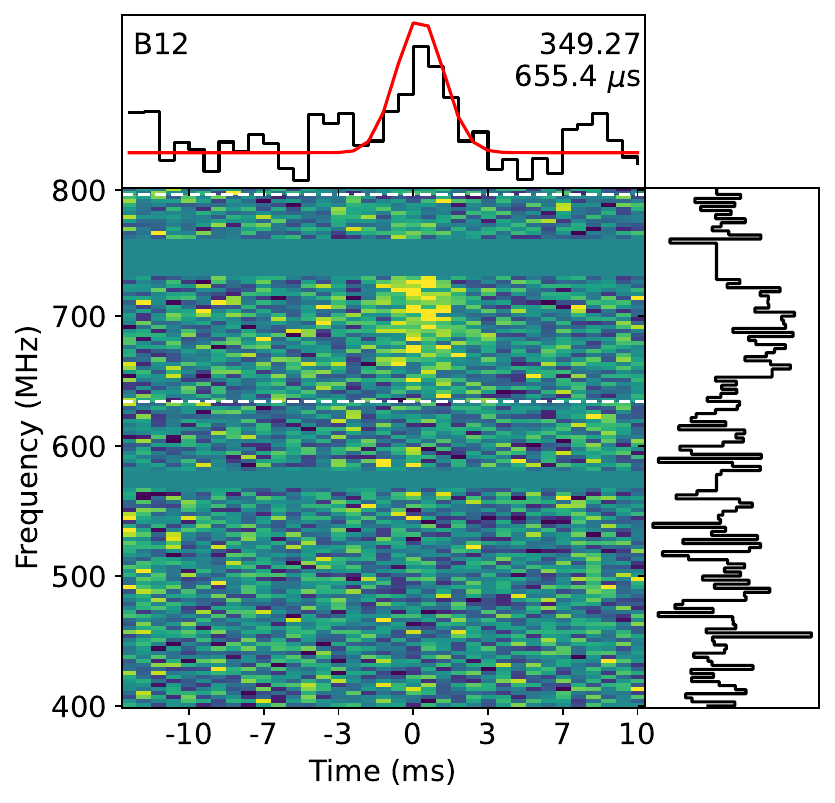}{0.2\textwidth}{}
          \fig{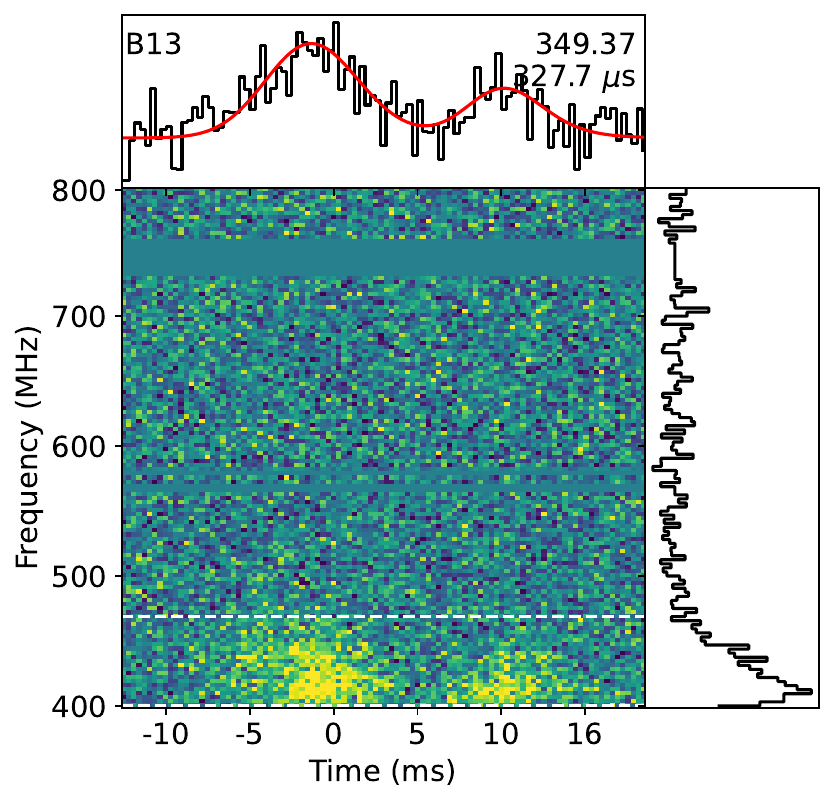}{0.2\textwidth}{}
          \fig{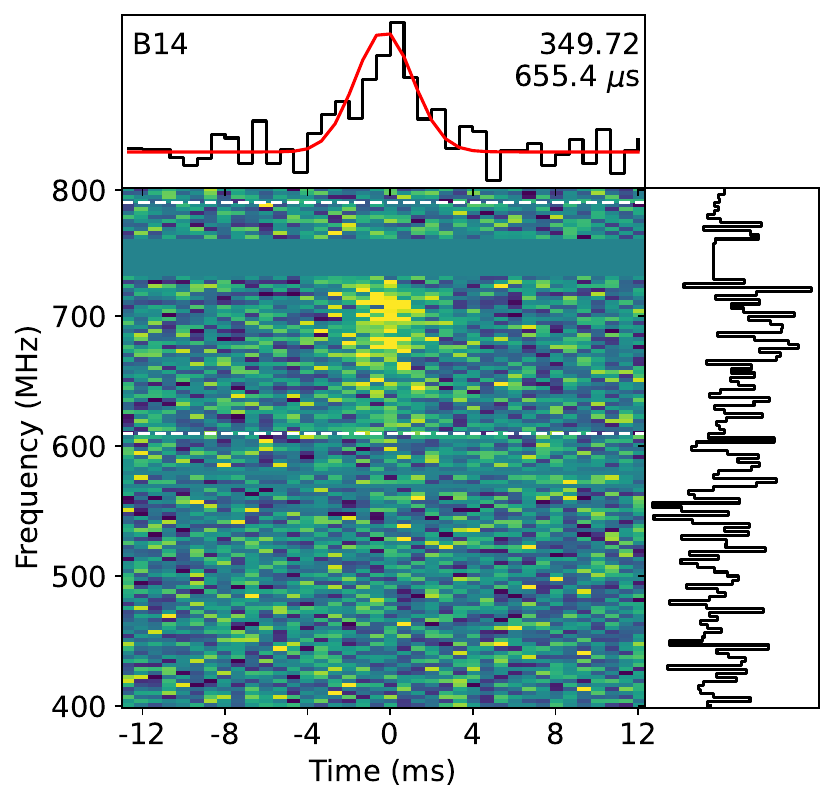}{0.2\textwidth}{}
          \fig{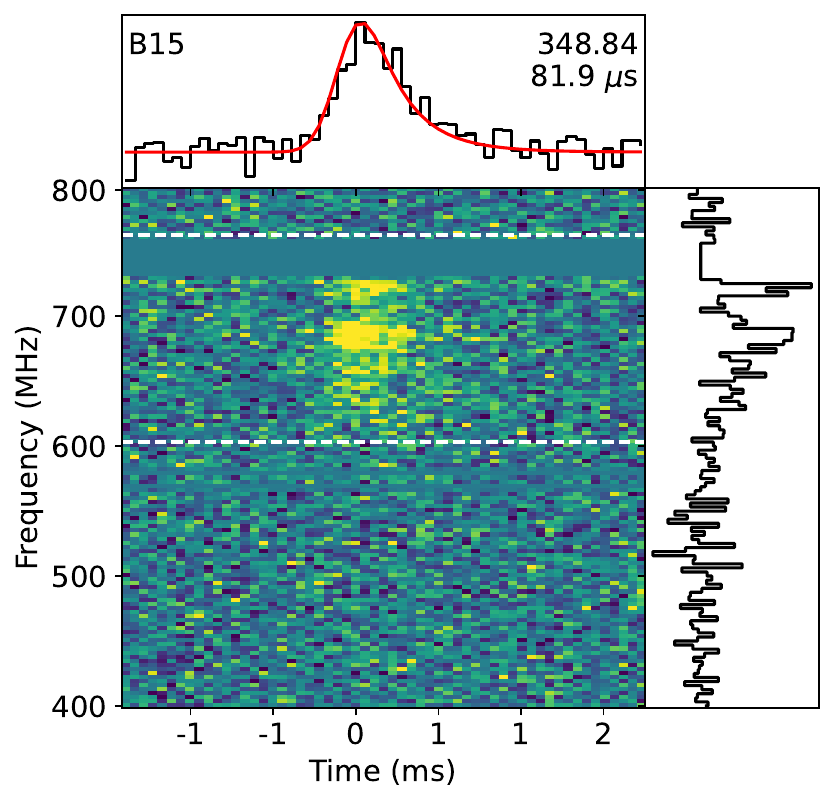}{0.2\textwidth}{}
          }
\gridline{\fig{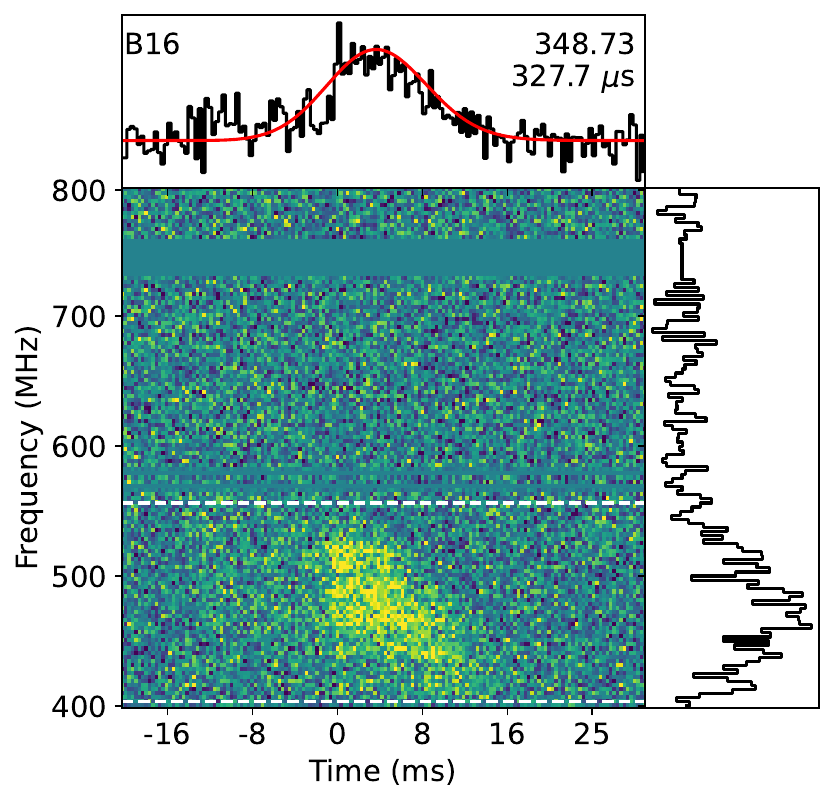}{0.2\textwidth}{}
          \fig{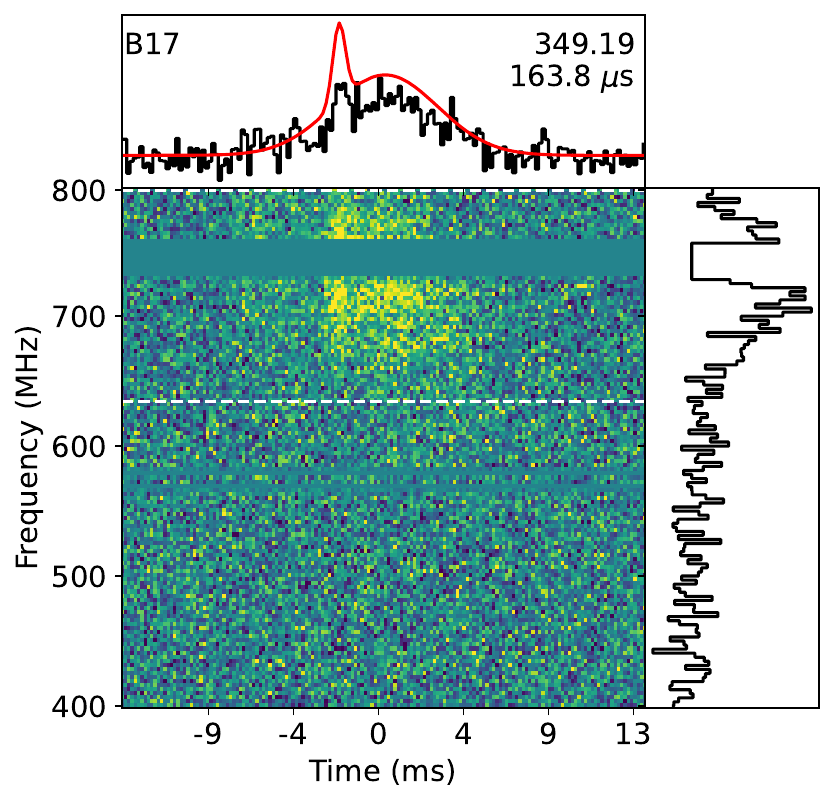}{0.2\textwidth}{}
          \fig{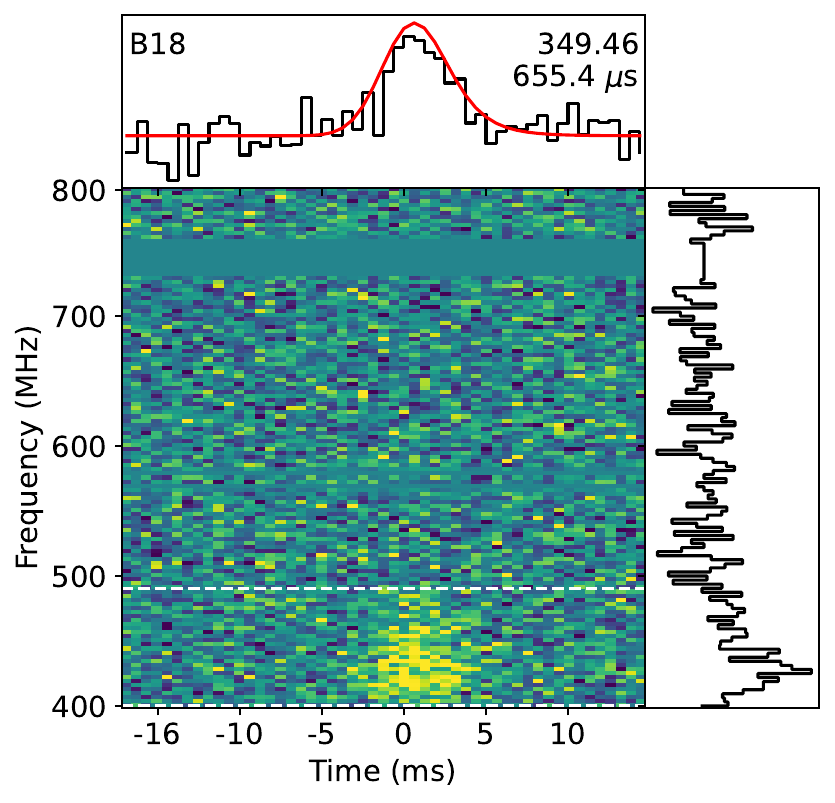}{0.2\textwidth}{}
          \fig{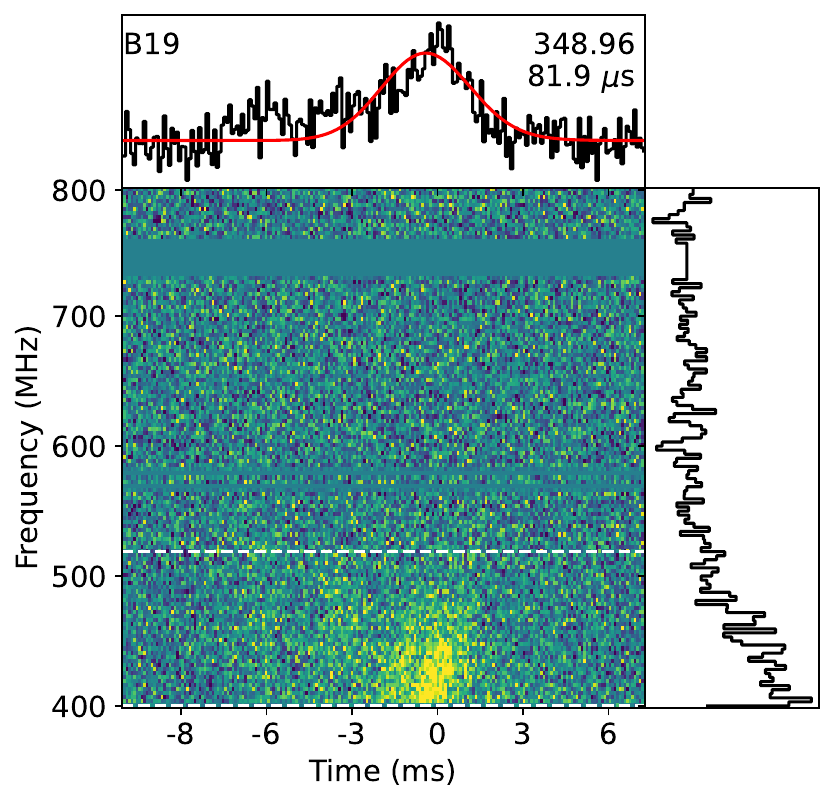}{0.2\textwidth}{}
          \fig{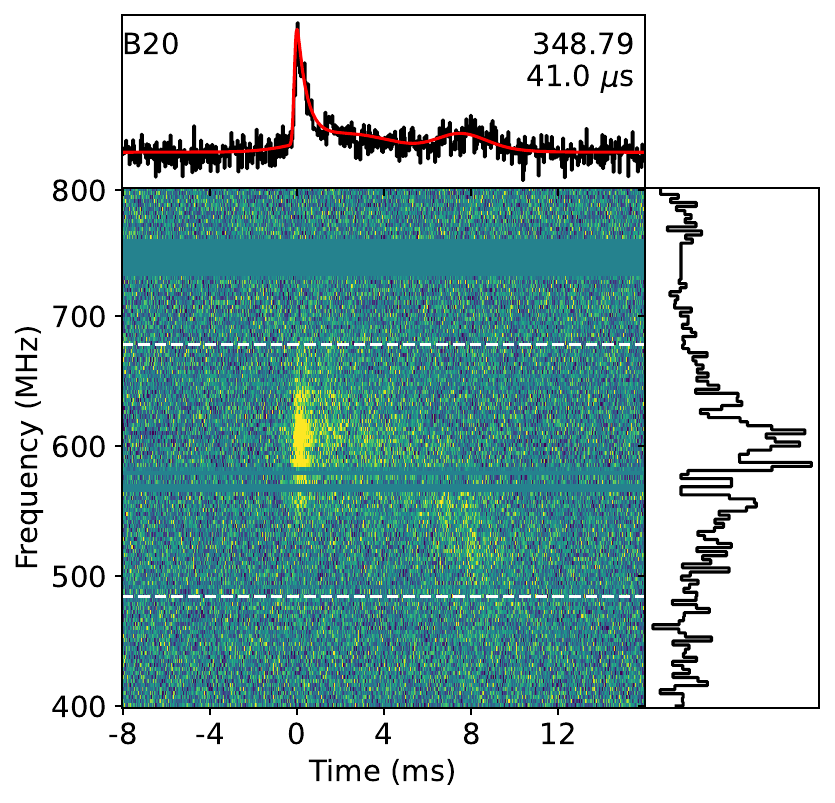}{0.2\textwidth}{}
          }
\gridline{\fig{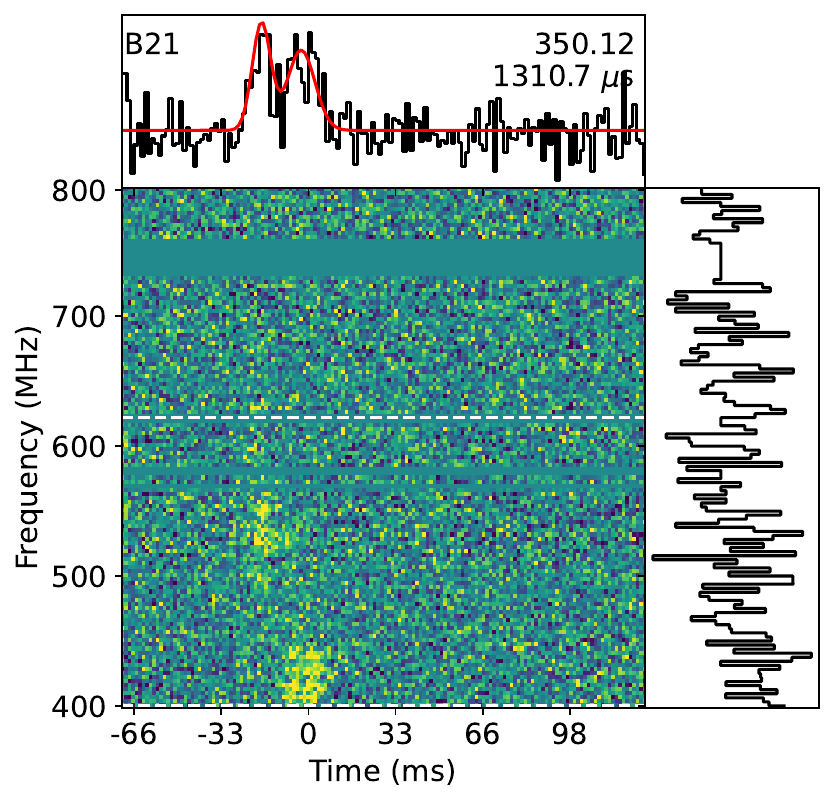}{0.2\textwidth}{}
          \fig{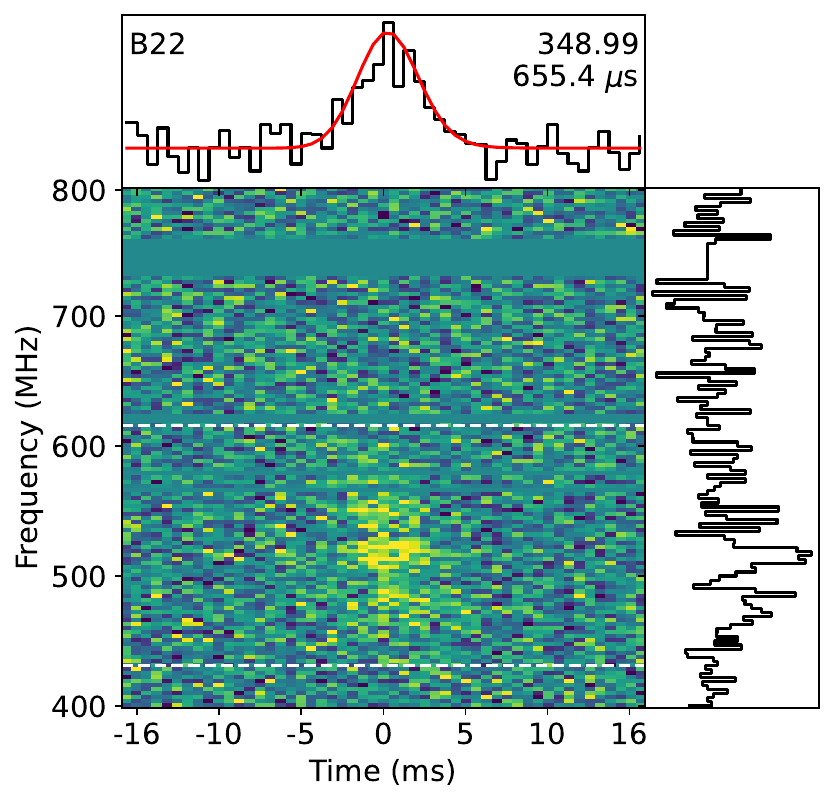}{0.2\textwidth}{}
          \fig{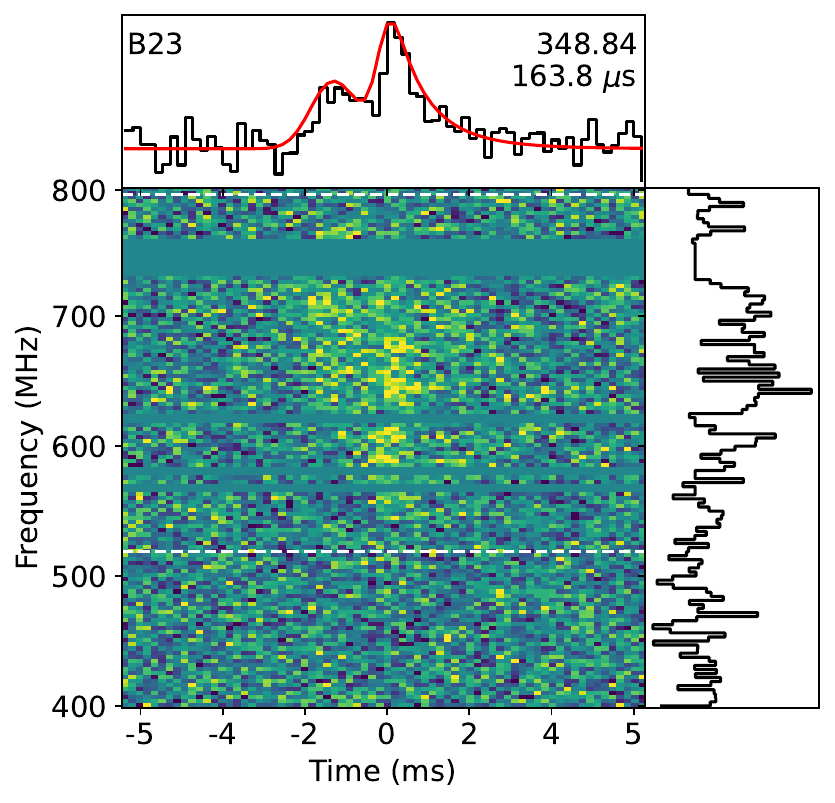}{0.2\textwidth}{}
          \fig{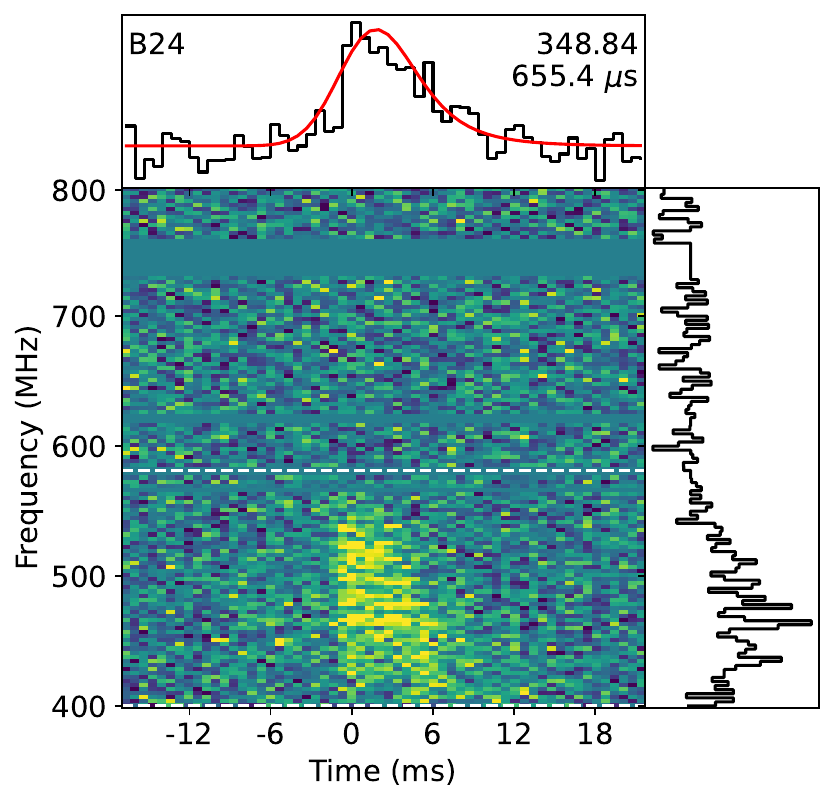}{0.2\textwidth}{}
          \fig{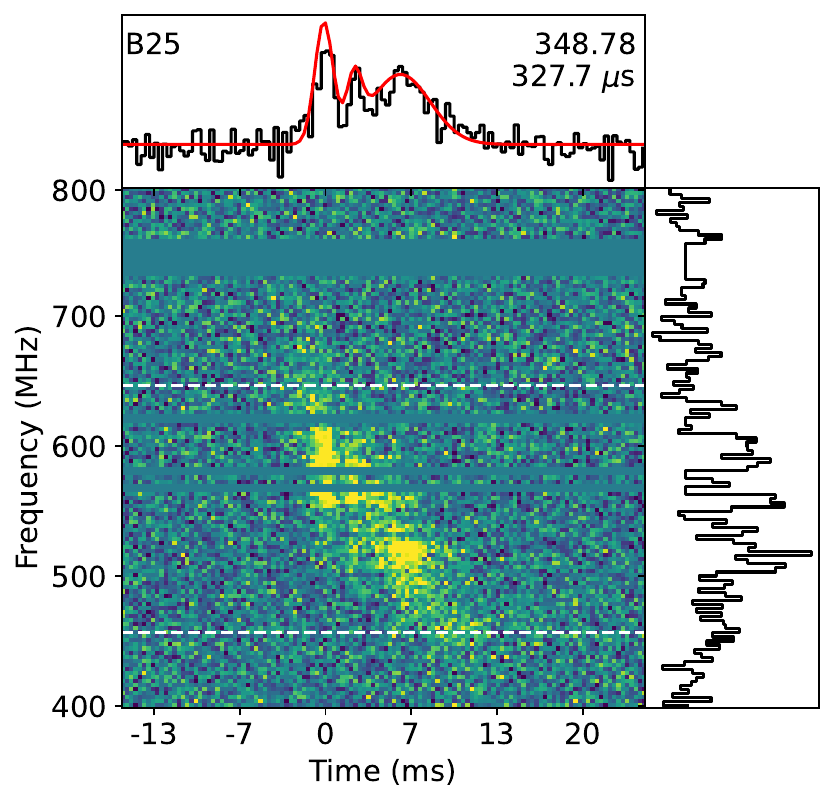}{0.2\textwidth}{}
          }

    \caption{Figure showing bursts detected by the baseband system from \frb dedispersed at their structure maximizing DM. In each sub-figure the top panel shows the time-series with red line showing the best fit, the panel below it shows the dynamic spectra and finally to the right of it is the frequency distribution of the power. The white-dotted lines show the extent of bandwidth. In top right we see the DM it has been dedispersed to and the time-resolutions its been plotted in microseconds, the top left shows the burst number. The number of channels here is 128. }
\label{fig:BB1}
\end{figure}

\begin{figure}
\centering
\gridline{\fig{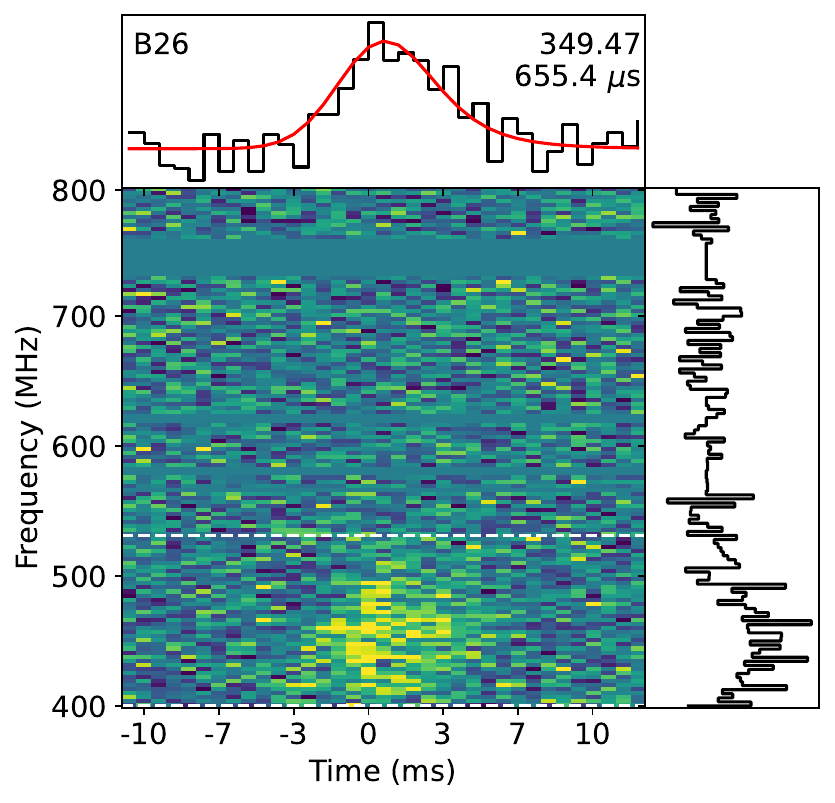}{0.2\textwidth}{}
          \fig{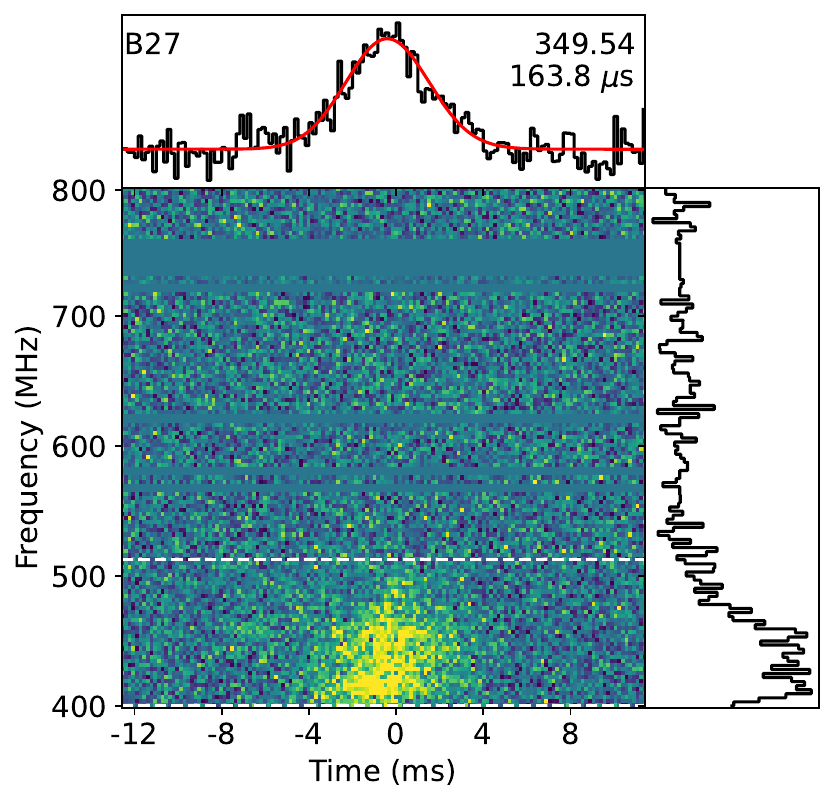}{0.2\textwidth}{}
          \fig{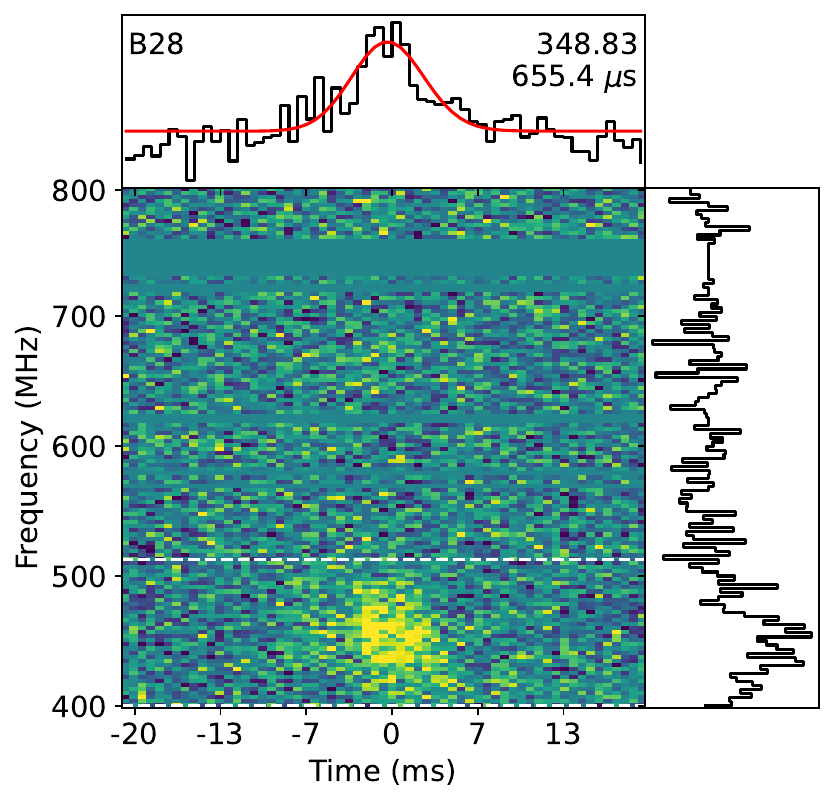}{0.2\textwidth}{}
          \fig{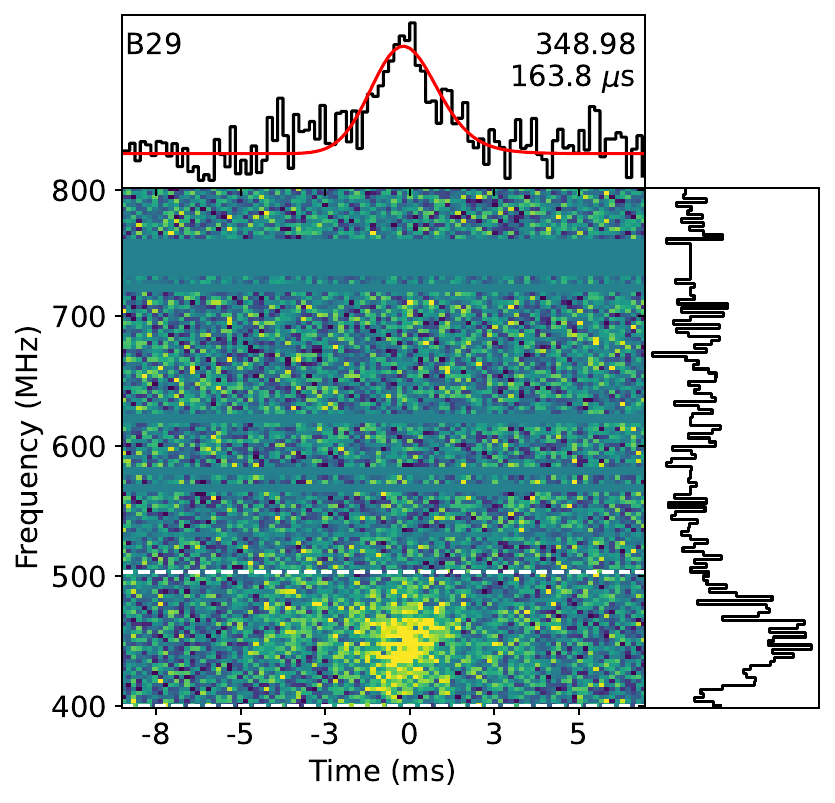}{0.2\textwidth}{}
          \fig{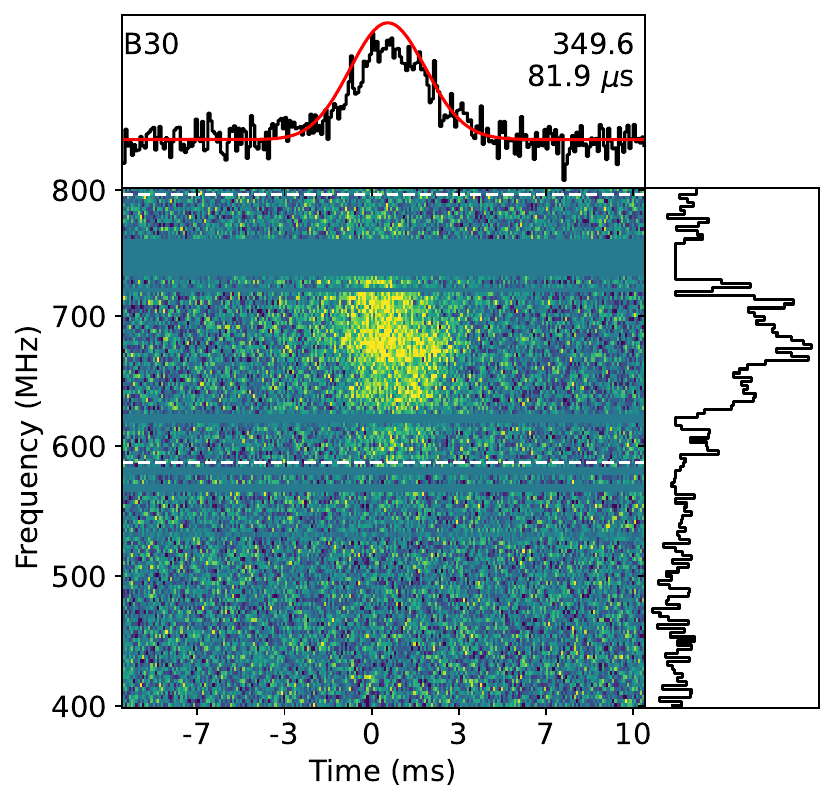}{0.2\textwidth}{}
          }
\gridline{\fig{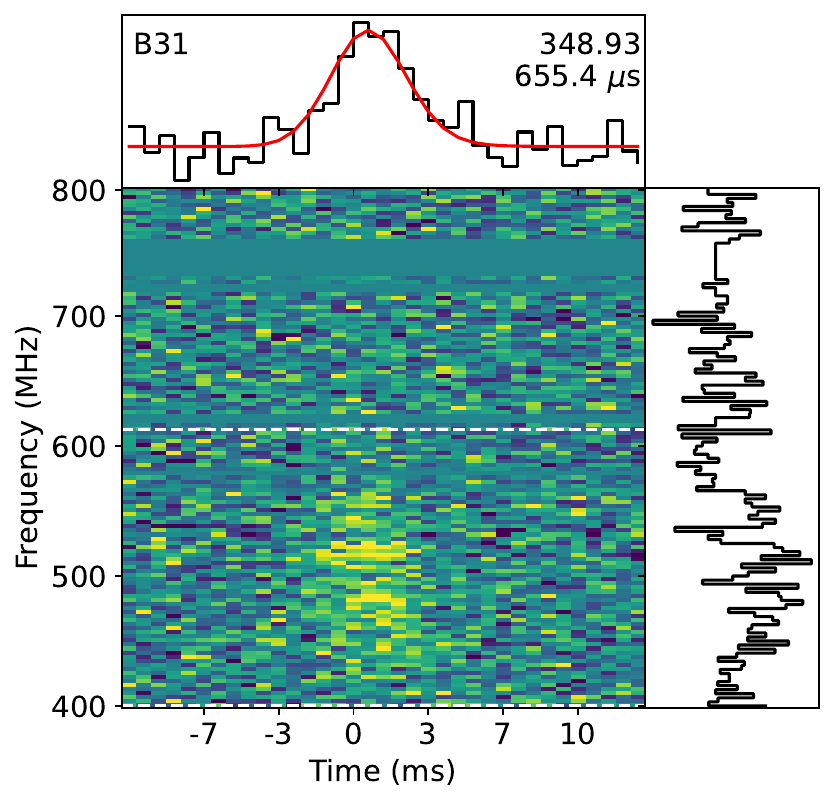}{0.2\textwidth}{}
          \fig{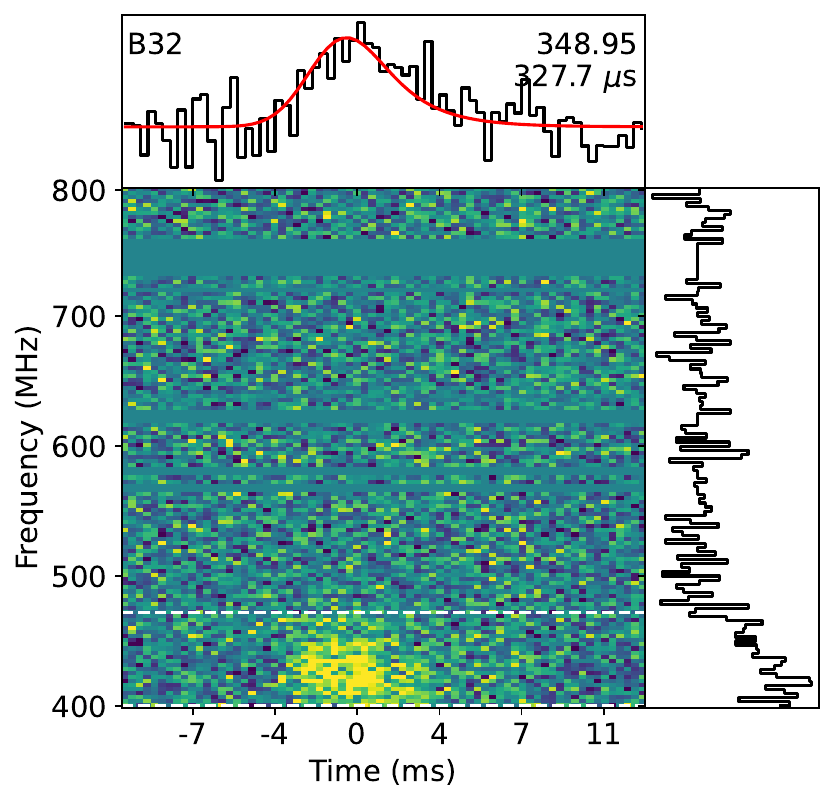}{0.2\textwidth}{}
          \fig{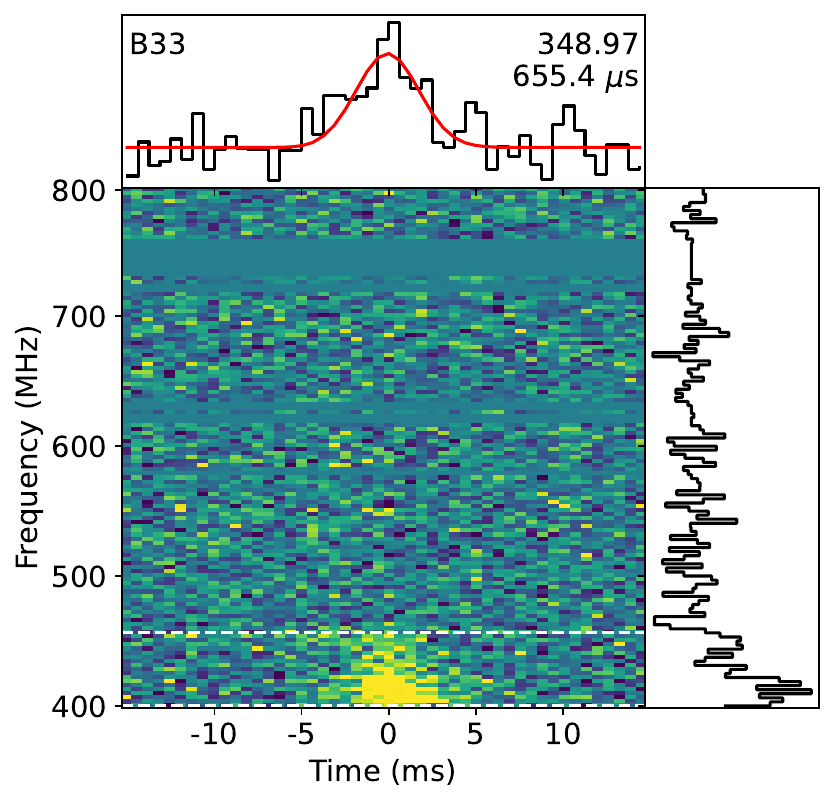}{0.2\textwidth}{}
          \fig{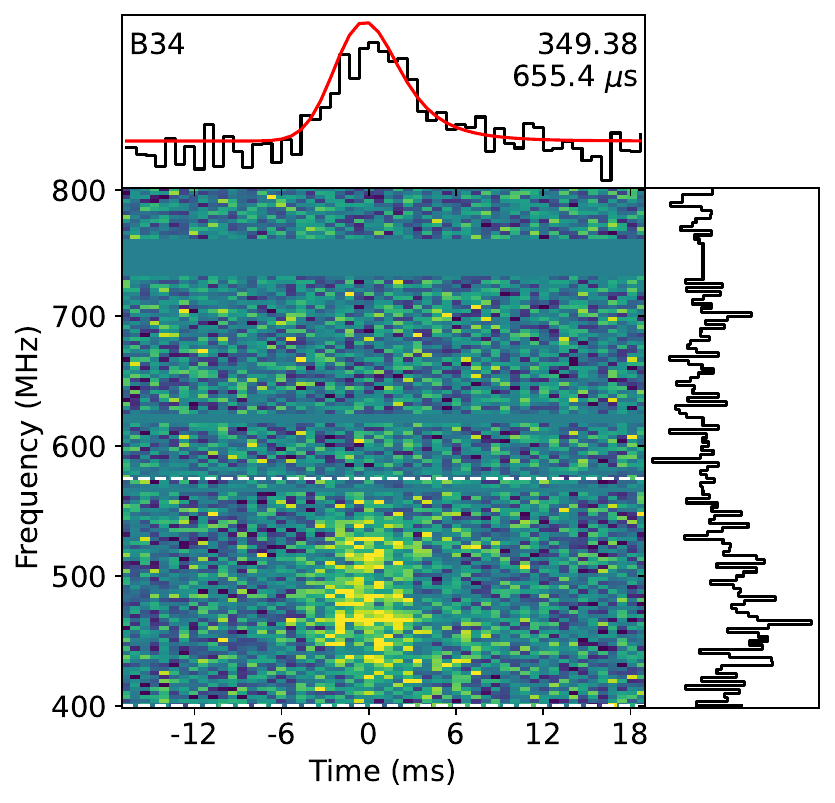}{0.2\textwidth}{}
          \fig{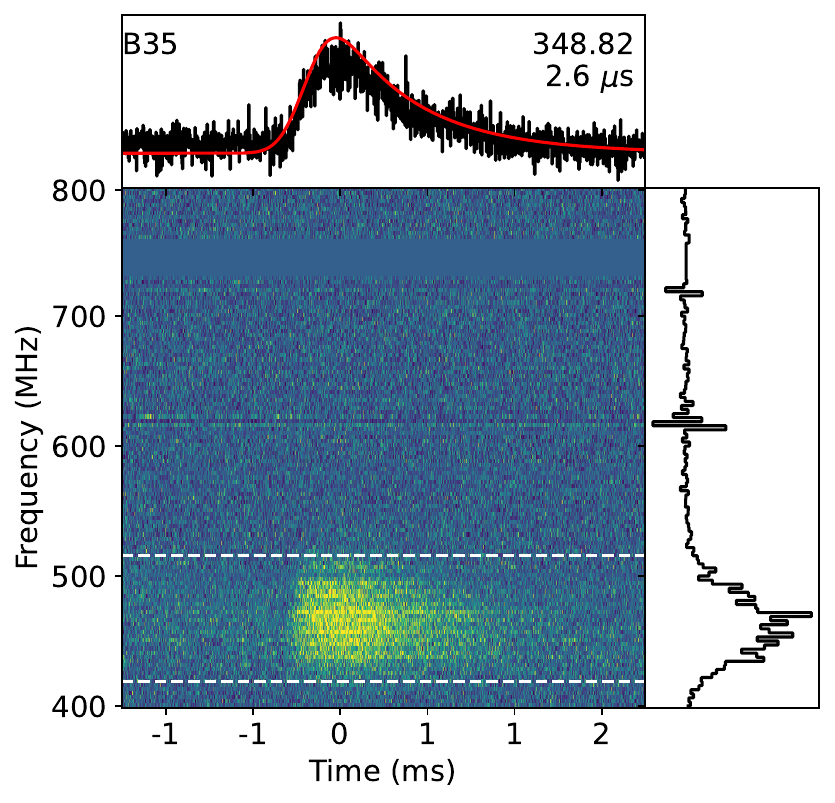}{0.2\textwidth}{}
          }
\gridline{\fig{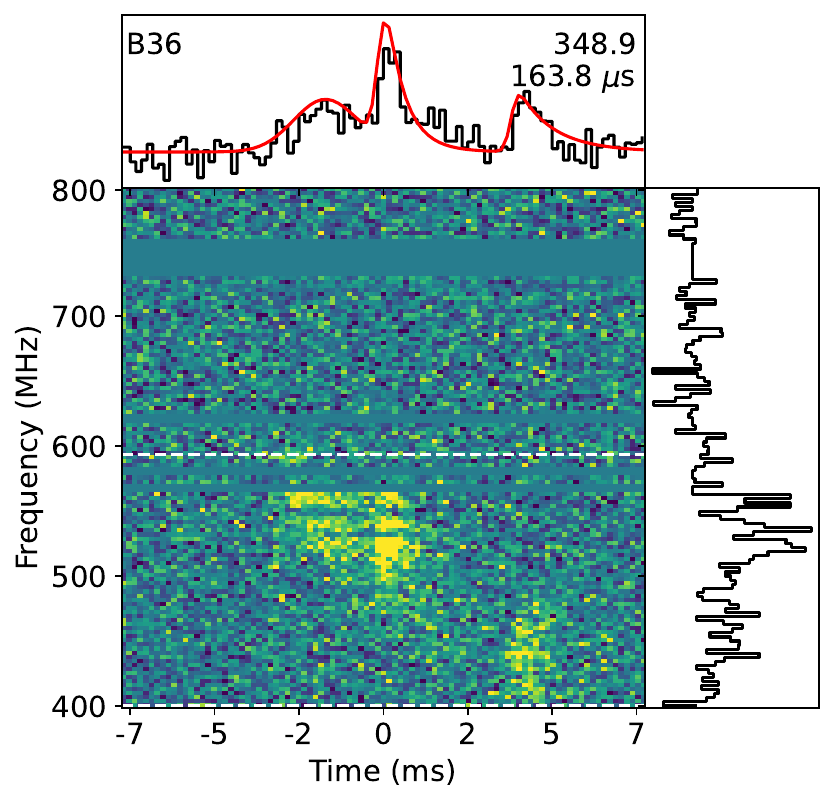}{0.2\textwidth}{}
          \fig{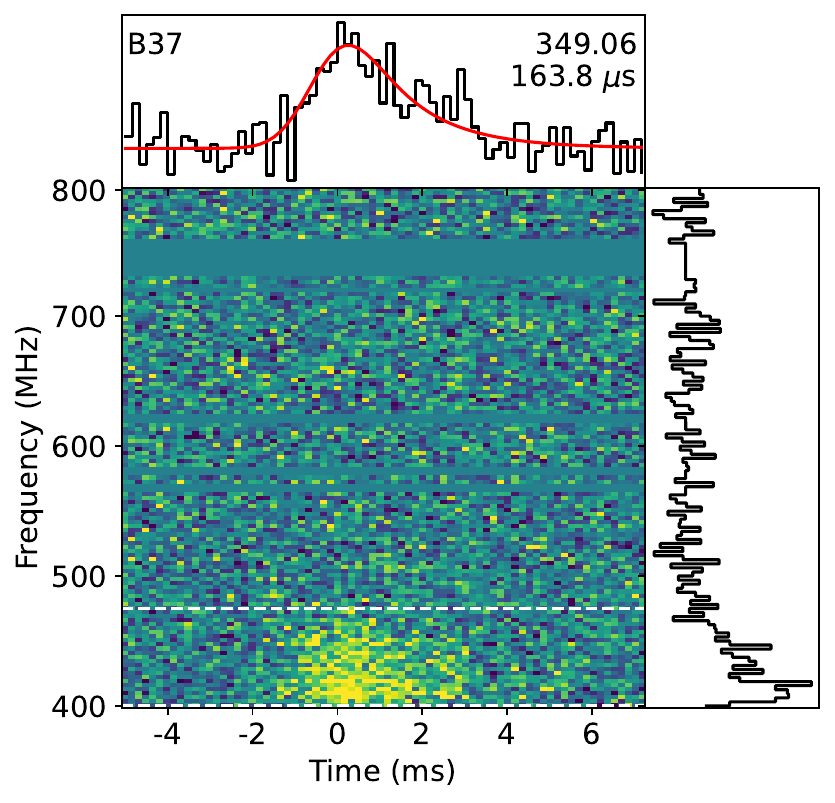}{0.2\textwidth}{}
          \fig{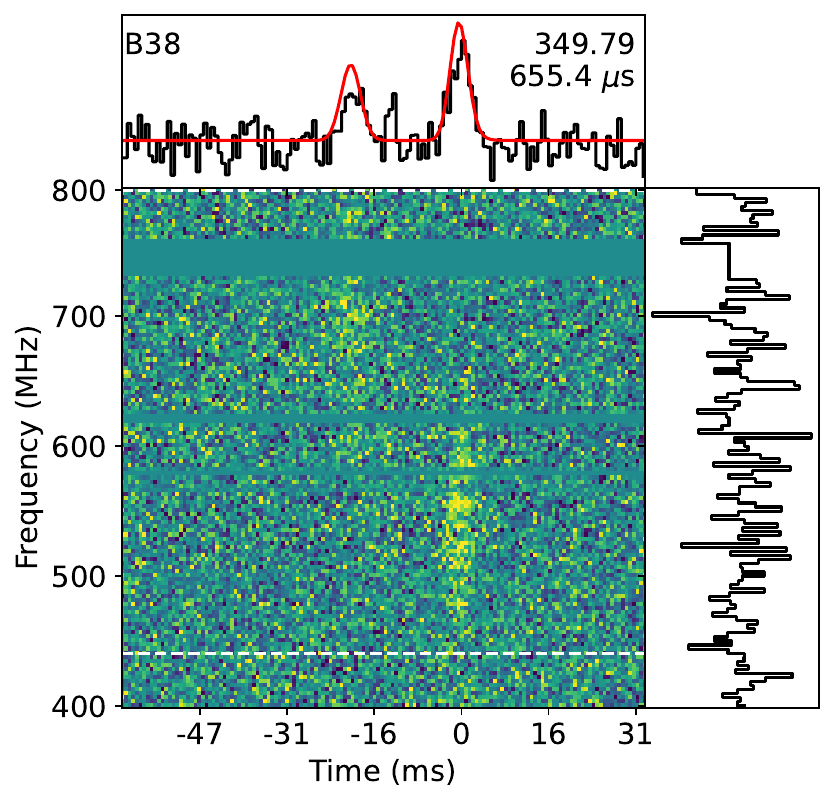}{0.2\textwidth}{}
          \fig{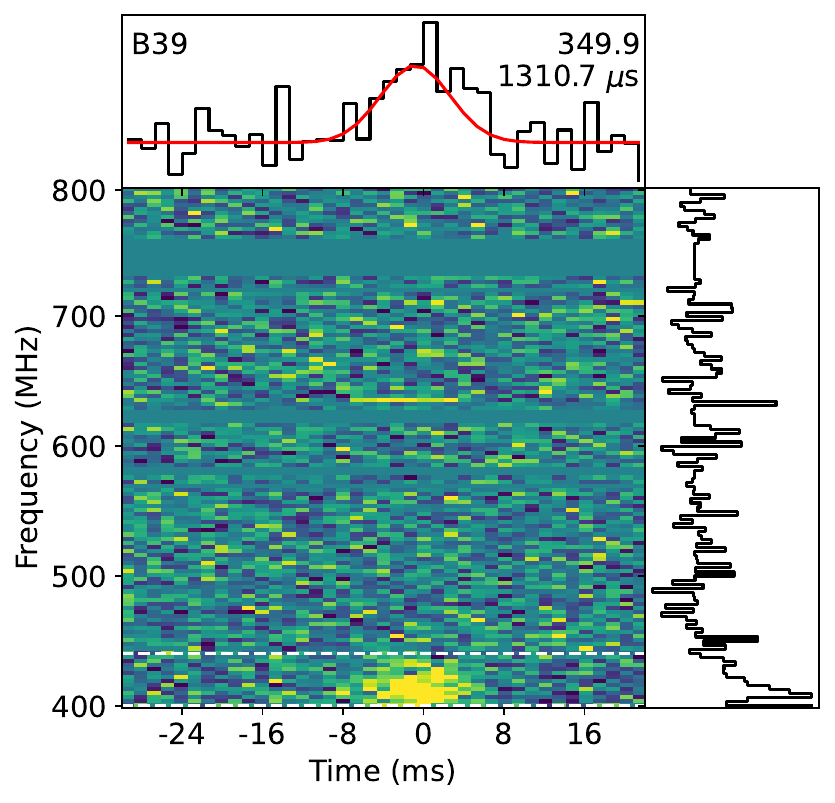}{0.2\textwidth}{}
          \fig{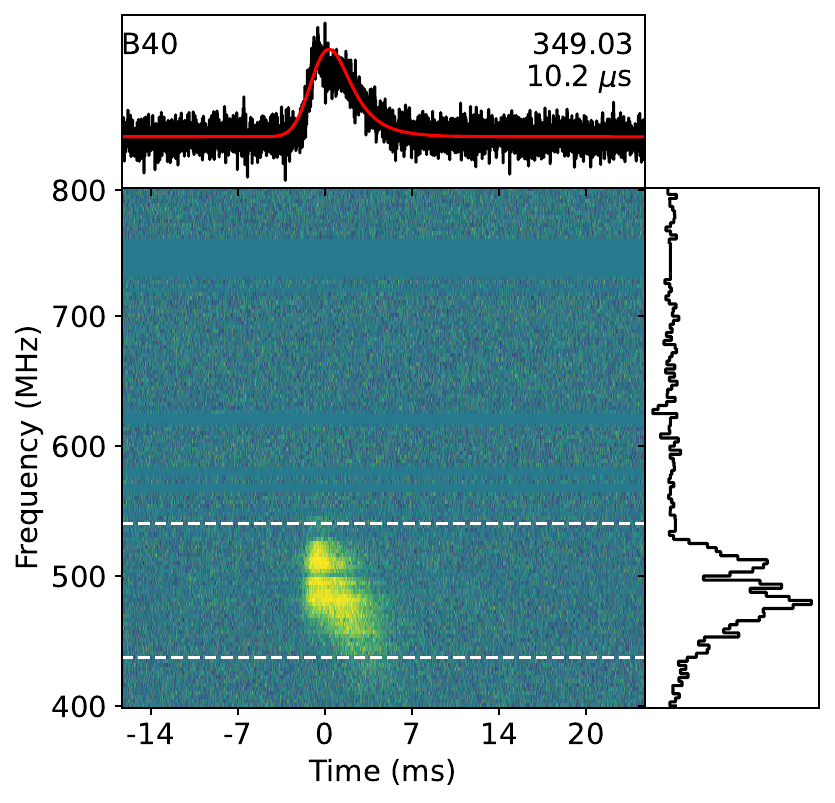}{0.2\textwidth}{}
          }
\gridline{\fig{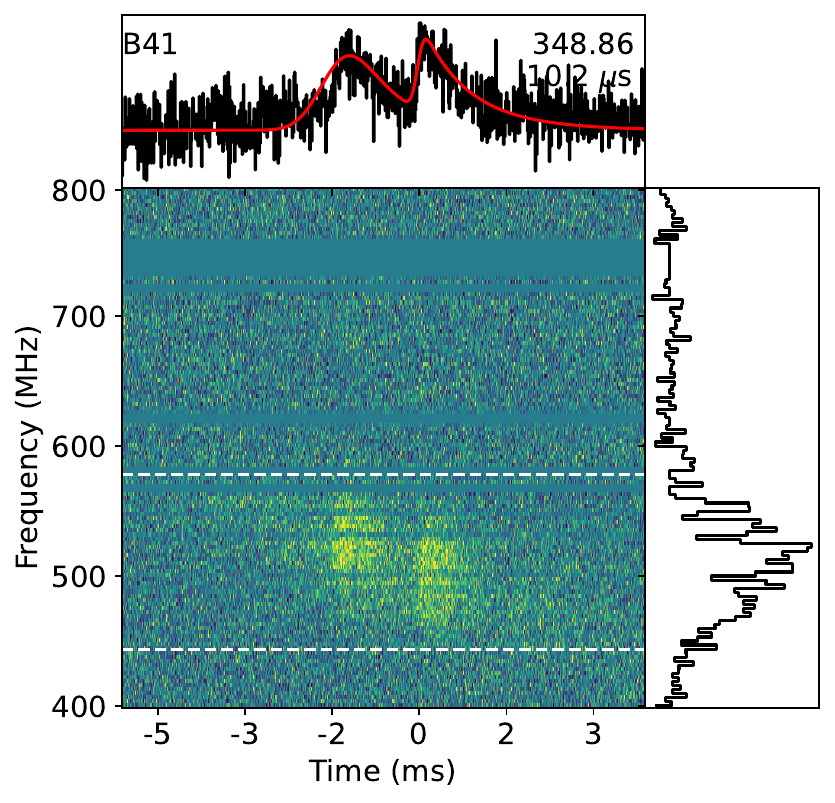}{0.2\textwidth}{}
          \fig{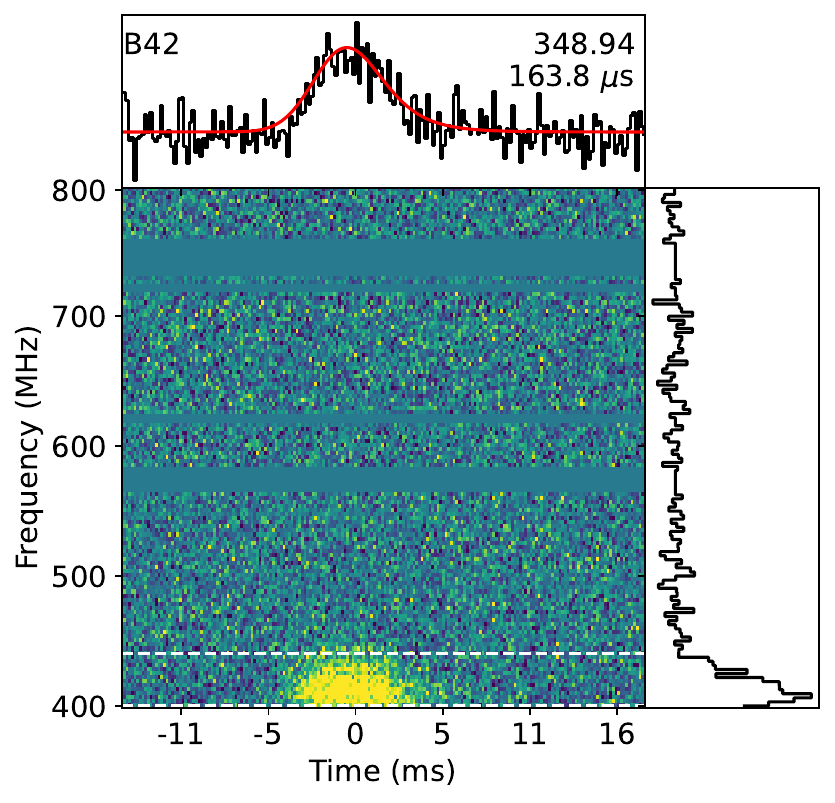}{0.2\textwidth}{}
          \fig{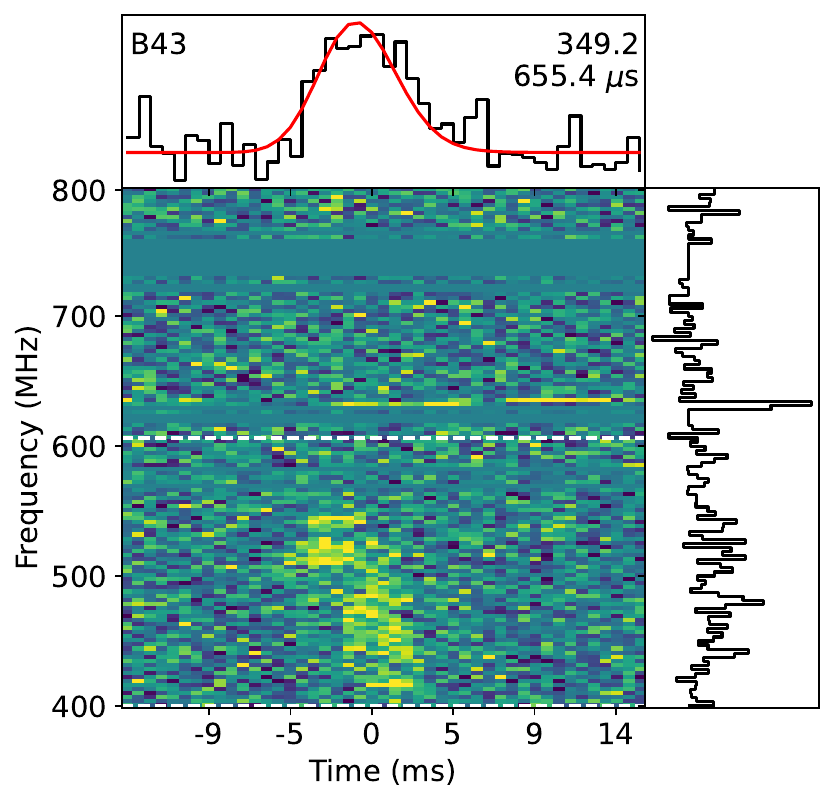}{0.2\textwidth}{}
          \fig{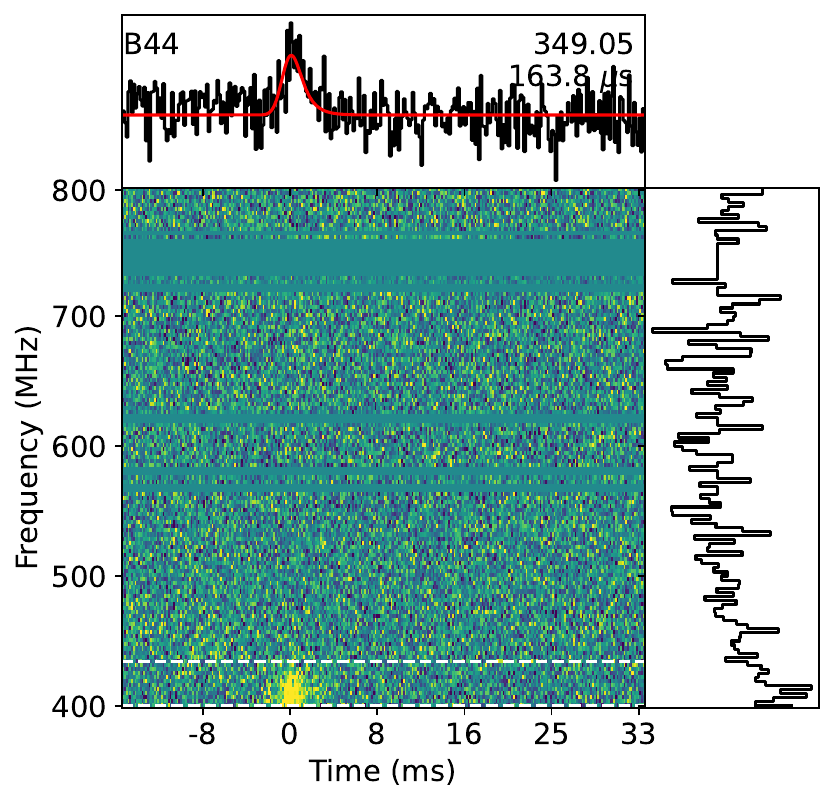}{0.2\textwidth}{}
          \fig{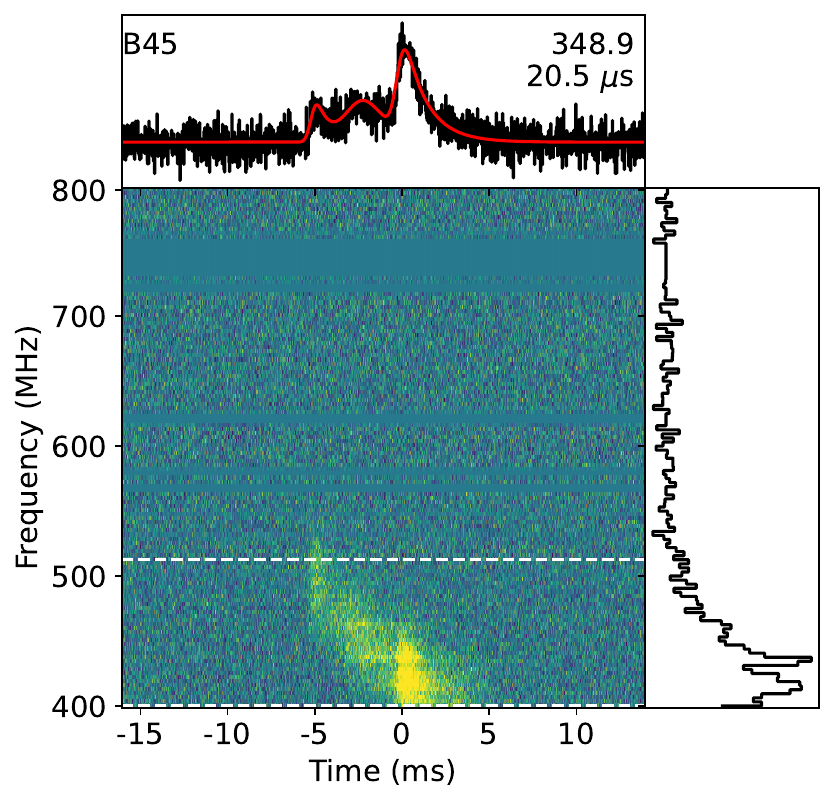}{0.2\textwidth}{}
          }

    \caption{Continued from Figure \ref{fig:BB1} }
\label{fig:BB2}
\end{figure}

\begin{figure}
\centering
\gridline{\fig{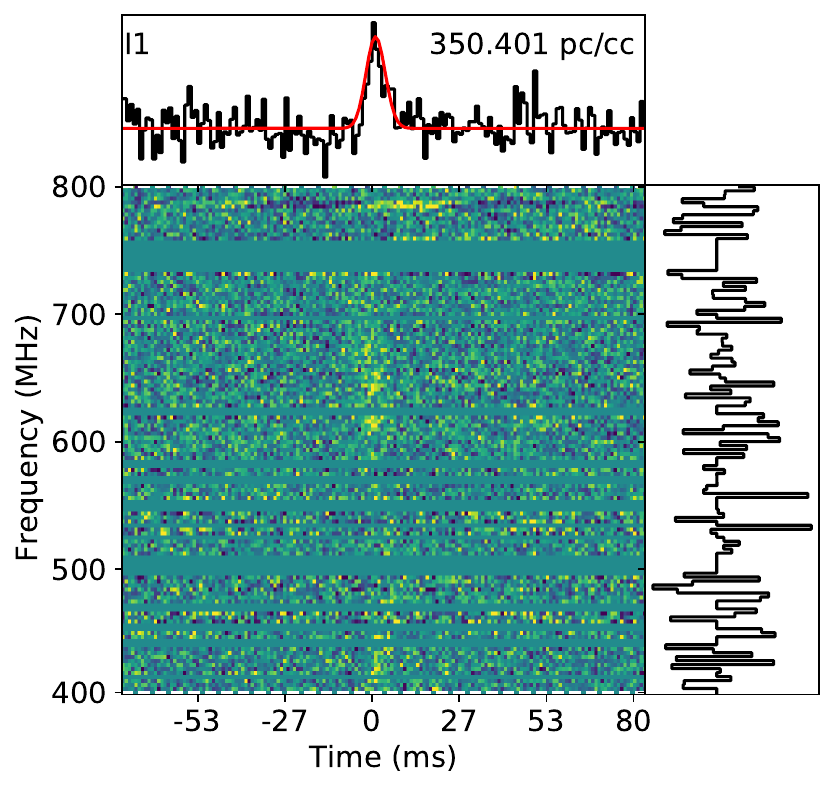}{0.2\textwidth}{}
          \fig{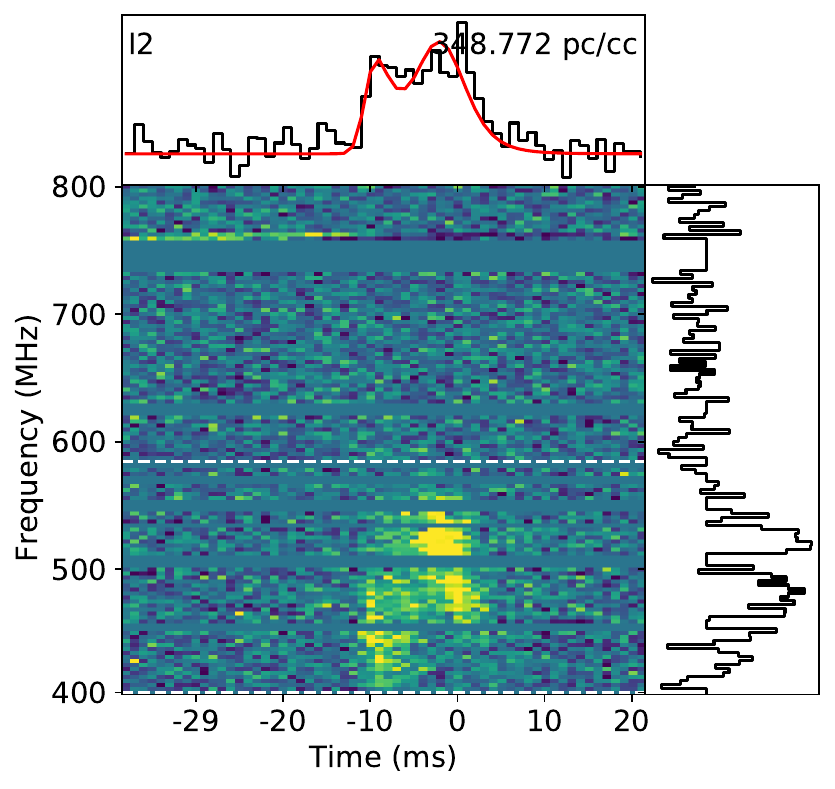}{0.2\textwidth}{}
          \fig{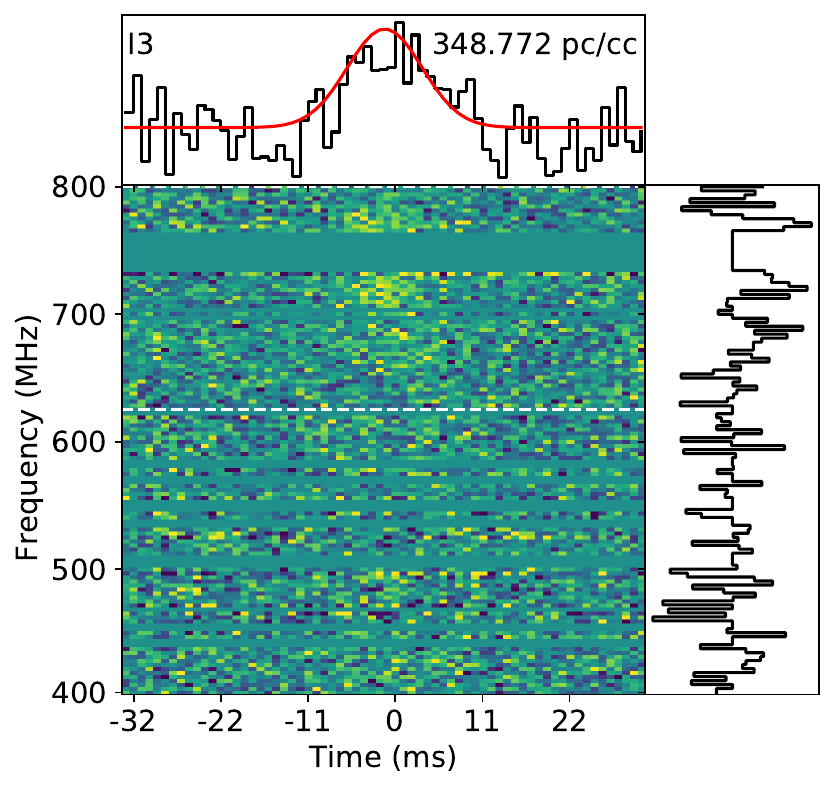}{0.2\textwidth}{}
          \fig{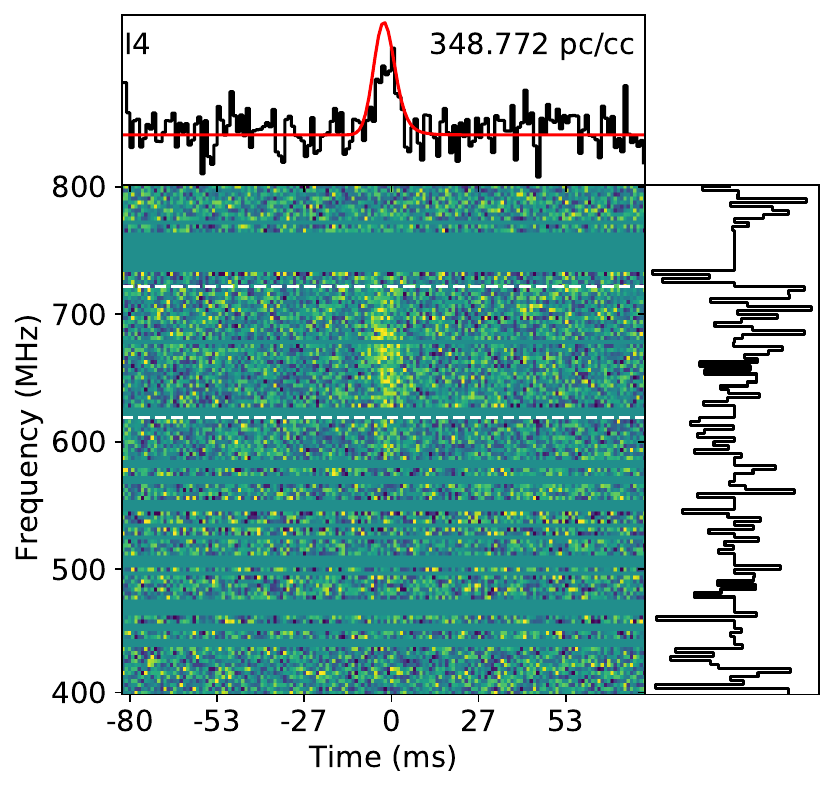}{0.2\textwidth}{}
          \fig{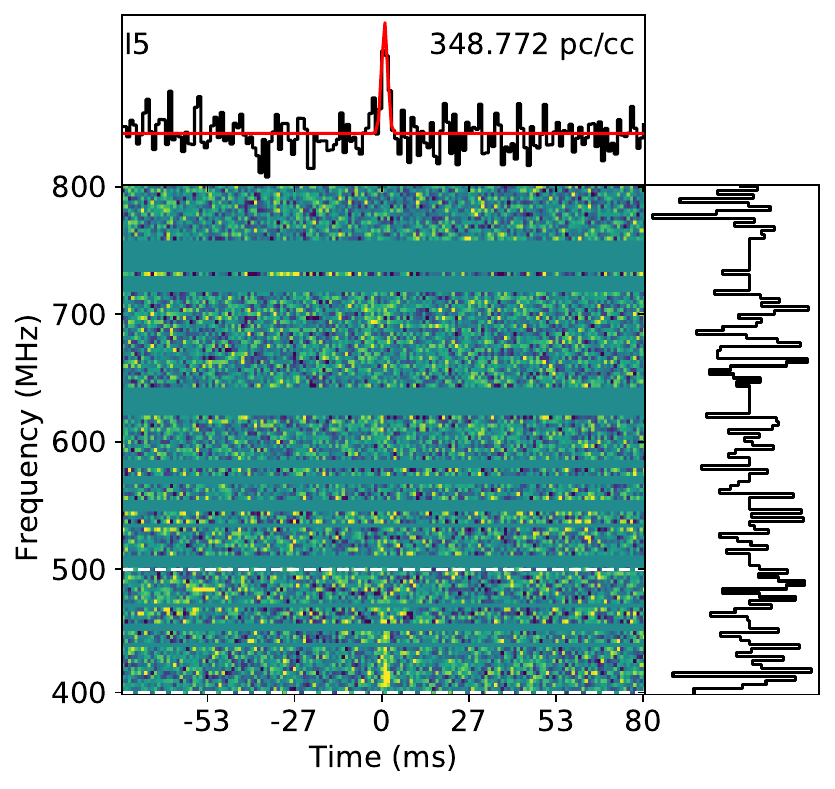}{0.2\textwidth}{}
          }
\gridline{\fig{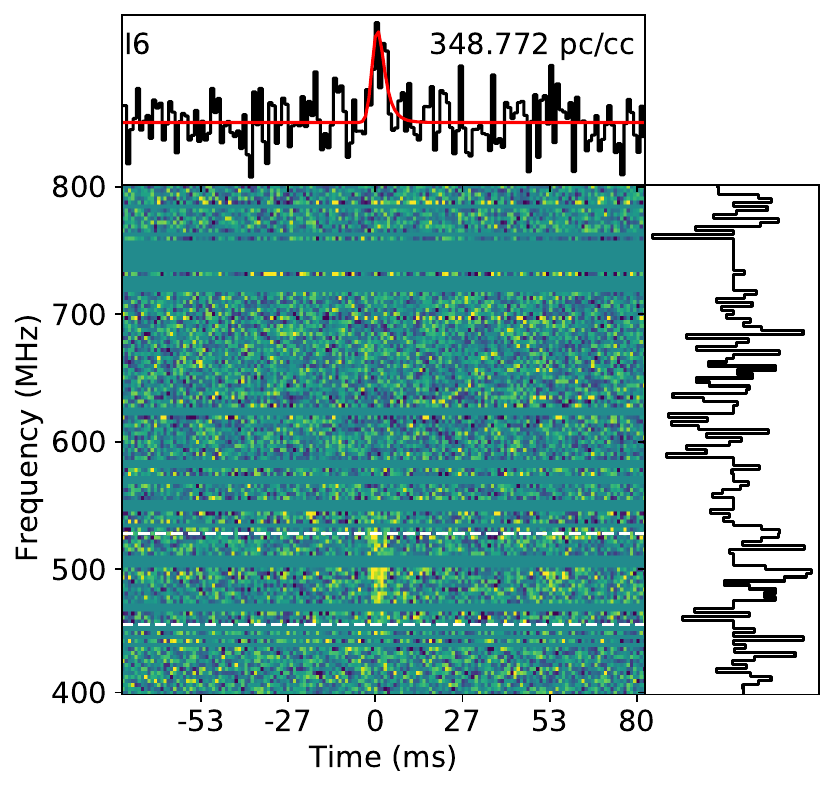}{0.2\textwidth}{}
          \fig{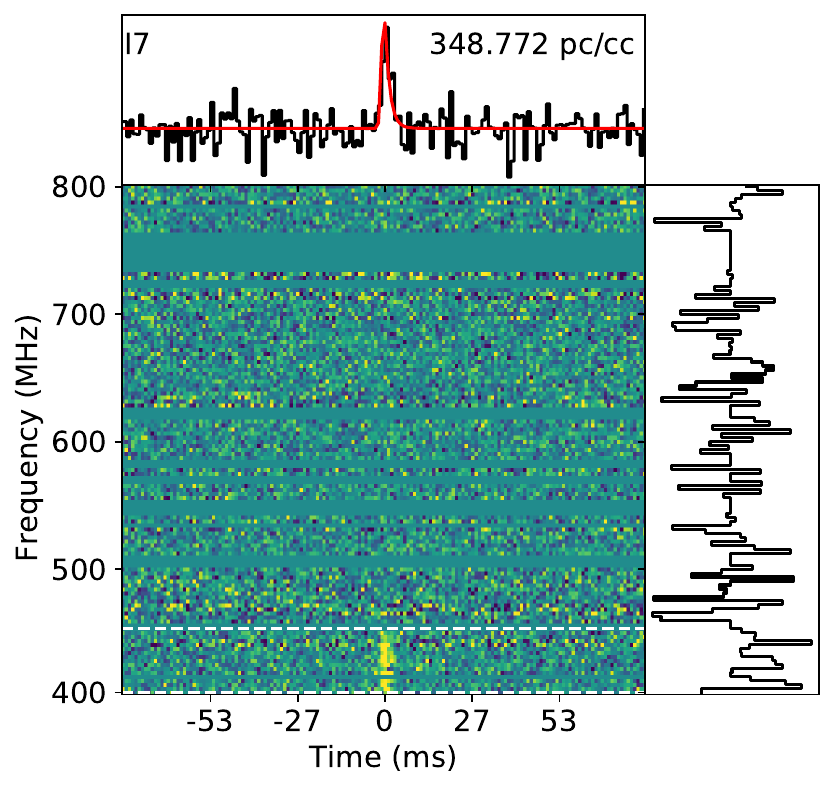}{0.2\textwidth}{}
          \fig{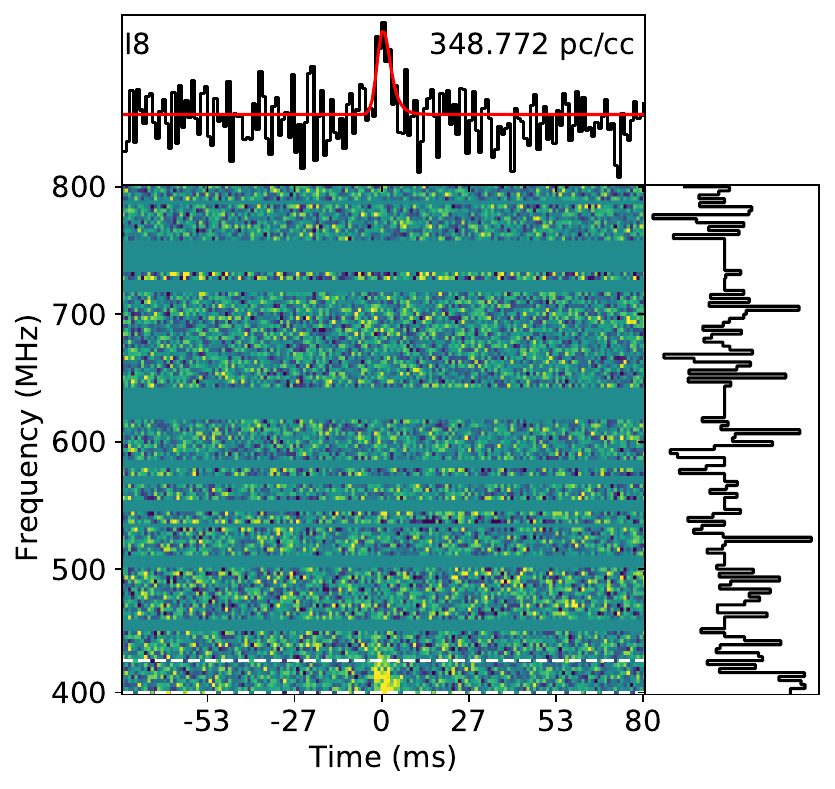}{0.2\textwidth}{}
          \fig{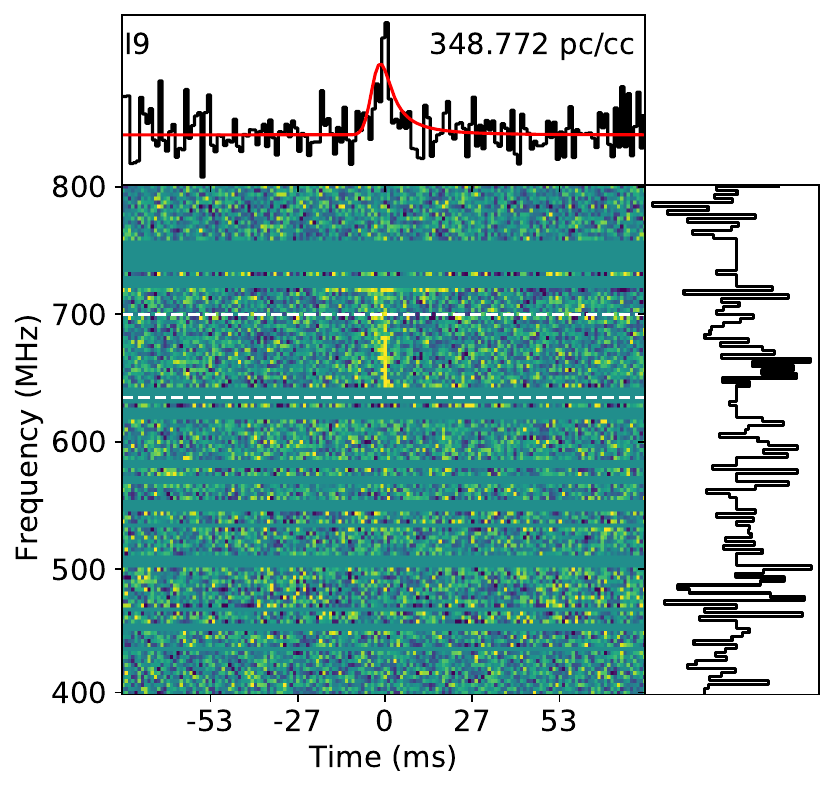}{0.2\textwidth}{}
          \fig{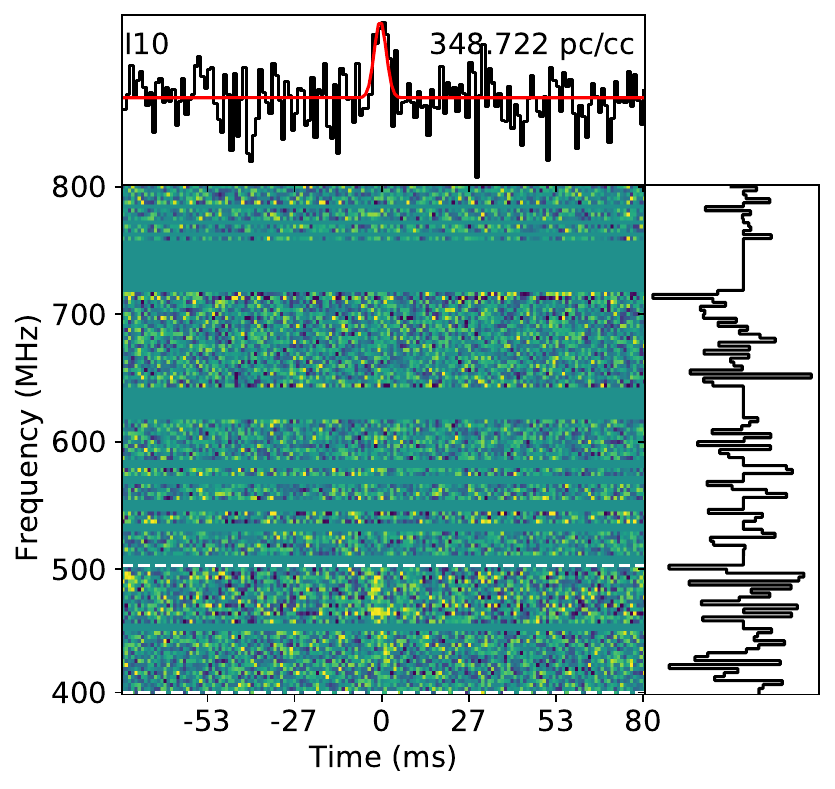}{0.2\textwidth}{}
          }
        
\gridline{\fig{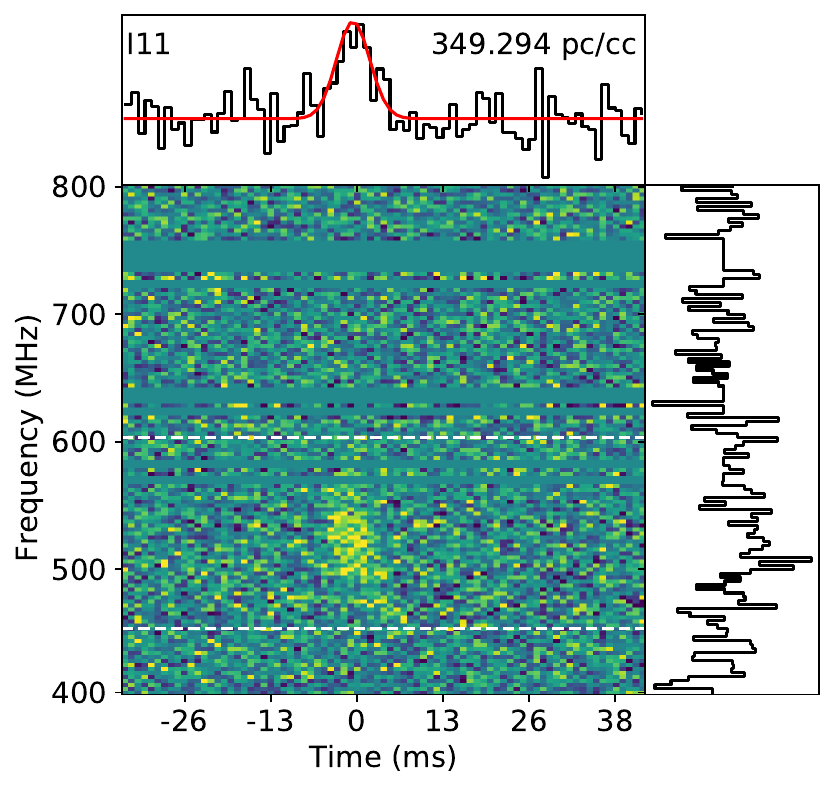}{0.2\textwidth}{}
          \fig{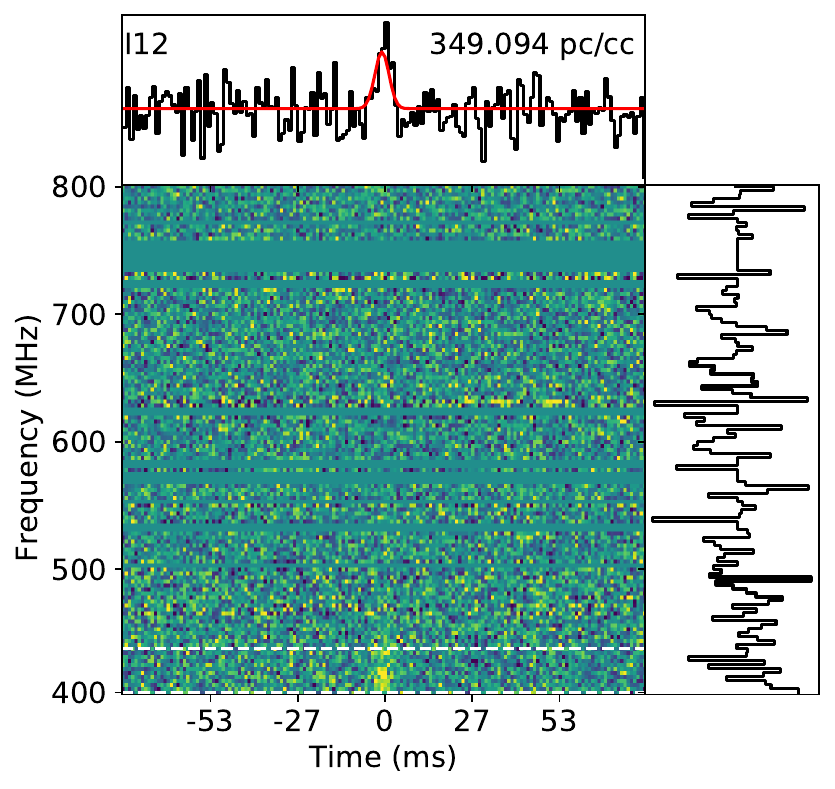}{0.2\textwidth}{}
          \fig{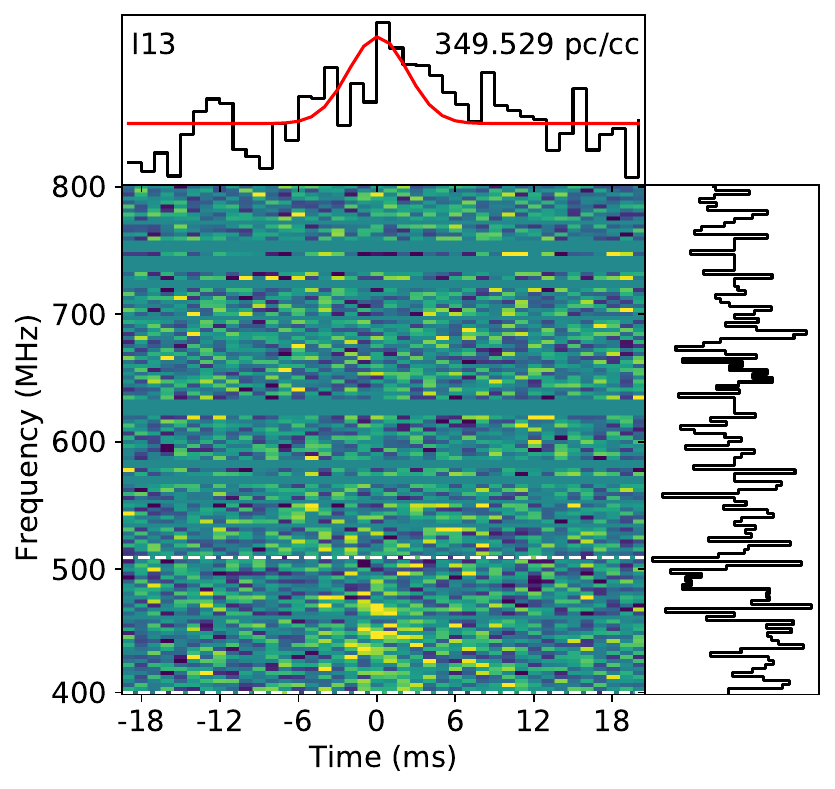}{0.2\textwidth}{}
          \fig{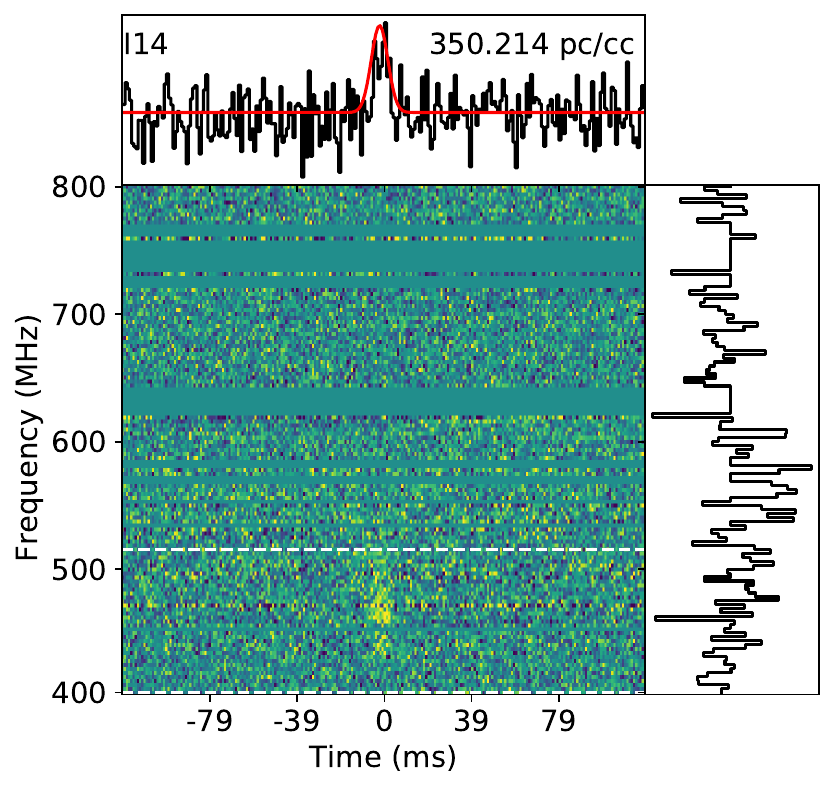}{0.2\textwidth}{}
          \fig{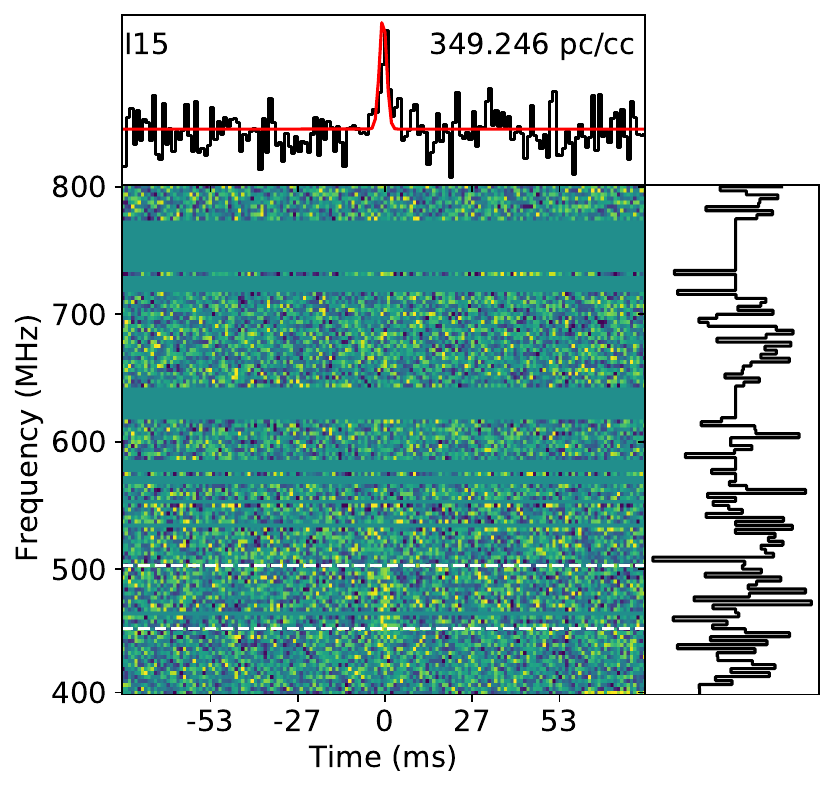}{0.2\textwidth}{}
          }

    \caption{Waterfall figure similar to Figure \ref{fig:BB1} showing bursts detected from \frb only with Intensity data from 20th September 2020 till 31st December 2021. All of these bursts are shown at a resolution of 0.983 ms and 128 channels}
\label{fig:Int}
\end{figure}

\begin{table}[ht!]
\begin{center}
\hspace{-1.in}
\resizebox{1.1\textwidth}{!}{
    \begin{tabular}{ccccccccc}
    \hline
        Burst & TNS & TOA & P & D & DM & $\tau$ & $\sigma$  & $\Delta\nu$ \\ 
         &  & (MJD) &  &  & (pc\,cm$^{-3}$) & (ms) & (ms) & (MHz) \\ \hline \hline
        B1 & 20181225A & 58477.161851 & 0.595 & 32 & 348.812(20) & 0.802(22) & 0.198(24)& 400.0-503.2 \\ 
         & & & & & & & 0.434(59) &  \\
        B2 & 20181226A & 58478.15521 & 0.656 & 16 & 348.801(28) & 0.564(230) & 0.165(8)& 463.3-615.4 \\ 
         & & & & & & & 0.186(25) & \\ 
        B3 & 20190604B & 58638.716433 & 0.482 & 16 & 348.919(32) & 0.253(1) & 1.044(41) & 514.2-754.3 \\ 
         & & & & &  &  & 0.873(3) & \\
         & & & & & & & 1.496(41) & \\
         & & & & & & & 0.996(66) & \\
        B4 & 20190605A & 58639.705612 & 0.543 & 64 & 349.845(152) & 0.290(3) & 2.214(5) & 400.0-600.6 \\ 
        B5 & 20190605B & 58639.710081 & 0.543 & 64 & 348.690(83) & 1.546(111) & 4.396(577) & 400.0-533.7 \\ 
         & & & & & & & 0.262(83) & \\ 
         & & & & & & & 1.568(54) & \\ 
         & & & & & & & 0.612(60) & \\ 
        B6 & 20191030B & 58786.320749 & 0.515 & 256 & 348.861(494) & 0.270(13) & 2.863(156) & 400.0-495.8 \\ 
        B7 & 20191218A & 58835.173236 & 0.505 & 256 & 349.243(235) & 0.712(6) & 4.538(16) & 400.0-616.2 \\ 
         & & & & & & & 0.924(141) & \\
        B8 & 20191219A & 58836.171977 & 0.566 & 256 & 349.540(10) & 0.137(3) & 2.476(59) & 400.0-468.0 \\ 
        B9 & 20200104E & 58852.136284 & 0.543 & 256 & 349.390(234) & 2.233(138) & 1.058(162) & 611.9-779.3 \\ 
        B10 & 20200104D & 58852.137732 & 0.543 & 64 & 349.350(84) & 0.575(38) & 0.586(101) & 410.2-586.9 \\ 
         & & & & & & & 1.372(82) & \\
        B11 & 20200120A & 58868.075857 & 0.519 & 128 & 348.883(146) & 2.209(99) & 1.295(159) & 507.9-709.7 \\ 
         & & & & & & & 0.105(37) & \\
        B12 & 20200203A & 58882.046806 & 0.374 & 256 & 349.292(276) & 0.360(6) & 1.045(8) & 635.4-796.5 \\ 
        B13 & 20200204B & 58883.03995 & 0.435 & 128 & 349.438(264) & 0.358(55) & 2.718(215) & 400.0-473.1 \\ 
         & & & & & & & 2.143(372) & \\
        B14 & 20200204D & 58883.044053 & 0.435 & 256 & 349.699(56) & 0.596(10) & 1.372(22) & 609.6-791.8 \\ 
        B15 & 20200204E & 58883.05372 & 0.435 & 32 & 348.785(67) & 0.570(73) & 0.169(18) & 603.3-766.4 \\ 
        B16 & 20200220A & 58899.007059 & 0.412 & 128 & 349.549(253) & 0.448(4) & 4.019(15) & 402.0-560.3 \\ 
        B17 & 20200512A & 58981.776623 & 0.477 & 64 & 349.264(85) & 0.135(135) & 0.306(3) & 633.4-800.0 \\ 
         & & & & & & & 2.651(84) & \\
        B18 & 20200513A & 58982.771575 & 0.538 & 256 & 349.350(638) & 0.451(34) & 1.834(217) & 400.0-494.6 \\ 
        B19 & 20200613A & 59013.692871 & 0.43 & 32 & 348.923(36) & 0.158(8) & 1.376(34) & 400.0-522.0 \\ 
        B20 & 20200614A & 59014.685325 & 0.491 & 16 & 348.781(9) & 0.499(3) & 0.078(5) & 485.2-681.1 \\ 
         & & & & & & & 2.014(6) & \\
         & & & & & & & 1.058(85) & \\
        B21 & 20200919B & 59111.40643 & 0.41 & 512 & 350.265(342) & 0.191(3) & 3.482(184) & 400.0-624.4 \\ 
         & & & & & & & 5.115(30) & \\
        B22 & 20210111E & 59225.104279 & 0.369 & 256 & 349.222(172) & 0.333(12) & 1.763(101) & 430.9-619.0 \\ 
        B23 & 20210127E & 59241.060714 & 0.345 & 64 & 348.783(33) & 1.011(46) & 0.417(80) & 519.3-796.1 \\ 
         & & & & & & & 0.193(48) & \\
        B24 & 20210130I & 59244.043996 & 0.528 & 256 & 349.17(75) & 0.876(60) & 2.119(170) & 400.0-583.0 \\ 
        B25 & 20210130H & 59244.062294 & 0.529 & 128 & 348.772(3) & 0.190(2) & 0.672(32) & 455.5-648.3 \\ 
         & & & & & & & 0.543(59) & \\
         & & & & & & & 2.268(4) & \\
        B26 & 20210131A & 59245.058327 & 0.59 & 256 & 349.175(369) & 0.649(48) & 1.669(206) & 400.0-534.5 \\ 
        B27 & 20210302A & 59275.968773 & 0.482 & 64 & 349.366(54) & 0.160(3) & 1.783(14) & 400.0-514.6 \\ 
        B28 & 20210303B & 59276.96257 & 0.542 & 256 & 349.163(462) & 0.254(8) & 2.557(76) & 400.0-516.9 \\ \hline
    \end{tabular}}
\end{center}
\end{table}

\begin{table}[ht!]
\begin{center}
\hspace{-1.in}
\resizebox{1.1\textwidth}{!}{
    \begin{tabular}{ccccccccc}
    \hline
        Burst & TNS & TOA & P & D & DM & $\tau$ & $\sigma$ & $\Delta\nu$ \\ 
         &  & (MJD) &  &  & (pc\,cm$^{-3}$) & (ms) & (ms) & (MHz) \\ \hline \hline
        B29 & 20210304A & 59277.960337 & 0.603 & 64 & 348.913(43) & 0.201(22) & 0.908(70) & 400.0-505.6 \\ 
        B30 & 20210402B & 59306.8901 & 0.374 & 32 & 349.600(170) & 0.495(2) & 1.430(5) & 586.5-796.9 \\ 
        B31 & 20210521C & 59355.747585 & 0.364 & 256 & 349.030(392) & 0.246(11) & 1.564(18) & 400.0-614.3 \\ 
        B32 & 20210523A & 59357.746804 & 0.486 & 128 & 349.087(390) & 0.461(32) & 1.378(135) & 400.0-475.9 \\ 
        B33 & 20210625A & 59390.652027 & 0.5 & 256 & 349.100(211) & 0.088(1) & 1.872(3) & 400.0-459.4 \\ 
        B34 & 20210711A & 59406.600248 & 0.476 & 256 & 349.330(186) & 0.884(55) & 1.784(172) & 400.0-576.7 \\ 
        B35 & 20210712A & 59407.599363 & 0.537 & 1 & 348.814(11) & 0.180(8) & 0.126(3) & 418.0-520.0 \\ 
        B36 & 20210814C & 59440.507259 & 0.551 & 64 & 348.849(52) & 0.324(21) & 0.868(75) & 400.0-597.8 \\ 
         & & & & & & & 0.161(21) & \\
         & & & & & & & 0.108(31) & \\
        B37 & 20210814B & 59440.515848 & 0.552 & 64 & 349.035(36) & 0.356(16) & 0.689(67) & 400.0-479.4 \\ 
        B38 & 20210929C & 59486.389633 & 0.359 & 256 & 349.731(793) & 0.443(4) & 1.761(20) & 439.9-800.0 \\ 
         & & & & & & & 1.514(95) & \\
        B39 & 20211101D & 59519.304104 & 0.374 & 512 & 349.823(1.9) & 0.157(5) & 3.423(43) & 400.0-444.6 \\ 
        B40 & 20211101C & 59519.304697 & 0.374 & 4 & 349.429(28) & 0.660(9) & 0.976(12) & 437.5-543.5 \\ 
        B41 & 20211101B & 59519.309086 & 0.374 & 4 & 348.867(19) & 0.415(164) & 0.378(17) & 442.6-581.8 \\ 
         & & & & & & & 0.087(11) & \\
        B42 & 20211103B & 59521.288708 & 0.495 & 64 & 348.971(234) & 0.403(29) & 1.808(86) & 400.0-444.2 \\ 
        B43 & 20211204D & 59552.211828 & 0.388 & 256 & 349.562(276) & 0.256(11) & 1.546(43) & 400.0-610.0 \\ 
        B44 & 20211221A & 59569.162656 & 0.425 & 64 & 348.951(206) & 0.220(14) & 0.711(72) & 400.0-438.3 \\ 
        B45 & 20211222B & 59570.154978 & 0.486 & 8 & 348.950(32) & 0.336(8) & 0.239(20) & 400.0-514.6 \\
         & & & & & & & 0.858(36) & \\
         & & & & & & & 0.334(11) & \\ \hline
    \end{tabular}}
    \caption{Table showing the morphological properties of all the bursts detected by the baseband system. The first column shows the burst number. Then we present their TNS names. Topocentric TOAs are in Modified Julian Date (MJD) format, referenced at 400 MHz with $\sim$ 1 second precision. P corresponds to the phase of the burst folded at 16.34 days with reference MJD 58369.40. D is the downsampling factor at which we measured the properties (i.e. time resolution = D $\times$ 2.56 $\upmu$s). DM is the dispersion measure reported by \texttt{fitburst}. $\tau$ is the scattering timescale referenced at 600 MHz. $\sigma$ is the intrinsic widths of all the sub-bursts at 600 MHz. $\Delta \nu$ corresponds to the bandwidth of the detection. See Sec. \ref{sec: Base_morph} for details.}\label{Baseband_Tab}
\end{center}
\end{table}

\begin{table}[t]
\begin{center}
\hspace{-1.in}
\resizebox{1.1\textwidth}{!}{
    \begin{tabular}{cccccccc}
    \hline
        Burst & TNS & TOA & P & DM & $\tau$ & $\sigma$ & $\Delta\nu$ \\
         & & (MJD) &  & (pc\,cm$^{-3}$) & (ms) & (ms) & (MHz)\\ \hline 
        I1 & 20201023B & 59145.325356 & 0.515 & 350.401(23) & $<$ 5.287 & 2.644(298) & 400.2-800.0 \\ 
        I2 & 20201123D & 59176.244474 & 0.409 & 348.772(0) & 0.943(109) & 0.833(188), 2.233(198) & 400.2-581.2 \\ 
        I3 & 20210129C & 59243.059205 & 0.5 & 348.772(0) & $<$ 9.284 & 4.642(602) & 501.1-800.0 \\ 
        I4 & 20210214D & 59259.00883 & 0.477 & 348.772(0) & 2.662(7) & 1.779(8) & 616.1-720.0 \\ 
        I5 & 20210301B & 59274.9735 & 0.455 & 348.772(0) & 2.303(617) & 1.832(338) & 400.0-499.0 \\ 
        I6 & 20210302G & 59275.978095 & 0.516 & 348.772(0) & 2.092(658) & 1.603(260) & 453.1-526.8 \\ 
        I7 & 20210320A & 59293.904126 & 0.614 & 348.772(0) & 2.080(477) & 0.574(143) & 419.4-449.1 \\ 
        I8 & 20210626A & 59391.661298 & 0.6 & 348.772(0) & 2.173(876) & 1.364(268) & 400.0-424.5 \\ 
        I9 & 20210711B & 59406.606126 & 0.515 & 348.772(0) & 0.562(3) & 0.364(5) & 631.8-699.2 \\ 
        I10 & 20210713C & 59408.592801 & 0.637 & 348.722(32) & $<$ 3.155 & 1.577(303) & 400.2-500.2 \\ 
        I11 & 183057477 & 59455.466284 & 0.507 & 349.294(23) & $<$ 4.463 & 2.232(244) & 450.2-602.0 \\
        I12 & 20210930E & 59487.382143 & 0.462 & 349.095(292) & $<$ 4.653 & 2.327(406) & 400.2-435.4 \\ 
        I13 & 20210930D & 59487.397668 & 0.463 & 349.530(55) & $<$ 4.349 & 2.174(278) & 400.2-506.4 \\ 
        I14 & 20211118B & 59536.265914 & 0.455 & 350.215(47) & $<$ 7.370 & 3.685(375) & 400.2-512.2 \\ 
        I15 & 201096712 & 59553.209720 & 0.493 & 349.246(16) & $<$ 2.000 & 1.000(186) & 450.2-500.2 \\ \hline
    \end{tabular}}
    \caption{Table showing the morphological properties of all the bursts detected with only total intensity data. The first column shows the burst number. Then we present their Transient Name Server (TNS) names. Topocentric TOAs provided here are in Modified Julian Date (MJD) format, referenced at 400 MHz. P corresponds to the phase of the burst folded at 16.34 days with reference MJD 58369.40. DM is the dispersion measure reported by \texttt{fitburst}. $\tau$ is the scattering timescale referenced at 600 MHz. $\sigma$ is the intrinsic widths of all the sub-bursts at 600 MHz. $\Delta \nu$ corresponds to the bandwidth of the detection. Some burst were too weak to get a reliable DM estimated  hence fixed their DM value to 347.772 pc\,cm$^{-3}$ \citep{nimmo2021_micro} and extracted their best fit results.}\label{Intensity_Tab}
\end{center}
\end{table}


\bibliography{sample631,R3_baseband,Models}{}

\begin{thebibliography}{}
\expandafter\ifx\csname natexlab\endcsname\relax\def\natexlab#1{#1}\fi
\providecommand{\url}[1]{\href{#1}{#1}}
\providecommand{\dodoi}[1]{doi:~\href{http://doi.org/#1}{\nolinkurl{#1}}}
\providecommand{\doeprint}[1]{\href{http://ascl.net/#1}{\nolinkurl{http://ascl.net/#1}}}
\providecommand{\doarXiv}[1]{\href{https://arxiv.org/abs/#1}{\nolinkurl{https://arxiv.org/abs/#1}}}

\bibitem[{{Amiri} {et~al.}(2023){Amiri}, {Bandura}, {Chen}, {Deng}, {Dobbs},
  {Fandino}, {Foreman}, {Halpern}, {Hill}, {Hinshaw}, {H{\"o}fer}, {Kania},
  {Landecker}, {MacEachern}, {Masui}, {Mena-Parra}, {Milutinovic},
  {Mirhosseini}, {Newburgh}, {Ordog}, {Pen}, {Pinsonneault-Marotte}, {Polzin},
  {Reda}, {Renard}, {Shaw}, {Siegel}, {Singh}, {Vanderlinde}, {Wang}, {Wiebe},
  {Wulf}, \& {CHIME Collaboration}}]{CHIME_stacking}
{Amiri}, M., {Bandura}, K., {Chen}, T., {et~al.} 2023, \apj, 947, 16,
  \dodoi{10.3847/1538-4357/acb13f}

\bibitem[{{Andersen} {et~al.}(2023){Andersen}, {Fonseca}, {McKee}, {Meyers},
  {Luo}, {Tan}, {Stairs}, {Kaspi}, {Kerkwijk}, {Bhardwaj}, {Boyle}, {Crowter},
  {Demorest}, {Dong}, {Good}, {Kaczmarek}, {Leung}, {Masui}, {Naidu}, {Ng},
  {Patel}, {Pearlman}, {Pleunis}, {Rafiei-Ravandi}, {Rahman}, {Ransom},
  {Smith}, \& {Tendulkar}}]{snowbird}
{Andersen}, B.~C., {Fonseca}, E., {McKee}, J.~W., {et~al.} 2023, \apj, 943, 57,
  \dodoi{10.3847/1538-4357/aca485}

\bibitem[{{Anna-Thomas} {et~al.}(2022){Anna-Thomas}, {Connor}, {Burke-Spolaor},
  {Beniamini}, {Aggarwal}, {Law}, {Lynch}, {Li}, {Feng}, {Ocker}, {Cruces},
  {Chatterjee}, {Yu}, {Niu}, \& {Xue}}]{thomas_reversal_FRB20190520B}
{Anna-Thomas}, R., {Connor}, L., {Burke-Spolaor}, S., {et~al.} 2022, arXiv
  e-prints, arXiv:2202.11112, \dodoi{10.48550/arXiv.2202.11112}

\bibitem[{{Astropy Collaboration} {et~al.}(2013){Astropy Collaboration},
  {Robitaille}, {Tollerud}, {Greenfield}, {Droettboom}, {Bray}, {Aldcroft},
  {Davis}, {Ginsburg}, {Price-Whelan}, {Kerzendorf}, {Conley}, {Crighton},
  {Barbary}, {Muna}, {Ferguson}, {Grollier}, {Parikh}, {Nair}, {Unther},
  {Deil}, {Woillez}, {Conseil}, {Kramer}, {Turner}, {Singer}, {Fox}, {Weaver},
  {Zabalza}, {Edwards}, {Azalee Bostroem}, {Burke}, {Casey}, {Crawford},
  {Dencheva}, {Ely}, {Jenness}, {Labrie}, {Lim}, {Pierfederici}, {Pontzen},
  {Ptak}, {Refsdal}, {Servillat}, \& {Streicher}}]{astropy1}
{Astropy Collaboration}, {Robitaille}, T.~P., {Tollerud}, E.~J., {et~al.} 2013,
  \aap, 558, A33, \dodoi{10.1051/0004-6361/201322068}

\bibitem[{{Astropy Collaboration} {et~al.}(2018){Astropy Collaboration},
  {Price-Whelan}, {Sip{\H{o}}cz}, {G{\"u}nther}, {Lim}, {Crawford}, {Conseil},
  {Shupe}, {Craig}, {Dencheva}, {Ginsburg}, {VanderPlas}, {Bradley},
  {P{\'e}rez-Su{\'a}rez}, {de Val-Borro}, {Aldcroft}, {Cruz}, {Robitaille},
  {Tollerud}, {Ardelean}, {Babej}, {Bach}, {Bachetti}, {Bakanov}, {Bamford},
  {Barentsen}, {Barmby}, {Baumbach}, {Berry}, {Biscani}, {Boquien}, {Bostroem},
  {Bouma}, {Brammer}, {Bray}, {Breytenbach}, {Buddelmeijer}, {Burke},
  {Calderone}, {Cano Rodr{\'\i}guez}, {Cara}, {Cardoso}, {Cheedella}, {Copin},
  {Corrales}, {Crichton}, {D'Avella}, {Deil}, {Depagne}, {Dietrich}, {Donath},
  {Droettboom}, {Earl}, {Erben}, {Fabbro}, {Ferreira}, {Finethy}, {Fox},
  {Garrison}, {Gibbons}, {Goldstein}, {Gommers}, {Greco}, {Greenfield},
  {Groener}, {Grollier}, {Hagen}, {Hirst}, {Homeier}, {Horton}, {Hosseinzadeh},
  {Hu}, {Hunkeler}, {Ivezi{\'c}}, {Jain}, {Jenness}, {Kanarek}, {Kendrew},
  {Kern}, {Kerzendorf}, {Khvalko}, {King}, {Kirkby}, {Kulkarni}, {Kumar},
  {Lee}, {Lenz}, {Littlefair}, {Ma}, {Macleod}, {Mastropietro}, {McCully},
  {Montagnac}, {Morris}, {Mueller}, {Mumford}, {Muna}, {Murphy}, {Nelson},
  {Nguyen}, {Ninan}, {N{\"o}the}, {Ogaz}, {Oh}, {Parejko}, {Parley}, {Pascual},
  {Patil}, {Patil}, {Plunkett}, {Prochaska}, {Rastogi}, {Reddy Janga},
  {Sabater}, {Sakurikar}, {Seifert}, {Sherbert}, {Sherwood-Taylor}, {Shih},
  {Sick}, {Silbiger}, {Singanamalla}, {Singer}, {Sladen}, {Sooley},
  {Sornarajah}, {Streicher}, {Teuben}, {Thomas}, {Tremblay}, {Turner},
  {Terr{\'o}n}, {van Kerkwijk}, {de la Vega}, {Watkins}, {Weaver}, {Whitmore},
  {Woillez}, {Zabalza}, \& {Astropy Contributors}}]{astropy2}
{Astropy Collaboration}, {Price-Whelan}, A.~M., {Sip{\H{o}}cz}, B.~M., {et~al.}
  2018, \aj, 156, 123, \dodoi{10.3847/1538-3881/aabc4f}

\bibitem[{{Astropy Collaboration} {et~al.}(2022){Astropy Collaboration},
  {Price-Whelan}, {Lim}, {Earl}, {Starkman}, {Bradley}, {Shupe}, {Patil},
  {Corrales}, {Brasseur}, {N{\"o}the}, {Donath}, {Tollerud}, {Morris},
  {Ginsburg}, {Vaher}, {Weaver}, {Tocknell}, {Jamieson}, {van Kerkwijk},
  {Robitaille}, {Merry}, {Bachetti}, {G{\"u}nther}, {Aldcroft},
  {Alvarado-Montes}, {Archibald}, {B{\'o}di}, {Bapat}, {Barentsen},
  {Baz{\'a}n}, {Biswas}, {Boquien}, {Burke}, {Cara}, {Cara}, {Conroy},
  {Conseil}, {Craig}, {Cross}, {Cruz}, {D'Eugenio}, {Dencheva}, {Devillepoix},
  {Dietrich}, {Eigenbrot}, {Erben}, {Ferreira}, {Foreman-Mackey}, {Fox},
  {Freij}, {Garg}, {Geda}, {Glattly}, {Gondhalekar}, {Gordon}, {Grant},
  {Greenfield}, {Groener}, {Guest}, {Gurovich}, {Handberg}, {Hart},
  {Hatfield-Dodds}, {Homeier}, {Hosseinzadeh}, {Jenness}, {Jones}, {Joseph},
  {Kalmbach}, {Karamehmetoglu}, {Ka{\l}uszy{\'n}ski}, {Kelley}, {Kern},
  {Kerzendorf}, {Koch}, {Kulumani}, {Lee}, {Ly}, {Ma}, {MacBride}, {Maljaars},
  {Muna}, {Murphy}, {Norman}, {O'Steen}, {Oman}, {Pacifici}, {Pascual},
  {Pascual-Granado}, {Patil}, {Perren}, {Pickering}, {Rastogi}, {Roulston},
  {Ryan}, {Rykoff}, {Sabater}, {Sakurikar}, {Salgado}, {Sanghi}, {Saunders},
  {Savchenko}, {Schwardt}, {Seifert-Eckert}, {Shih}, {Jain}, {Shukla}, {Sick},
  {Simpson}, {Singanamalla}, {Singer}, {Singhal}, {Sinha}, {Sip{\H{o}}cz},
  {Spitler}, {Stansby}, {Streicher}, {{\v{S}}umak}, {Swinbank}, {Taranu},
  {Tewary}, {Tremblay}, {Val-Borro}, {Van Kooten}, {Vasovi{\'c}}, {Verma}, {de
  Miranda Cardoso}, {Williams}, {Wilson}, {Winkel}, {Wood-Vasey}, {Xue},
  {Yoachim}, {Zhang}, {Zonca}, \& {Astropy Project Contributors}}]{astropy3}
{Astropy Collaboration}, {Price-Whelan}, A.~M., {Lim}, P.~L., {et~al.} 2022,
  \apj, 935, 167, \dodoi{10.3847/1538-4357/ac7c74}

\bibitem[{{Beniamini} {et~al.}(2020){Beniamini}, {Wadiasingh}, \&
  {Metzger}}]{Paz_period_magnetar}
{Beniamini}, P., {Wadiasingh}, Z., \& {Metzger}, B.~D. 2020, \mnras, 496, 3390,
  \dodoi{10.1093/mnras/staa1783}

\bibitem[{{Bethapudi} {et~al.}(2022){Bethapudi}, {Spitler}, {Main}, {Li}, \&
  {Wharton}}]{surya2022}
{Bethapudi}, S., {Spitler}, L.~G., {Main}, R.~A., {Li}, D.~Z., \& {Wharton},
  R.~S. 2022, arXiv e-prints, arXiv:2207.13669.
\newblock \doarXiv{2207.13669}

\bibitem[{Bochenek {et~al.}(2020)Bochenek, Ravi, Belov, Hallinan, Kocz,
  Kulkarni, \& McKenna}]{bochenek2020fast}
Bochenek, C.~D., Ravi, V., Belov, K.~V., {et~al.} 2020, Nature, 587, 59

\bibitem[{Chawla {et~al.}(2020)Chawla, Andersen, Bhardwaj, Fonseca, Josephy,
  Kaspi, Michilli, Pleunis, Bandura, Bassa, {et~al.}}]{chawla2020detection}
Chawla, P., Andersen, B., Bhardwaj, M., {et~al.} 2020, The Astrophysical
  Journal Letters, 896, L41

\bibitem[{{Chawla} {et~al.}(2022){Chawla}, {Kaspi}, {Ransom}, {Bhardwaj},
  {Boyle}, {Breitman}, {Cassanelli}, {Cubranic}, {Dong}, {Fonseca}, {Gaensler},
  {Giri}, {Josephy}, {Kaczmarek}, {Leung}, {Masui}, {Mena-Parra}, {Merryfield},
  {Michilli}, {M{\"u}nchmeyer}, {Ng}, {Patel}, {Pearlman}, {Petroff},
  {Pleunis}, {Rahman}, {Sanghavi}, {Shin}, {Smith}, {Stairs}, \&
  {Tendulkar}}]{chawla_scattering_dm}
{Chawla}, P., {Kaspi}, V.~M., {Ransom}, S.~M., {et~al.} 2022, \apj, 927, 35,
  \dodoi{10.3847/1538-4357/ac49e1}

\bibitem[{{Chen} {et~al.}(2022){Chen}, {Gu}, {Fu}, {Weng}, {Wang}, \&
  {Sun}}]{Chen_ULX_FRB}
{Chen}, H.-Y., {Gu}, W.-M., {Fu}, J.-B., {et~al.} 2022, \apj, 937, 9,
  \dodoi{10.3847/1538-4357/ac8b7f}

\bibitem[{{CHIME Collaboration} {et~al.}(2022){CHIME Collaboration}, {Amiri},
  {Bandura}, {Boskovic}, {Chen}, {Cliche}, {Deng}, {Denman}, {Dobbs},
  {Fandino}, {Foreman}, {Halpern}, {Hanna}, {Hill}, {Hinshaw}, {H{\"o}fer},
  {Kania}, {Klages}, {Landecker}, {MacEachern}, {Masui}, {Mena-Parra},
  {Milutinovic}, {Mirhosseini}, {Newburgh}, {Nitsche}, {Ordog}, {Pen},
  {Pinsonneault-Marotte}, {Polzin}, {Reda}, {Renard}, {Shaw}, {Siegel},
  {Singh}, {Smegal}, {Tretyakov}, {van Gassen}, {Vanderlinde}, {Wang}, {Wiebe},
  {Willis}, \& {Wulf}}]{chime_cosmo_overview}
{CHIME Collaboration}, {Amiri}, M., {Bandura}, K., {et~al.} 2022, \apjs, 261,
  29, \dodoi{10.3847/1538-4365/ac6fd9}

\bibitem[{{CHIME/FRB Collaboration} {et~al.}(2018){CHIME/FRB Collaboration},
  {Amiri}, {Bandura}, {Berger}, {Bhardwaj}, {Boyce}, {Boyle}, {Brar},
  {Burhanpurkar}, {Chawla}, {Chowdhury}, {Cliche}, {Cranmer}, {Cubranic},
  {Deng}, {Denman}, {Dobbs}, {Fandino}, {Fonseca}, {Gaensler}, {Giri},
  {Gilbert}, {Good}, {Guliani}, {Halpern}, {Hinshaw}, {H{\"o}fer}, {Josephy},
  {Kaspi}, {Landecker}, {Lang}, {Liao}, {Masui}, {Mena-Parra}, {Naidu},
  {Newburgh}, {Ng}, {Patel}, {Pen}, {Pinsonneault-Marotte}, {Pleunis}, {Rafiei
  Ravandi}, {Ransom}, {Renard}, {Scholz}, {Sigurdson}, {Siegel}, {Smith},
  {Stairs}, {Tendulkar}, {Vanderlinde}, \& {Wiebe}}]{chime18_overview}
{CHIME/FRB Collaboration}, {Amiri}, M., {Bandura}, K., {et~al.} 2018, \apj,
  863, 48, \dodoi{10.3847/1538-4357/aad188}

\bibitem[{{CHIME/FRB Collaboration} {et~al.}(2019){CHIME/FRB Collaboration},
  {Andersen}, {Bandura}, {Bhardwaj}, {Boubel}, {Boyce}, {Boyle}, {Brar},
  {Cassanelli}, {Chawla}, {Cubranic}, {Deng}, {Dobbs}, {Fandino}, {Fonseca},
  {Gaensler}, {Gilbert}, {Giri}, {Good}, {Halpern}, {Hill}, {Hinshaw},
  {H{\"o}fer}, {Josephy}, {Kaspi}, {Kothes}, {Landecker}, {Lang}, {Li}, {Lin},
  {Masui}, {Mena-Parra}, {Merryfield}, {Mckinven}, {Michilli}, {Milutinovic},
  {Naidu}, {Newburgh}, {Ng}, {Patel}, {Pen}, {Pinsonneault-Marotte}, {Pleunis},
  {Rafiei-Ravandi}, {Rahman}, {Ransom}, {Renard}, {Scholz}, {Siegel}, {Singh},
  {Smith}, {Stairs}, {Tendulkar}, {Tretyakov}, {Vanderlinde}, {Yadav}, \&
  {Zwaniga}}]{chimefrb_RN1_2019}
{CHIME/FRB Collaboration}, {Andersen}, B.~C., {Bandura}, K., {et~al.} 2019,
  \apjl, 885, L24, \dodoi{10.3847/2041-8213/ab4a80}

\bibitem[{{CHIME/FRB Collaboration} {et~al.}(2020{\natexlab{a}}){CHIME/FRB
  Collaboration}, Andersen, Bandura, Bhardwaj, Bij, Boyce, Boyle, Brar,
  Cassanelli, Chawla, Chen, {et~al.}}]{andersen2020bright}
{CHIME/FRB Collaboration}, Andersen, B., Bandura, K., {et~al.}
  2020{\natexlab{a}}, Nature, 587, 54

\bibitem[{{CHIME/FRB Collaboration} {et~al.}(2020{\natexlab{b}}){CHIME/FRB
  Collaboration}, Amiri, Andersen, Bandura, Bhardwaj, Boyle, Brar, Chawla,
  Chen, Cliche, Cubranic, {et~al.}}]{periodic_chime2020}
{CHIME/FRB Collaboration}, Amiri, M., Andersen, B., {et~al.}
  2020{\natexlab{b}}, Nature, 582, 351

\bibitem[{{CHIME/FRB Collaboration} {et~al.}(2021){CHIME/FRB Collaboration},
  {Amiri}, {Andersen}, {Bandura}, {Berger}, {Bhardwaj}, {Boyce}, {Boyle},
  {Brar}, {Breitman}, {Cassanelli}, {Chawla}, {Chen}, {Cliche}, {Cook},
  {Cubranic}, {Curtin}, {Deng}, {Dobbs}, {Dong}, {Eadie}, {Fandino}, {Fonseca},
  {Gaensler}, {Giri}, {Good}, {Halpern}, {Hill}, {Hinshaw}, {Josephy},
  {Kaczmarek}, {Kader}, {Kania}, {Kaspi}, {Landecker}, {Lang}, {Leung}, {Li},
  {Lin}, {Masui}, {McKinven}, {Mena-Parra}, {Merryfield}, {Meyers}, {Michilli},
  {Milutinovic}, {Mirhosseini}, {M{\"u}nchmeyer}, {Naidu}, {Newburgh}, {Ng},
  {Patel}, {Pen}, {Petroff}, {Pinsonneault-Marotte}, {Pleunis},
  {Rafiei-Ravandi}, {Rahman}, {Ransom}, {Renard}, {Sanghavi}, {Scholz}, {Shaw},
  {Shin}, {Siegel}, {Sikora}, {Singh}, {Smith}, {Stairs}, {Tan}, {Tendulkar},
  {Vanderlinde}, {Wang}, {Wulf}, \& {Zwaniga}}]{chimefrb_first_cat_2021}
{CHIME/FRB Collaboration}, {Amiri}, M., {Andersen}, B.~C., {et~al.} 2021,
  \apjs, 257, 59, \dodoi{10.3847/1538-4365/ac33ab}

\bibitem[{{CHIME/FRB Collaboration} {et~al.}(2023){CHIME/FRB Collaboration},
  {Andersen}, {Bandura}, {Bhardwaj}, {Boyle}, {Brar}, {Cassanelli},
  {Chatterjee}, {Chawla}, {Cook}, {Curtin}, {Dobbs}, {Dong}, {Faber},
  {Fandino}, {Fonseca}, {Gaensler}, {Giri}, {Herrera-Martin}, {Hill}, {Ibik},
  {Josephy}, {Kaczmarek}, {Kader}, {Kaspi}, {Landecker}, {Lanman}, {Lazda},
  {Leung}, {Lin}, {Masui}, {McKinven}, {Mena-Parra}, {Meyers}, {Michilli},
  {Ng}, {Pandhi}, {Pearlman}, {Pen}, {Petroff}, {Pleunis}, {Rafiei-Ravandi},
  {Rahman}, {Ransom}, {Renard}, {Sand}, {Sanghavi}, {Scholz}, {Shah}, {Shin},
  {Siegel}, {Smith}, {Stairs}, {Su}, {Tendulkar}, {Vanderlinde}, {Wang},
  {Wulf}, \& {Zwaniga}}]{RN3}
{CHIME/FRB Collaboration}, {Andersen}, B.~C., {Bandura}, K., {et~al.} 2023,
  \apj, 947, 83, \dodoi{10.3847/1538-4357/acc6c1}

\bibitem[{Cordes \& Chatterjee(2019)}]{cordes2019fast}
Cordes, J.~M., \& Chatterjee, S. 2019, Annual Review of Astronomy and
  Astrophysics, 57

\bibitem[{{Cordes} \& {Lazio}(2002)}]{Cor02}
{Cordes}, J.~M., \& {Lazio}, T.~J.~W. 2002, arXiv e-prints, astro,
  \dodoi{10.48550/arXiv.astro-ph/0207156}

\bibitem[{{Cordes} {et~al.}(2016){Cordes}, {Wharton}, {Spitler}, {Chatterjee},
  \& {Wasserman}}]{Cordes_2016}
{Cordes}, J.~M., {Wharton}, R.~S., {Spitler}, L.~G., {Chatterjee}, S., \&
  {Wasserman}, I. 2016, arXiv e-prints, arXiv:1605.05890.
\newblock \doarXiv{1605.05890}

\bibitem[{{Cruces} {et~al.}(2021){Cruces}, {Spitler}, {Scholz}, {Lynch},
  {Seymour}, {Hessels}, {Gouiff{\'e}s}, {Hilmarsson}, {Kramer}, \&
  {Munjal}}]{cruces_R1_period}
{Cruces}, M., {Spitler}, L.~G., {Scholz}, P., {et~al.} 2021, \mnras, 500, 448,
  \dodoi{10.1093/mnras/staa3223}

\bibitem[{{Dai} {et~al.}(2022){Dai}, {Feng}, {Yang}, {Zhang}, {Li}, {Niu},
  {Wang}, {Xue}, {Zhang}, {Burke-Spolaor}, {Law}, {Lynch}, {Connor},
  {Anna-Thomas}, {Zhang}, {Duan}, {Yao}, {Tsai}, {Zhu}, {Cruces}, {Hobbs},
  {Miao}, {Niu}, {Filipovic}, \& {Zhu}}]{dai_reversal_FRB20190520B}
{Dai}, S., {Feng}, Y., {Yang}, Y.~P., {et~al.} 2022, arXiv e-prints,
  arXiv:2203.08151, \dodoi{10.48550/arXiv.2203.08151}

\bibitem[{{Dai} \& {Zhong}(2020)}]{Dai_asteroid}
{Dai}, Z.~G., \& {Zhong}, S.~Q. 2020, \apjl, 895, L1,
  \dodoi{10.3847/2041-8213/ab8f2d}

\bibitem[{De~Luca {et~al.}(2006)De~Luca, Caraveo, Mereghetti, Tiengo, \&
  Bignami}]{de2006long}
De~Luca, A., Caraveo, P.~A., Mereghetti, S., Tiengo, A., \& Bignami, G.~F.
  2006, Science, 313, 814

\bibitem[{{Deng} {et~al.}(2021){Deng}, {Zhong}, \& {Dai}}]{deng_binary}
{Deng}, C.-M., {Zhong}, S.-Q., \& {Dai}, Z.-G. 2021, \apj, 922, 98,
  \dodoi{10.3847/1538-4357/ac30db}

\bibitem[{Dickey \& Fuller(1979)}]{dickey1979distribution}
Dickey, D.~A., \& Fuller, W.~A. 1979, Journal of the American statistical
  association, 74, 427

\bibitem[{{Feng} {et~al.}(2023){Feng}, {Li}, {Zhang}, {Tsai}, {Wang}, {Yang},
  {Qu}, {Wang}, {Zhou}, {Niu}, {Miao}, {Yuan}, {Xu}, {Lynch}, {Armentrout},
  {Gregory}, {Meng}, {Wang}, {Chen}, {Dai}, {Niu}, {Xue}, {Yao}, {Zhang},
  {Zhang}, {Zhu}, \& {Zhu}}]{Feng_2023_R117}
{Feng}, Y., {Li}, D., {Zhang}, Y.-K., {et~al.} 2023, arXiv e-prints,
  arXiv:2304.14671, \dodoi{10.48550/arXiv.2304.14671}

\bibitem[{Fonseca {et~al.}(2020)Fonseca, Andersen, Bhardwaj, Chawla, Good,
  Josephy, Kaspi, Masui, Mckinven, Michilli, {et~al.}}]{Fonseca2020_RN2}
Fonseca, E., Andersen, B., Bhardwaj, M., {et~al.} 2020, The Astrophysical
  Journal Letters, 891, L6

\bibitem[{{Foreman-Mackey} {et~al.}(2013){Foreman-Mackey}, {Hogg}, {Lang}, \&
  {Goodman}}]{emcee}
{Foreman-Mackey}, D., {Hogg}, D.~W., {Lang}, D., \& {Goodman}, J. 2013, \pasp,
  125, 306, \dodoi{10.1086/670067}

\bibitem[{{Gajjar} {et~al.}(2018){Gajjar}, {Siemion}, {Price}, {Law},
  {Michilli}, {Hessels}, {Chatterjee}, {Archibald}, {Bower}, {Brinkman},
  {Burke-Spolaor}, {Cordes}, {Croft}, {Enriquez}, {Foster}, {Gizani},
  {Hellbourg}, {Isaacson}, {Kaspi}, {Lazio}, {Lebofsky}, {Lynch}, {MacMahon},
  {McLaughlin}, {Ransom}, {Scholz}, {Seymour}, {Spitler}, {Tendulkar},
  {Werthimer}, \& {Zhang}}]{Gaj18}
{Gajjar}, V., {Siemion}, A.~P.~V., {Price}, D.~C., {et~al.} 2018, \apj, 863, 2,
  \dodoi{10.3847/1538-4357/aad005}

\bibitem[{{Gavriil} {et~al.}(2008){Gavriil}, {Gonzalez}, {Gotthelf}, {Kaspi},
  {Livingstone}, \& {Woods}}]{Gavriil_2008}
{Gavriil}, F.~P., {Gonzalez}, M.~E., {Gotthelf}, E.~V., {et~al.} 2008, Science,
  319, 1802, \dodoi{10.1126/science.1153465}

\bibitem[{{Gavriil} {et~al.}(2002){Gavriil}, {Kaspi}, \& {Woods}}]{2002gavriil}
{Gavriil}, F.~P., {Kaspi}, V.~M., \& {Woods}, P.~M. 2002, \nat, 419, 142,
  \dodoi{10.1038/nature01011}

\bibitem[{{Geyer} {et~al.}(2017){Geyer}, {Karastergiou}, {Kondratiev},
  {Zagkouris}, {Kramer}, {Stappers}, {Grie{\ss}meier}, {Hessels}, {Michilli},
  {Pilia}, \& {Sobey}}]{geyer_2017}
{Geyer}, M., {Karastergiou}, A., {Kondratiev}, V.~I., {et~al.} 2017, \mnras,
  470, 2659, \dodoi{10.1093/mnras/stx1151}

\bibitem[{{Gopinath} {et~al.}(2023){Gopinath}, {Bassa}, {Pleunis}, {Hessels},
  {Chawla}, {Keane}, {Kondratiev}, {Michilli5}, \& {Nimmo}}]{Gopinath_R3_2023}
{Gopinath}, A., {Bassa}, C.~G., {Pleunis}, Z., {et~al.} 2023, arXiv e-prints,
  arXiv:2305.06393, \dodoi{10.48550/arXiv.2305.06393}

\bibitem[{{G{\'o}rski} {et~al.}(2005){G{\'o}rski}, {Hivon}, {Banday},
  {Wandelt}, {Hansen}, {Reinecke}, \& {Bartelmann}}]{ghb+05_healpix}
{G{\'o}rski}, K.~M., {Hivon}, E., {Banday}, A.~J., {et~al.} 2005, \apj, 622,
  759, \dodoi{10.1086/427976}

\bibitem[{{Gu} {et~al.}(2020){Gu}, {Yi}, \& {Liu}}]{Gu_NS_WD_binary}
{Gu}, W.-M., {Yi}, T., \& {Liu}, T. 2020, \mnras, 497, 1543,
  \dodoi{10.1093/mnras/staa1914}

\bibitem[{Harris {et~al.}(2020)Harris, Millman, van~der Walt, Gommers,
  Virtanen, Cournapeau, Wieser, Taylor, Berg, Smith, Kern, Picus, Hoyer, van
  Kerkwijk, Brett, Haldane, del R{'{\i}}o, Wiebe, Peterson,
  G{'{e}}rard-Marchant, Sheppard, Reddy, Weckesser, Abbasi, Gohlke, \&
  Oliphant}]{numpy}
Harris, C.~R., Millman, K.~J., van~der Walt, S.~J., {et~al.} 2020, Nature, 585,
  357, \dodoi{10.1038/s41586-020-2649-2}

\bibitem[{{Hessels} {et~al.}(2019){Hessels}, {Spitler}, {Seymour}, {Cordes},
  {Michilli}, {Lynch}, {Gourdji}, {Archibald}, {Bassa}, {Bower}, {Chatterjee},
  {Connor}, {Crawford}, {Deneva}, {Gajjar}, {Kaspi}, {Keimpema}, {Law},
  {Marcote}, {McLaughlin}, {Paragi}, {Petroff}, {Ransom}, {Scholz}, {Stappers},
  \& {Tendulkar}}]{Hes2019}
{Hessels}, J.~W.~T., {Spitler}, L.~G., {Seymour}, A.~D., {et~al.} 2019, \apjl,
  876, L23, \dodoi{10.3847/2041-8213/ab13ae}

\bibitem[{Hilmarsson {et~al.}(2021)Hilmarsson, Michilli, Spitler, Wharton,
  Demorest, Desvignes, Gourdji, Hackstein, Hessels, Nimmo,
  {et~al.}}]{hilmarsson2021rotation}
Hilmarsson, G., Michilli, D., Spitler, L., {et~al.} 2021, The Astrophysical
  journal letters, 908, L10

\bibitem[{{Hilmarsson} {et~al.}(2021){Hilmarsson}, {Spitler}, {Main}, \&
  {Li}}]{Hilmarsson_R67}
{Hilmarsson}, G.~H., {Spitler}, L.~G., {Main}, R.~A., \& {Li}, D.~Z. 2021,
  \mnras, 508, 5354, \dodoi{10.1093/mnras/stab2936}

\bibitem[{Hunter(2007)}]{matplotlib}
Hunter, J.~D. 2007, Computing in Science \& Engineering, 9, 90,
  \dodoi{10.1109/MCSE.2007.55}

\bibitem[{{Ioka} \& {Zhang}(2020)}]{Ioka_comb}
{Ioka}, K., \& {Zhang}, B. 2020, \apjl, 893, L26,
  \dodoi{10.3847/2041-8213/ab83fb}

\bibitem[{{Johnston} {et~al.}(2001){Johnston}, {Wex}, {Nicastro}, {Manchester},
  \& {Lyne}}]{jwn+01}
{Johnston}, S., {Wex}, N., {Nicastro}, L., {Manchester}, R.~N., \& {Lyne},
  A.~G. 2001, \mnras, 326, 643, \dodoi{10.1046/j.1365-8711.2001.04615.x}

\bibitem[{Josephy {et~al.}(2019)Josephy, Chawla, Fonseca, Ng, Patel, Pleunis,
  Scholz, Andersen, Bandura, Bhardwaj, {et~al.}}]{josephy2019chime}
Josephy, A., Chawla, P., Fonseca, E., {et~al.} 2019, The Astrophysical Journal
  Letters, 882, L18

\bibitem[{{Kaspi} {et~al.}(2000){Kaspi}, {Lackey}, \& {Chakrabarty}}]{klc00}
{Kaspi}, V.~M., {Lackey}, J.~R., \& {Chakrabarty}, D. 2000, \apjl, 537, L31,
  \dodoi{10.1086/312758}

\bibitem[{{Kaspi} {et~al.}(1996){Kaspi}, {Tauris}, \&
  {Manchester}}]{kaspi_binary_wind}
{Kaspi}, V.~M., {Tauris}, T.~M., \& {Manchester}, R.~N. 1996, \apj, 459, 717,
  \dodoi{10.1086/176936}

\bibitem[{{Katz}(2021)}]{katz21}
{Katz}, J.~I. 2021, \mnras, 502, 4664, \dodoi{10.1093/mnras/stab399}

\bibitem[{{Kumar} {et~al.}(2021){Kumar}, {Shannon}, {Flynn}, {Os{\l}owski},
  {Bhandari}, {Day}, {Deller}, {Farah}, {Kaczmarek}, {Kerr}, {Phillips},
  {Price}, {Qiu}, \& {Thyagarajan}}]{kumar_narrow}
{Kumar}, P., {Shannon}, R.~M., {Flynn}, C., {et~al.} 2021, \mnras, 500, 2525,
  \dodoi{10.1093/mnras/staa3436}

\bibitem[{{Lanman} {et~al.}(2022){Lanman}, {Andersen}, {Chawla}, {Josephy},
  {Noble}, {Kaspi}, {Bandura}, {Bhardwaj}, {Boyle}, {Brar}, {Breitman},
  {Cassanelli}, {Dong}, {Fonseca}, {Gaensler}, {Good}, {Kaczmarek}, {Leung},
  {Masui}, {Meyers}, {Ng}, {Patel}, {Pearlman}, {Petroff}, {Pleunis},
  {Rafiei-Ravandi}, {Rahman}, {Sanghavi}, {Scholz}, {Shin}, {Stairs},
  {Tendulkar}, \& {Zwaniga}}]{lanman_20201124A}
{Lanman}, A.~E., {Andersen}, B.~C., {Chawla}, P., {et~al.} 2022, \apj, 927, 59,
  \dodoi{10.3847/1538-4357/ac4bc7}

\bibitem[{{Levin} {et~al.}(2020){Levin}, {Beloborodov}, \&
  {Bransgrove}}]{levin_precess}
{Levin}, Y., {Beloborodov}, A.~M., \& {Bransgrove}, A. 2020, \apjl, 895, L30,
  \dodoi{10.3847/2041-8213/ab8c4c}

\bibitem[{{Lewandowski} {et~al.}(2015){Lewandowski}, {Kowali{\'n}ska}, \&
  {Kijak}}]{lewandowski2015}
{Lewandowski}, W., {Kowali{\'n}ska}, M., \& {Kijak}, J. 2015, \mnras, 449,
  1570, \dodoi{10.1093/mnras/stv385}

\bibitem[{{Li} \& {Zanazzi}(2021)}]{Li_emission_chromatic}
{Li}, D., \& {Zanazzi}, J.~J. 2021, \apjl, 909, L25,
  \dodoi{10.3847/2041-8213/abeaa4}

\bibitem[{{Li} {et~al.}(2021{\natexlab{a}}){Li}, {Wang}, {Zhu}, {Zhang},
  {Zhang}, {Duan}, {Zhang}, {Feng}, {Tang}, {Chatterjee}, {Cordes}, {Cruces},
  {Dai}, {Gajjar}, {Hobbs}, {Jin}, {Kramer}, {Lorimer}, {Miao}, {Niu}, {Niu},
  {Pan}, {Qian}, {Spitler}, {Werthimer}, {Zhang}, {Wang}, {Xie}, {Yue},
  {Zhang}, {Zhi}, \& {Zhu}}]{FAST_121102A}
{Li}, D., {Wang}, P., {Zhu}, W.~W., {et~al.} 2021{\natexlab{a}}, \nat, 598,
  267, \dodoi{10.1038/s41586-021-03878-510.48550/arXiv.2107.08205}

\bibitem[{{Li} {et~al.}(2021{\natexlab{b}}){Li}, {Yang}, {Wang}, {Xu}, {Shao},
  {Liu}, \& {Dai}}]{Li_Be/X-ray_bin}
{Li}, Q.-C., {Yang}, Y.-P., {Wang}, F.~Y., {et~al.} 2021{\natexlab{b}}, \apjl,
  918, L5, \dodoi{10.3847/2041-8213/ac1922}

\bibitem[{{Lu} {et~al.}(2020){Lu}, {Kumar}, \& {Zhang}}]{lu_emission}
{Lu}, W., {Kumar}, P., \& {Zhang}, B. 2020, \mnras, 498, 1397,
  \dodoi{10.1093/mnras/staa2450}

\bibitem[{{Lyutikov} {et~al.}(2020){Lyutikov}, {Barkov}, \&
  {Giannios}}]{lyutikov_ob_binary}
{Lyutikov}, M., {Barkov}, M.~V., \& {Giannios}, D. 2020, \apjl, 893, L39,
  \dodoi{10.3847/2041-8213/ab87a4}

\bibitem[{{Marcote} {et~al.}(2020){Marcote}, {Nimmo}, {Hessels}, {Tendulkar},
  {Bassa}, {Paragi}, {Keimpema}, {Bhardwaj}, {Karuppusamy}, {Kaspi}, {Law},
  {Michilli}, {Aggarwal}, {Andersen}, {Archibald}, {Bandura}, {Bower}, {Boyle},
  {Brar}, {Burke-Spolaor}, {Butler}, {Cassanelli}, {Chawla}, {Demorest},
  {Dobbs}, {Fonseca}, {Giri}, {Good}, {Gourdji}, {Josephy}, {Kirichenko},
  {Kirsten}, {Landecker}, {Lang}, {Lazio}, {Li}, {Lin}, {Linford}, {Masui},
  {Mena-Parra}, {Naidu}, {Ng}, {Patel}, {Pen}, {Pleunis}, {Rafiei-Ravandi},
  {Rahman}, {Renard}, {Scholz}, {Siegel}, {Smith}, {Stairs}, {Vanderlinde}, \&
  {Zwaniga}}]{mar2020}
{Marcote}, B., {Nimmo}, K., {Hessels}, J.~W.~T., {et~al.} 2020, \nat, 577, 190,
  \dodoi{10.1038/s41586-019-1866-z}

\bibitem[{{Margalit} \& {Metzger}(2018)}]{mar_metz_R1}
{Margalit}, B., \& {Metzger}, B.~D. 2018, \apjl, 868, L4,
  \dodoi{10.3847/2041-8213/aaedad}

\bibitem[{Massey~Jr(1951)}]{massey1951kolmogorov}
Massey~Jr, F.~J. 1951, Journal of the American statistical Association, 46, 68

\bibitem[{{Masui}(2017)}]{mas17_bitshuffle}
{Masui}, K. 2017, {Bitshuffle: Filter for improving compression of typed binary
  data}, Astrophysics Source Code Library, record ascl:1712.004.
\newblock \doeprint{1712.004}

\bibitem[{{McKee} {et~al.}(2018){McKee}, {Lyne}, {Stappers}, {Bassa}, \&
  {Jordan}}]{Mckee_Crab}
{McKee}, J.~W., {Lyne}, A.~G., {Stappers}, B.~W., {Bassa}, C.~G., \& {Jordan},
  C.~A. 2018, \mnras, 479, 4216, \dodoi{10.1093/mnras/sty1727}

\bibitem[{{Mckinven} \& {CHIME/FRB Collaboration}(2022)}]{Mckinven_R117}
{Mckinven}, R., \& {CHIME/FRB Collaboration}. 2022, The Astronomer's Telegram,
  15679, 1

\bibitem[{{Mckinven} {et~al.}(2021){Mckinven}, {Michilli}, {Masui}, {Cubranic},
  {Gaensler}, {Ng}, {Bhardwaj}, {Leung}, {Boyle}, {Brar}, {Cassanelli}, {Li},
  {Mena-Parra}, {Rahman}, \& {Stairs}}]{mckinven_pol_pipe}
{Mckinven}, R., {Michilli}, D., {Masui}, K., {et~al.} 2021, \apj, 920, 138,
  \dodoi{10.3847/1538-4357/ac126a}

\bibitem[{{Mckinven} {et~al.}(2023){Mckinven}, {Gaensler}, {Michilli}, {Masui},
  {Kaspi}, {Bhardwaj}, {Cassanelli}, {Chawla}, {Dong}, {Fonseca}, {Leung},
  {Li}, {Ng}, {Patel}, {Petroff}, {Pearlman}, {Pleunis}, {Rafiei-Ravandi},
  {Rahman}, {Sand}, {Shin}, {Scholz}, {Stairs}, {Smith}, {Su}, \&
  {Tendulkar}}]{mckinven_2022_r3}
{Mckinven}, R., {Gaensler}, B.~M., {Michilli}, D., {et~al.} 2023, \apj, 950,
  12, \dodoi{10.3847/1538-4357/acc65f}

\bibitem[{{Merryfield} {et~al.}(2023){Merryfield}, {Tendulkar}, {Shin},
  {Andersen}, {Josephy}, {Good}, {Dong}, {Masui}, {Lang}, {M{\"u}nchmeyer},
  {Brar}, {Cassanelli}, {Dobbs}, {Fonseca}, {Kaspi}, {Mena-Parra}, {Pleunis},
  {Rafiei-Ravandi}, {Sand}, {Scholz}, {Smith}, \& {Stairs}}]{marcus_injection}
{Merryfield}, M., {Tendulkar}, S.~P., {Shin}, K., {et~al.} 2023, \aj, 165, 152,
  \dodoi{10.3847/1538-3881/ac9ab5}

\bibitem[{{Metzger} {et~al.}(2019){Metzger}, {Margalit}, \& {Sironi}}]{mms19}
{Metzger}, B.~D., {Margalit}, B., \& {Sironi}, L. 2019, \mnras, 485, 4091,
  \dodoi{10.1093/mnras/stz700}

\bibitem[{{Michilli} {et~al.}(2018){Michilli}, {Seymour}, {Hessels}, {Spitler},
  {Gajjar}, {Archibald}, {Bower}, {Chatterjee}, {Cordes}, {Gourdji}, {Heald},
  {Kaspi}, {Law}, {Sobey}, {Adams}, {Bassa}, {Bogdanov}, {Brinkman},
  {Demorest}, {Fernandez}, {Hellbourg}, {Lazio}, {Lynch}, {Maddox}, {Marcote},
  {McLaughlin}, {Paragi}, {Ransom}, {Scholz}, {Siemion}, {Tendulkar}, {van
  Rooy}, {Wharton}, \& {Whitlow}}]{mic2018}
{Michilli}, D., {Seymour}, A., {Hessels}, J.~W.~T., {et~al.} 2018, \nat, 553,
  182, \dodoi{10.1038/nature25149}

\bibitem[{{Michilli} {et~al.}(2021){Michilli}, {Masui}, {Mckinven}, {Cubranic},
  {Bruneault}, {Brar}, {Patel}, {Boyle}, {Stairs}, {Renard}, {Bandura},
  {Berger}, {Breitman}, {Cassanelli}, {Dobbs}, {Kaspi}, {Leung}, {Mena-Parra},
  {Pleunis}, {Russell}, {Scholz}, {Siegel}, {Tendulkar}, \&
  {Vanderlinde}}]{Baseband}
{Michilli}, D., {Masui}, K.~W., {Mckinven}, R., {et~al.} 2021, \apj, 910, 147,
  \dodoi{10.3847/1538-4357/abe626}

\bibitem[{{Michilli} {et~al.}(2022){Michilli}, {Bhardwaj}, {Brar}, {Patel},
  {Gaensler}, {Kaspi}, {Kirichenko}, {Masui}, {Sand}, {Scholz}, {Shin},
  {Stairs}, {Cassanelli}, {Cook}, {Dobbs}, {Dong}, {Fonseca}, {Ibik},
  {Kaczmarek}, {Leung}, {Pearlman}, {Petroff}, {Pleunis}, {Rafiei-Ravandi},
  {Sanghavi}, \& {Tendulkar}}]{RN12_basebandloc}
{Michilli}, D., {Bhardwaj}, M., {Brar}, C., {et~al.} 2022, arXiv e-prints,
  arXiv:2212.11941, \dodoi{10.48550/arXiv.2212.11941}

\bibitem[{{Nimmo} {et~al.}(2021){Nimmo}, {Hessels}, {Keimpema}, {Archibald},
  {Cordes}, {Karuppusamy}, {Kirsten}, {Li}, {Marcote}, \&
  {Paragi}}]{nimmo2021_micro}
{Nimmo}, K., {Hessels}, J.~W.~T., {Keimpema}, A., {et~al.} 2021, Nature
  Astronomy, 5, 594, \dodoi{10.1038/s41550-021-01321-3}

\bibitem[{Nimmo {et~al.}(2022)Nimmo, Hessels, Kirsten, Keimpema, Cordes,
  Snelders, Hewitt, Karuppusamy, Archibald, Bezrukovs,
  {et~al.}}]{nimmo2022burst}
Nimmo, K., Hessels, J., Kirsten, F., {et~al.} 2022, Nature Astronomy, 1

\bibitem[{{Nimmo} {et~al.}(2023){Nimmo}, {Hessels}, {Snelders}, {Karuppusamy},
  {Hewitt}, {Kirsten}, {Marcote}, {Bach}, {Bansod}, {Barr}, {Behrend},
  {Bezrukovs}, {Buttaccio}, {Feiler}, {Gawro{\'n}ski}, {Lindqvist}, {Orbidans},
  {Puchalska}, {Wang}, {Winchen}, {Wolak}, {Wu}, \& {Yuan}}]{nimmo_burst_storm}
{Nimmo}, K., {Hessels}, J.~W.~T., {Snelders}, M.~P., {et~al.} 2023, \mnras,
  \dodoi{10.1093/mnras/stad269}

\bibitem[{{Ocker} {et~al.}(2023){Ocker}, {Cordes}, {Chatterjee}, {Li}, {Niu},
  {McKee}, {Law}, \& {Anna-Thomas}}]{ocker_20190520B}
{Ocker}, S.~K., {Cordes}, J.~M., {Chatterjee}, S., {et~al.} 2023, \mnras, 519,
  821, \dodoi{10.1093/mnras/stac3547}

\bibitem[{{Olausen} \& {Kaspi}(2014)}]{Olausen_and_Kaspi}
{Olausen}, S.~A., \& {Kaspi}, V.~M. 2014, \apjs, 212, 6,
  \dodoi{10.1088/0067-0049/212/1/6}

\bibitem[{Pastor-Marazuela {et~al.}(2021)Pastor-Marazuela, Connor, van Leeuwen,
  Maan, Ter~Veen, Bilous, Oostrum, Petroff, Straal, Vohl,
  {et~al.}}]{pastormarazuela2020chromatic_180916}
Pastor-Marazuela, I., Connor, L., van Leeuwen, J., {et~al.} 2021, Nature, 596,
  505

\bibitem[{{Petroff} {et~al.}(2022){Petroff}, {Hessels}, \&
  {Lorimer}}]{petroff_dawn_2022}
{Petroff}, E., {Hessels}, J.~W.~T., \& {Lorimer}, D.~R. 2022, \aapr, 30, 2,
  \dodoi{10.1007/s00159-022-00139-w}

\bibitem[{Platts {et~al.}(2019)Platts, Weltman, Walters, Tendulkar, Gordin, \&
  Kandhai}]{platts2019living}
Platts, E., Weltman, A., Walters, A., {et~al.} 2019, Physics Reports, 821, 1

\bibitem[{{Plavin} {et~al.}(2022){Plavin}, {Paragi}, {Marcote}, {Keimpema},
  {Hessels}, {Nimmo}, {Vedantham}, \& {Spitler}}]{plavin_r1}
{Plavin}, A., {Paragi}, Z., {Marcote}, B., {et~al.} 2022, \mnras, 511, 6033,
  \dodoi{10.1093/mnras/stac500}

\bibitem[{Pleunis {et~al.}(2021)Pleunis, Michilli, Bassa, Hessels, Naidu,
  Andersen, Chawla, Fonseca, Gopinath, Kaspi, \& et~al.}]{Pleunis_2021_r3}
Pleunis, Z., Michilli, D., Bassa, C.~G., {et~al.} 2021, The Astrophysical
  Journal Letters, 911, L3, \dodoi{10.3847/2041-8213/abec72}

\bibitem[{{Pleunis} {et~al.}(2021){Pleunis}, {Good}, {Kaspi}, {Mckinven},
  {Ransom}, {Scholz}, {Bandura}, {Bhardwaj}, {Boyle}, {Brar}, {Cassanelli},
  {Chawla}, {(Adam) Dong}, {Fonseca}, {Gaensler}, {Josephy}, {Kaczmarek},
  {Leung}, {Lin}, {Masui}, {Mena-Parra}, {Michilli}, {Ng}, {Patel},
  {Rafiei-Ravandi}, {Rahman}, {Sanghavi}, {Shin}, {Smith}, {Stairs}, \&
  {Tendulkar}}]{pleunismorph}
{Pleunis}, Z., {Good}, D.~C., {Kaspi}, V.~M., {et~al.} 2021, \apj, 923, 1,
  \dodoi{10.3847/1538-4357/ac33ac}

\bibitem[{{Rafiei-Ravandi} \& {Smith}(2022)}]{masoudrfi}
{Rafiei-Ravandi}, M., \& {Smith}, K.~M. 2022, arXiv e-prints, arXiv:2206.07292.
\newblock \doarXiv{2206.07292}

\bibitem[{Rajwade {et~al.}(2020)Rajwade, Mickaliger, Stappers, Morello,
  Agarwal, Bassa, Breton, Caleb, Karastergiou, Keane, \&
  et~al.}]{Rajwade_2020_121102_period}
Rajwade, K.~M., Mickaliger, M.~B., Stappers, B.~W., {et~al.} 2020, Monthly
  Notices of the Royal Astronomical Society, 495, 3551–3558,
  \dodoi{10.1093/mnras/staa1237}

\bibitem[{{Sand} {et~al.}(2022){Sand}, {Faber}, {Gajjar}, {Michilli},
  {Andersen}, {Joshi}, {Kudale}, {Pilia}, {Brzycki}, {Cassanelli}, {Croft},
  {Dey}, {John}, {Leung}, {Mckinven}, {Ng}, {Pearlman}, {Petroff}, {Price},
  {Siemion}, {Smith}, \& {Tendulkar}}]{sand2022}
{Sand}, K.~R., {Faber}, J.~T., {Gajjar}, V., {et~al.} 2022, \apj, 932, 98,
  \dodoi{10.3847/1538-4357/ac6cee}

\bibitem[{Scholz \& Stephens(1987)}]{scholz1987k}
Scholz, F.~W., \& Stephens, M.~A. 1987, Journal of the American Statistical
  Association, 82, 918

\bibitem[{Seabold \& Perktold(2010)}]{seabold2010statsmodels}
Seabold, S., \& Perktold, J. 2010, in 9th Python in Science Conference

\bibitem[{{Seymour} {et~al.}(2019){Seymour}, {Michilli}, \&
  {Pleunis}}]{dm_phase}
{Seymour}, A., {Michilli}, D., \& {Pleunis}, Z. 2019, {DM\_phase: Algorithm for
  correcting dispersion of radio signals}.
\newblock \doeprint{1910.004}

\bibitem[{{Sridhar} {et~al.}(2021){Sridhar}, {Metzger}, {Beniamini},
  {Margalit}, {Renzo}, {Sironi}, \& {Kovlakas}}]{sridhar_xray_bin}
{Sridhar}, N., {Metzger}, B.~D., {Beniamini}, P., {et~al.} 2021, \apj, 917, 13,
  \dodoi{10.3847/1538-4357/ac0140}

\bibitem[{{Tendulkar} {et~al.}(2021){Tendulkar}, {Gil de Paz}, {Kirichenko},
  {Hessels}, {Bhardwaj}, {{\'A}vila}, {Bassa}, {Chawla}, {Fonseca}, {Kaspi},
  {Keimpema}, {Kirsten}, {Lazio}, {Marcote}, {Masui}, {Nimmo}, {Paragi},
  {Rahman}, {Pay{\'a}}, {Scholz}, \& {Stairs}}]{ten2021_60pc}
{Tendulkar}, S.~P., {Gil de Paz}, A., {Kirichenko}, A.~Y., {et~al.} 2021,
  \apjl, 908, L12, \dodoi{10.3847/2041-8213/abdb38}

\bibitem[{{The HDF Group}(1997--2023)}]{hdf5}
{The HDF Group}. 1997--2023, {Hierarchical Data Format, version 5}

\bibitem[{{The pandas development team}(2020)}]{pandas}
{The pandas development team}. 2020, pandas-dev/pandas: Pandas 1.1.5, v1.1.5,
  Zenodo, \dodoi{10.5281/zenodo.4309786}

\bibitem[{{Tong} {et~al.}(2020){Tong}, {Wang}, \& {Wang}}]{tong_forced_precess}
{Tong}, H., {Wang}, W., \& {Wang}, H.-G. 2020, Research in Astronomy and
  Astrophysics, 20, 142, \dodoi{10.1088/1674-4527/20/9/142}

\bibitem[{{van Haaften} {et~al.}(2012){van Haaften}, {Nelemans}, {Voss},
  {Wood}, \& {Kuijpers}}]{UCB_haaften}
{van Haaften}, L.~M., {Nelemans}, G., {Voss}, R., {Wood}, M.~A., \& {Kuijpers},
  J. 2012, \aap, 537, A104, \dodoi{10.1051/0004-6361/201117880}

\bibitem[{{Vedantham} \& {Phinney}(2019)}]{vedantham_phinney_2019}
{Vedantham}, H.~K., \& {Phinney}, E.~S. 2019, \mnras, 483, 971,
  \dodoi{10.1093/mnras/sty2948}

\bibitem[{Virtanen {et~al.}(2020)Virtanen, Gommers, Oliphant, Haberland, Reddy,
  Cournapeau, Burovski, Peterson, Weckesser, Bright, \&
  et~al.}]{scipy_Virtanen_2020}
Virtanen, P., Gommers, R., Oliphant, T.~E., {et~al.} 2020, Nature Methods, 17,
  261–272, \dodoi{10.1038/s41592-019-0686-2}

\bibitem[{{Wada} {et~al.}(2021){Wada}, {Ioka}, \& {Zhang}}]{Wada_bin_comb}
{Wada}, T., {Ioka}, K., \& {Zhang}, B. 2021, \apj, 920, 54,
  \dodoi{10.3847/1538-4357/ac127a}

\bibitem[{{Wang} {et~al.}(2022){Wang}, {Zhang}, {Dai}, \&
  {Cheng}}]{Wang_binary}
{Wang}, F.~Y., {Zhang}, G.~Q., {Dai}, Z.~G., \& {Cheng}, K.~S. 2022, Nature
  Communications, 13, 4382, \dodoi{10.1038/s41467-022-31923-y}

\bibitem[{{Xu} {et~al.}(2022){Xu}, {Niu}, {Chen}, {Lee}, {Zhu}, {Dong},
  {Zhang}, {Jiang}, {Wang}, {Xu}, {Zhang}, {Fu}, {Filippenko}, {Peng}, {Zhou},
  {Zhang}, {Wang}, {Feng}, {Li}, {Brink}, {Li}, {Lu}, {Yang}, {Caballero},
  {Cai}, {Chen}, {Dai}, {Djorgovski}, {Esamdin}, {Gan}, {Guhathakurta}, {Han},
  {Hao}, {Huang}, {Jiang}, {Li}, {Li}, {Li}, {Li}, {Li}, {Liu}, {Luo}, {Men},
  {Niu}, {Peng}, {Qian}, {Song}, {Stern}, {Stockton}, {Sun}, {Wang}, {Wang},
  {Wang}, {Wang}, {Wu}, {Xiao}, {Xiong}, {Xu}, {Xu}, {Yang}, {Yang}, {Yao},
  {Yi}, {Yue}, {Yu}, {Yu}, {Yuan}, {Zhang}, {Zhang}, {Zhang}, {Zhao}, {Zheng},
  {Zhu}, \& {Zou}}]{FAST_R67}
{Xu}, H., {Niu}, J.~R., {Chen}, P., {et~al.} 2022, \nat, 609, 685,
  \dodoi{10.1038/s41586-022-05071-810.48550/arXiv.2111.11764}

\bibitem[{{Yang} \& {Zou}(2020)}]{Yang_orbit_precess}
{Yang}, H., \& {Zou}, Y.-C. 2020, \apjl, 893, L31,
  \dodoi{10.3847/2041-8213/ab800f}

\bibitem[{{Younes} {et~al.}(2020){Younes}, {G{\"u}ver}, {Kouveliotou},
  {Baring}, {Hu}, {Wadiasingh}, {Begi{\c{c}}arslan}, {Enoto},
  {G{\"o}{\u{g}}{\"u}{\c{s}}}, {Lin}, {Harding}, {van der Horst}, {Majid},
  {Guillot}, \& {Malacaria}}]{Younes_Sgr}
{Younes}, G., {G{\"u}ver}, T., {Kouveliotou}, C., {et~al.} 2020, \apjl, 904,
  L21, \dodoi{10.3847/2041-8213/abc94c}

\bibitem[{{Zanazzi} \& {Lai}(2020)}]{zanazzi_free_precess}
{Zanazzi}, J.~J., \& {Lai}, D. 2020, \apjl, 892, L15,
  \dodoi{10.3847/2041-8213/ab7cdd}

\bibitem[{{Zhang}(2018)}]{zhang_R1}
{Zhang}, B. 2018, \apjl, 854, L21, \dodoi{10.3847/2041-8213/aaadba}

\bibitem[{{Zhang} {et~al.}(2023){Zhang}, {Li}, {Zhang}, {Cao}, {Feng}, {Wang},
  {Qu}, {Niu}, {Zhu}, {Han}, {Jiang}, {Lee}, {Li}, {Luo}, {Niu}, {Tsai},
  {Wang}, {Wang}, {Wu}, {Xu}, {Yang}, {Zhang}, {Zhou}, \&
  {Zhu}}]{Zhang_2023_R117}
{Zhang}, Y.-K., {Li}, D., {Zhang}, B., {et~al.} 2023, arXiv e-prints,
  arXiv:2304.14665, \dodoi{10.48550/arXiv.2304.14665}

\bibitem[{{Zhao} {et~al.}(2023){Zhao}, {Zhang}, {Wang}, \&
  {Dai}}]{Zhao_RM_reversal}
{Zhao}, Z.~Y., {Zhang}, G.~Q., {Wang}, F.~Y., \& {Dai}, Z.~G. 2023, \apj, 942,
  102, \dodoi{10.3847/1538-4357/aca66b}

\bibitem[{Zonca {et~al.}(2019)Zonca, Singer, Lenz, Reinecke, Rosset, Hivon, \&
  Gorski}]{healpy}
Zonca, A., Singer, L., Lenz, D., {et~al.} 2019, Journal of Open Source
  Software, 4, 1298, \dodoi{10.21105/joss.01298}

\end{thebibliography}
\bibliographystyle{aasjournal}



\end{document}